\documentclass[traditabstract]{aa} % for the abstract without structuration 
\usepackage{graphicx}
\usepackage{txfonts}
\usepackage{natbib}
\usepackage{longtable}
\begin{document}
    \title{A line confusion limited millimeter survey of \object{Orion
 KL} (I):
 \\
     sulfur carbon chains\thanks{Appendix A (online Figures) and Appendix B (online Tables)
     are only available in
     electronic form via http://www.edpscience.org}}
%   \subtitle{I. Overviewing the $\kappa$-mechanism}

   \author{B. Tercero, J. Cernicharo,
  J. R. Pardo and 
%          \and
          J. R. Goicoechea
          }

   \institute{Centro de Astrobiolog\'ia (CSIC-INTA). Departamento de
              Astrof\'isica Molecular. Ctra. de Aljalvir Km 4, 28850
              Torrej\'on de Ardoz, Madrid, Spain \\
              \email{belen@damir.iem.csic.es; jcernicharo@cab.inta-csic.es;
              pardo@damir.iem.csic.es; jr.goicoechea@cab.inta-csic.es}
}
   \date{Received October 19, 2009; accepted April 15, 2010}

% \abstract{}{}{}{}{} 
% 5 {} token are mandatory
 
  \abstract
{We perform a sensitive (line confusion limited), single-side
band spectral survey towards Orion KL with the IRAM 30m
telescope, covering the following frequency ranges: 
80-115.5 GHz, 130-178 GHz, and 197-281 GHz. We detect 
more than 14 400 spectral 
features of which 10 040 have been identified up to date and attributed
to 43 different molecules, 
including 148 isotopologues and lines from vibrationally excited
states.
In this paper, we focus on the study of OCS, HCS$^+$, H$_2$CS, CS, CCS,
C$_3$S, and their isotopologues. 
In addition, we map the OCS $J$=18-17 line
and complete complementary observations of several
OCS lines at selected positions around \object{Orion IRc2} (the position
selected for the survey).
We report the first detection of
OCS $\nu_2$ = 1 and $\nu_3$ = 1 vibrationally
excited states in space and the first detection of C$_3$S in warm clouds.
Most of CCS, and almost all C$_3$S, line emission arises from
the hot core indicating an enhancement of their abundances in warm and
dense gas. 
Column densities and isotopic ratios have been calculated
using a large velocity gradient (LVG) excitation and radiative transfer
code (for the low density gas components) and a local thermal equilibrium 
(LTE) code (appropriate for the warm and dense hot core component), which takes 
into account the different 
cloud components known to exist towards Orion KL, 
the \textit{extended ridge},
\textit{compact ridge}, \textit{plateau}, and \textit{hot
core}.
The vibrational temperature derived from OCS $\nu_2$ = 1 and
$\nu_3$ = 1 levels is $\simeq$210 K, similar to the gas kinetic
temperature in the hot core. These OCS high energy levels are
probably  
%mainly 
pumped by absorption of IR dust photons.
We derive an upper limit to the OC$_3$S,
H$_2$CCS, HNCS,
HOCS$^+$, and NCS column densities.
Finally, we discuss the D/H abundance ratio and 
infer the following isotopic abundances:
$^{12}$C/$^{13}$C = 45$\pm$20, $^{32}$S/$^{34}$S
= 20$\pm$6, $^{32}$S/$^{33}$S = 75$\pm$29, 
and $^{16}$O/$^{18}$O = 250$\pm$135.}

   \keywords{Surveys -- Stars: formation --
                ISM: abundances -- ISM: clouds -- ISM: molecules --
                Radio lines: ISM
               }

\titlerunning{Survey towards Orion KL: sulfur carbon chains}  
\authorrunning{B. Tercero et al.}

\maketitle
%
%________________________________________________________________

\section{Introduction}
\label{sect_int}

The Orion KL (Kleinmann-Low) cloud is the closest ($\simeq$ 414 pc,
\citealt{men07})
and most well studied  
high mass star-forming region in our Galaxy (see, e. g.,
\citealt{gen89} for review).
The prevailing chemistry of 
the cloud is particularly complex as a result of the interaction of the
newly formed protostars, outflows, and their environment. The
evaporation of dust
mantles and the high gas temperatures
produce a wide variety of molecules
in the gas phase that are responsible for a spectacularly prolific and intense
line spectrum \citep{bla87, bro88, cha97}.

%\bf{
%Early studies
%soon realized that the large scale distribution of gas and
%dust was heavily influenced by violent phenomena such
%as the interaction of compact and large scale outflows
%with the quiescent gas producing strong line and continuum
%emission. 
Near- and mid-IR subarcsecond
resolution imaging and (sub)millimeter interferometric
observations have identified the main sources of 
luminosity, heating, and dynamics in the region. 
At first, IRc2 was believed to be the responsible for 
this complex environment.
%of luminosity, heating and dynamics of the region. However,
%Near- and mid-IR subarcsecond
%resolution imaging and (sub)millimeter interferometric
%observations have s our understanding
%of the region. First, 
However, the 8-12 $\mu$m emission
peak of IRc2 is not coincident with the
% Orion SiO maser
%origin (related to 
the origin of the outflow(s) (and the Orion SiO maser origin),
and its intrinsic IR luminosity (L$\approx$1000 L$\sun$) 
is only a fraction of the luminosity of the entire system (\citealt{gez98}).
In addition,
%gezari et al. 1998
3.6-22 $\mu$m images indicate that IRc2 is resolved into
four non self-luminous components.
Therefore, IRc2 is not presently the powerful engine
of Orion KL and its nature
remains unclear \citep{dou93, shu04, gre04}.
%(Dougados et al. 1993; Shuping et al.
%2004; Greenhill et al. 2004). 
%A new step forward was
%given by 
\citet{men95} identified
the very embedded radio continuum source I (a young star 
with a very high luminosity without an infrared
counterpart, $\simeq$10$^5$ L$_\odot$,  
\citealt{gez98, gre04}, located
0''.5 south of IRc2) as the source coinciding with the
centroid of the SiO maser distribution \citep{pla09, zap09a, god09b}.
They also detected the radio continuum emission of IR
source $n$, suggesting this source as another precursor
of the large-scale phenomena.
In addition, \citet{beu04} detected a sub-millimeter source without
IR and centimeter counterparts, SMA1, previously predicted
by \citet{dev02}, which may be the source
driving the high velocity outflow \citep{beu08}.
% and suggested that it could also contribute to
%the origin of some of the phenomena observed at larger
%scales.
Thus, the core of Orion KL
contains the compact HII regions $I$ and $n$ (in addition to BN,
which was resolved with high resolution at 7 mm by \citealt{rod09}),
which appear to be receding from a common point,
an originally massive stellar system that 
disintegrated $\simeq$500 years ago \citep{gom05, zap09b}.
%Thus, in addition to BN, the core of Orion KL
%contains at least two more compact HII regions (I and
%n) that seem to be running away from a common point,
%suggesting that BN, I and n were originally part of a
%common massive stellar system that disintegrated $\simeq$500
%years ago \citep{gom05}.
%(G¡äomez et al. 2005). 
Finally, submm aperture synthesis
line surveys provided the spatial location
and extent of many molecular species \citep{bla96, wri96,
liu02, beu05, god09b, pla09, zap09a}.
%(Blake et al. 1996;
%Wright et al. 1996; Liu et al. 2002; Beuther et al. 2005).

The chemical complexity of Orion KL has been demonstrated by 
several line surveys performed
at different frequency ranges: 72.2-91.1 GHz by \citet{joh84};
215-247 GHz by \citet{sut85}; 247-263 GHz by \citet{bla86}; 200.7-202.3, 
203.7-205.3 and 330-360 GHz by \citet{jew89}; 70-115 GHz by \citet{tur89}; 
257-273 GHz by \citet{gre91}; 150-160 GHz by \citet{ziu93}; 
325-360 GHz by \citet{sch97};
607-725 GHz by \citet{sch01}; 138-150 GHz by \citet{lee01}; 
159.7- 164.7 GHz by 
\citet{lee02}; 455-507 GHz by \citet{whi03}; 795-903 GHz 
by \citet{com05}; 44-188 $\mu$m by \citet{ler06}; 486-492, 541-577
GHz by \citet{olo07} and \citet{per07}; and 42.3-43.6 GHz by \citet{god09a}.

In spite of this large amount of data, no line confusion limited survey
has been carried out so far with a large single dish telescope. 
We performed such a line survey  
towards Orion IRc2 with the IRAM 30-m
telescope at wide frequency 
ranges (a total frequency coverage of $\simeq$ 168 GHz). 
Our main goal was to obtain a deep insight into the 
molecular content and chemistry of the Orion KL, an archetype high
mass star-forming region (SFR), and to improve our knowledge of its
prevailing physical conditions. It also allows us to search for new 
molecular species and new isotopologues, as well as the rotational
emission of vibrationally excited 
states of molecules already known to exist in this source. Since the
amount and complexity of the data is large, we divided
our analysis into families of molecules so that model development and discussions
could be more focused. In this paper, 
we concentrate 
on sulfur carbon chains, in particular carbonyl sulfide OCS 
(see previous studies by \citealt{gol81}; \citealt{eva91};
\citealt{wri96};
\citealt{cha97}), CS (previously analyzed by \citealt{has84}; 
\citealt{mur91}; \citealt{zen95}; \citealt{wri96};
\citealt{joh03}), H$_2$CS (\citealt{min91}; \citealt{gar85}), HCS$^+$, 
CCS, CCCS, and their isotopologues.  

Column density calculations, and
therefore the estimation of isotopic abundance ratios and 
molecular excitation, have improved, with respect to
previous works, due to the much larger number of available 
lines, their consistent calibration across the explored frequency
range, the up-to-date information about the physical properties of 
the region and molecular constants, and the use of a LVG radiative
transfer code to derive reliable physical and chemical parameters. 
Modeled brightness temperatures obtained from a fit to all observed
lines have been convolved with the telescope beam profile, assuming a
given size for each cloud component, to provide accurate
source-averaged, and not beam-averaged, molecular column densities.

After presenting the line survey (Sects. \ref{sect_obs} and
\ref{sect_sur}), 
this work concentrates 
on the detection of OCS, HCS$^+$, H$_2$CS, CS, CCS, and CCCS lines and
their analysis, as well as providing upper
limits to the abundance of several non-detected sulfur-carbon-chain bearing
molecules such us OC$_3$S, H$_2$CCS, HNCS,
HOCS$^+$, and NCS (Sects. \ref{sect_res} to \ref{sect_vib}). 
This is the first of a series of papers dedicated to the
analysis of the millimeter
emission from different molecular families towards Orion KL.
  
\section{Observations and data analysis}
\label{sect_obs}

%%%%%------------one column table
\begin{table}%t1
\begin{center}
\caption{$\eta_{MB}$ and HPBW along the covered frequency range\label{tab_eta_hpbw}}
\begin{tabular}{lll}
\hline 
\hline
 Frequency (GHz) & $\eta_{MB}$ & HPBW ('')\\
\hline \\
86 & 0.82 & 29.0\\
110 & 0.79 & 22.0\\
145 & 0.74 & 17.0\\
170 & 0.70 & 14.5\\
210 & 0.62 & 12.0\\
235 & 0.57 & 10.5\\
260 & 0.52 & 9.5\\
279 & 0.48 & 9.0\\
\hline
\end{tabular}
\end{center}
Note.-$\eta_{MB}$ and HPBW along the covered frequency range.
\end{table}
%%%%%%-----------------------------------------------------------

\begin{figure}%f1
\includegraphics[angle=0,scale=.5]{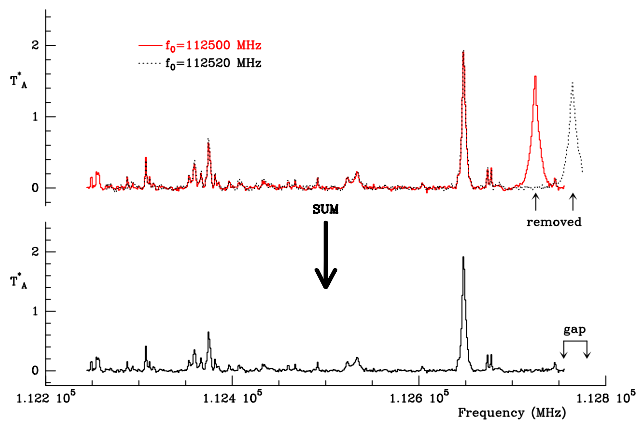}
\caption{The top panel shows two superimposed spectra corresponding to
different frequency settings (112 500 and
112 520 MHz). The 40 MHz displaced line is the 115.5 GHz CO line in the
image side band. The bottom panel shows the final spectrum resulting
from our procedure to eliminate the image side band (see text,
Sect. \ref{sect_obs}). We are confident that all lines above 0.05 K
have frequencies correctly assigned.}
\label{fig_isb}
\end{figure}

The observations were carried out using the IRAM 30m radiotelescope
during September 2004 (3 mm and 1.3 mm windows), March 2005 (full 2 mm
window), 
April 2005 (completion of 3 mm and 1.3 mm windows), and January 2007 (maps and  
pointed observations at particular positions). Four SiS receivers  
operating at 3, 2, and 1.3 mm were used 
simultaneously with image sideband rejections within 20-27 dB 
(3 mm receivers), 
12-16 dB (2 mm receivers) and $\simeq$13 dB (1.3 mm receivers). 
System temperatures were in the range 
100-350 K for the 3 mm receivers, 200-500 K for the 2 mm 
receivers, and 200-800 K for the 1.3 mm receivers, depending on the particular 
frequency, weather conditions, and source elevation. For frequencies
in the range 172-178 GHz, the system temperature was significantly
higher, 1000-4000 K, due to proximity of the atmospheric water line at 
183.31 GHz. The intensity scale was calibrated using two absorbers at
different temperatures and the atmospheric transmission model 
(ATM, \citealt{cer85}; \citealt{par01a}). 
Table \ref{tab_eta_hpbw} shows the variation in the main beam efficiency
($\eta_{MB}$) and the half power beam width
(HPBW) across the covered frequency range.
The error beam contribution to the observed line intensities
is negligible for heavy polyatomic molecules because their emission originates
in a compact region.
However, the error beam contribution of the low-J line extended emission of
abundant species (HCO+, HCN, CN or CS) can be significant (up to
$\simeq$10-20\% of the
observed intensities at 1mm). Deriving the correct brightness
temperature for these lines
requires large-scale mapping.
%{\bf{For most heavy species the contribution of the error beam
%%to the observed intensity will be small as these complex molecules emit 
%just in a small region. However, for low-J lines of abundant species
%(HCO$^+$, HCN, CN, CS),  the contribution of the extended emission,
%as seen through the error beam, to the observed intensities at 1mm 
%should be taken into account (up to $\simeq$10-20\%). Deriving the correct 
%brightness temperatures for these lines will require large scale maps 
%that are not available so far.}}

Pointing and focus were regularly
checked on the nearby quasars 0420-014 and 0528+134. Observations
were made in the balanced wobbler-switching mode, with a wobbling frequency 
of 0.5 Hz and a beam throw in azimuth of $\pm$240''.
No contamination from the off position affected our observations 
except for a marginal one at the lowest elevations ($\sim$ 25$^{\circ}$) for 
molecules showing low $J$ emission along the extended ridge.

Two filter banks with 512$\times$1 MHz channels and a 
correlator providing two 512 MHz bandwidths and
1.25 MHz resolution were used as backends. We pointed
towards IRc2 source at $\alpha$$_{2000.0}$ = 5$^h$ 35$^m$
14.5$^s$, $\delta$$_{2000.0}$ = $-5$$^{\circ}$ 22' 30.0'' (J2000.0) 
corresponding to
the 'survey position'. We observed two additional 
positions to target 
both the compact ridge ($\alpha$$_{2000.0}$ = 5$^h$
35$^m$ 14.3$^s$, 
$\delta$$_{2000.0}$ = -5$^{\circ}$ 22' 37.0'') and the ambient molecular cloud 
($\alpha$$_{2000.0}$ = 5$^h$ 35$^m$ 15.3$^s$, $\delta$$_{2000.0}$
 = -5$^{\circ}$ 22' 09.0''). Figure 15 of \citet{wri96} shows a 3 mm dust image
depicting all positions quoted above.

The spectra shown in all figures are in units of antenna temperature, $T^*_A$,
corrected for atmospheric absorption and spillover losses. In spite of the
good receiver image-band rejection, each setting was repeated 
at a slightly shifted frequency (10-20 MHz)
to manually identify and remove all features arising from the image side band.
The spectra from different frequency settings were used to identify all
potential contaminating lines from the image side band. In some
cases, it was impossible to eliminate the contribution of the image
side band and we removed the signal in those contaminated channels 
leaving holes in
the data. The total frequencies blanked this way represent less than
0.5 \% of the total frequency coverage. Figure \ref{fig_isb} shows our
procedure for removing the image side band lines.
We removed most of the features above a 0.05 K threshold.

\section{The line survey}
\label{sect_sur}
%%%%%------------two column table
\begin{table*}%t2
\begin{center}
\caption{The assumed Orion KL spectral components.\label{tab_prop}}
\begin{tabular}{lllll}
\hline 
\hline
 Parameter & Extended ridge & Compact ridge & Plateau & Hot core\\
 & (ER) & (CR) & (P) & (HC)\\
\hline \\
Source diameter ('') & 120 & 15 & 30 & 10 \\
Offset (from IRc2) ('') & 0 & 7 & 0 & 2 \\
n(H$_2$) (cm$^{-3}$) & 1.0$\times$10$^{5}$ &
1.0$\times$10$^{6}$ & 1.0$\times$10$^{6}$ & 5.0$\times$10$^{7}$ \\
T$_k$ (K) & 60 & 110 & 125 & 225 \\
$\Delta$v$_{FWHM}$ (km s$^{-1}$) & 4 & 4 & 25 & 10 \\
v$_{LSR}$ (km s$^{-1}$) & 9 & 7.5 & 6 & 5.5 \\
\hline
\end{tabular}
\end{center}
Note.-Obtained physical parameters for Orion KL.
\end{table*}
%%%%%%-----------------------------------------------------------
%%%--------------------------------figures two columns
\begin{figure*}%f2
\includegraphics[angle=270,scale=.7]{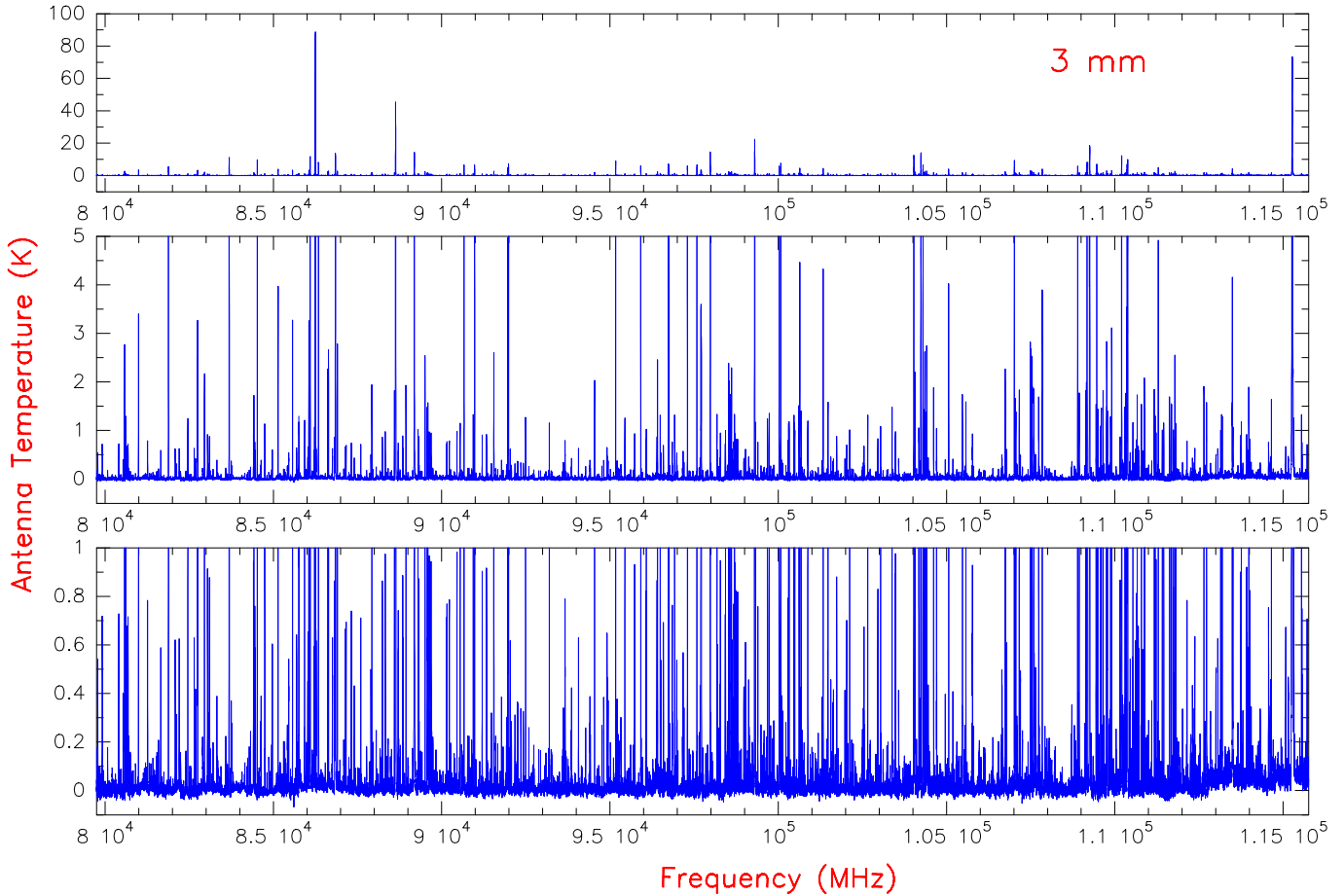}
\caption{Molecular line survey of Orion KL at 3 mm. The top panel
  shows the total intensity scale; the middle and the bottom panels
  show a zoom of the total intensity. A v$_{LSR}$ of 9 km s$^{-1}$ is
assumed.}
\label{fig_3mm}
\end{figure*}

\begin{figure*}%f3
\includegraphics[angle=270,scale=.7]{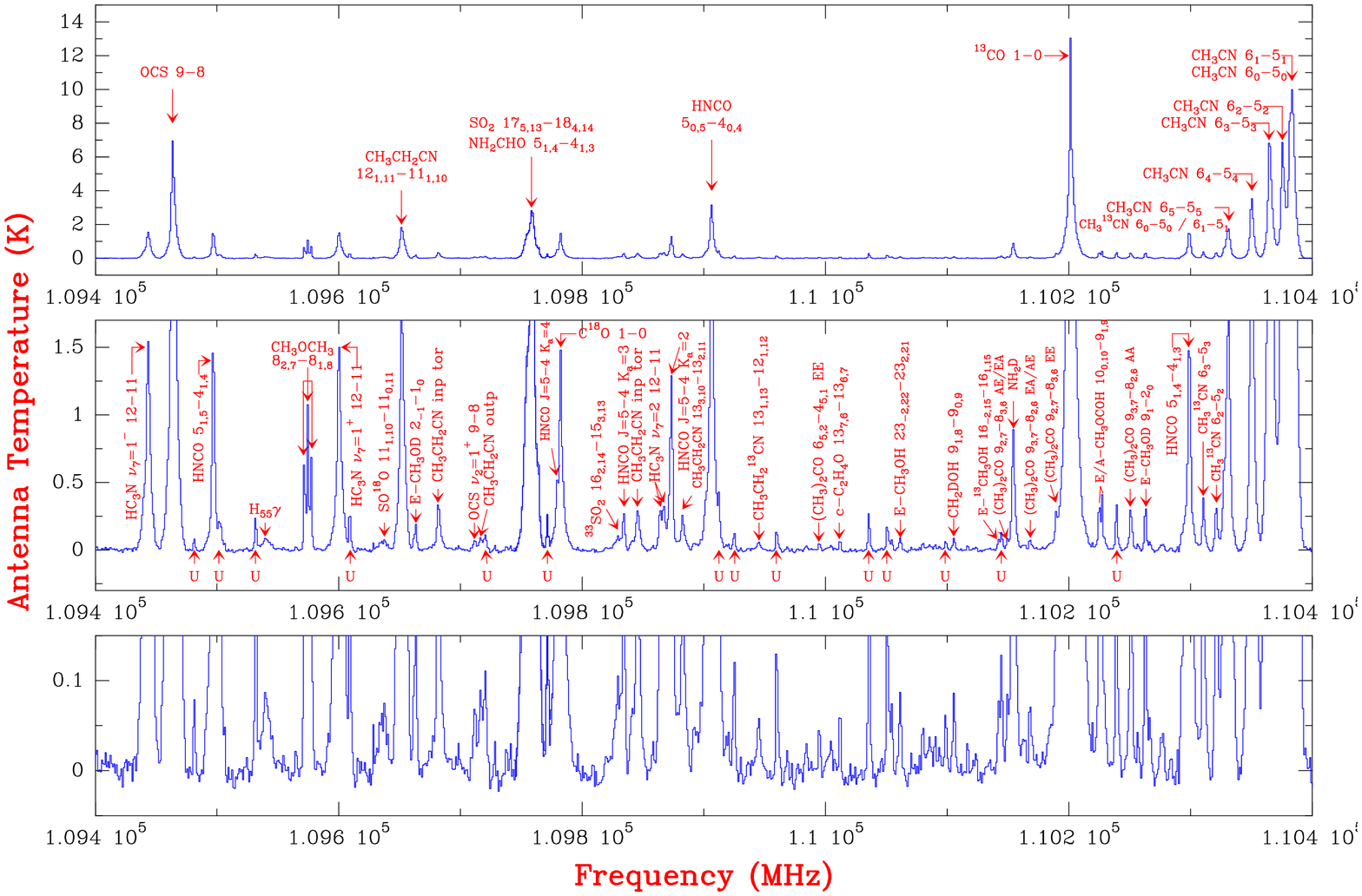}
\caption{Example of Orion's KL survey at 3 mm with 1 GHz
  bandwidth. The top panel
  shows the total intensity scale; the middle and the bottom panels
  show a zoom of the total intensity.
  Detected molecules are marked with labels and some unidentified
  features are marked as U.}
\label{fig_3mm_1G}
\end{figure*}

\begin{figure*}%f4
\includegraphics[angle=270,scale=.7]{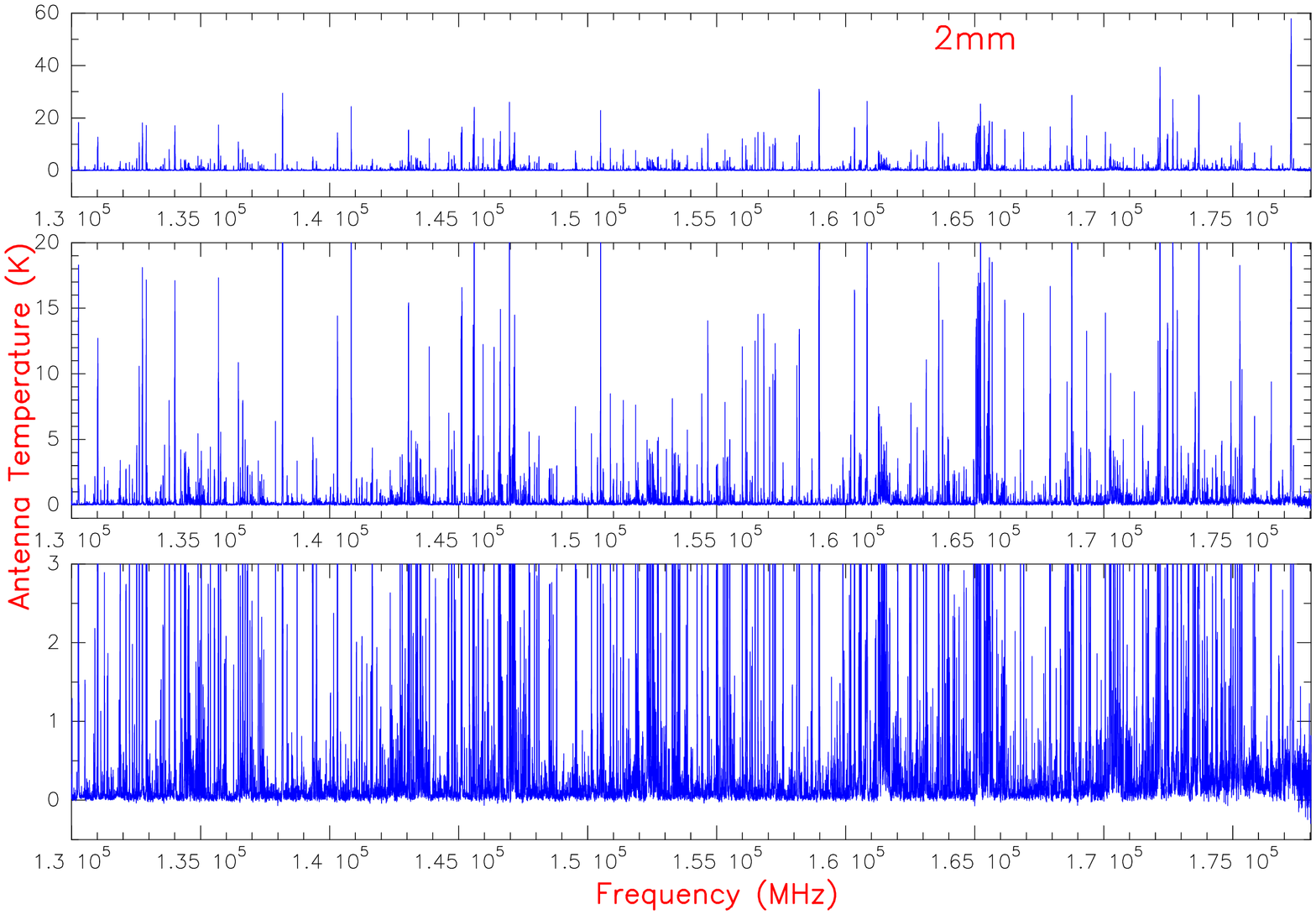}
\caption{Molecular line survey of Orion KL at 2 mm presented similarly 
to Fig. \ref{fig_3mm}.} 
\label{fig_2mm}
\end{figure*}

\begin{figure*}%f5
\includegraphics[angle=270,scale=.7]{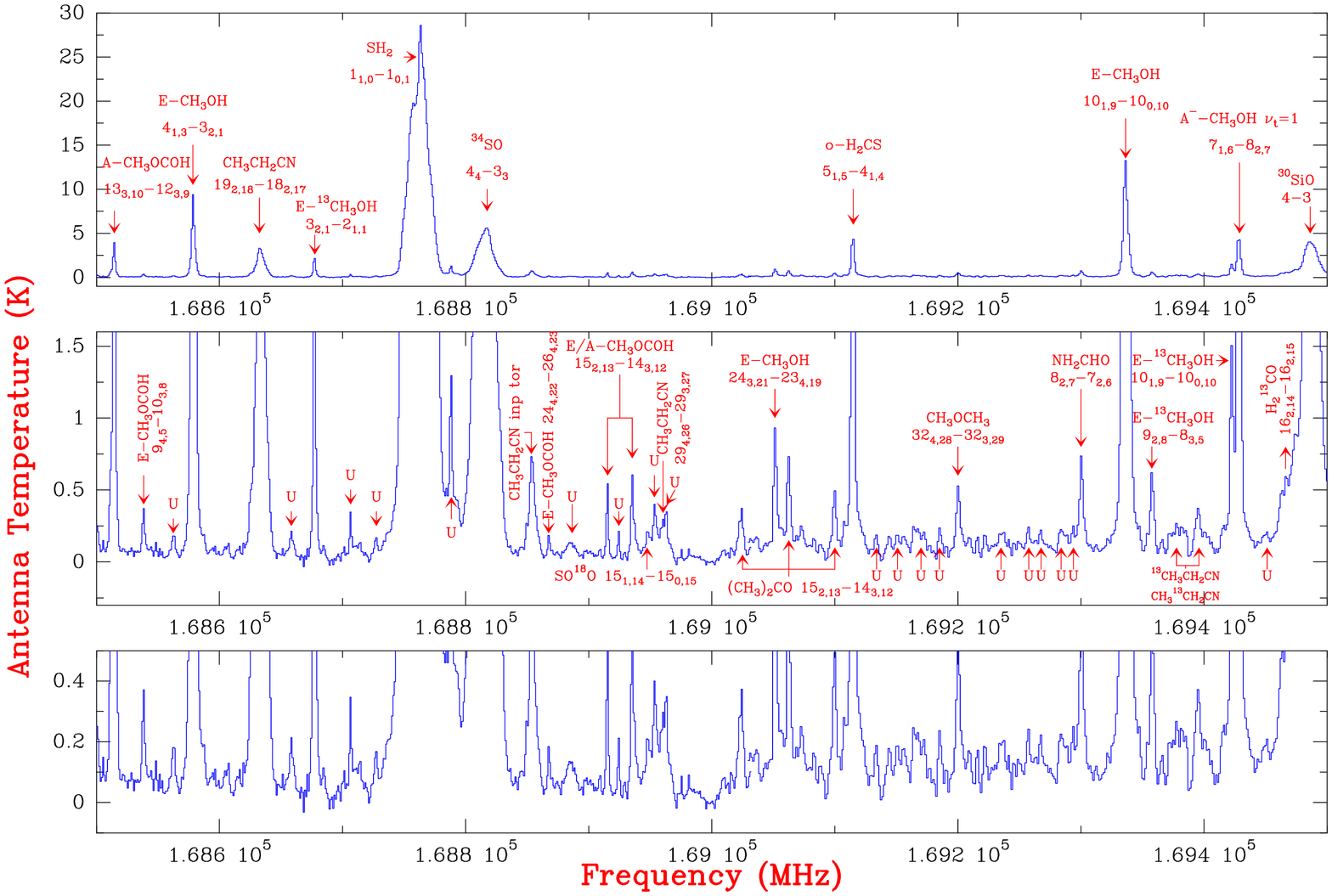}
\caption{Example of Orion's KL survey at 2 mm with 1 GHz
  bandwidth. The top panel
  shows the total intensity scale; the middle and the bottom panels
  show a zoom of the total intensity.
  Detected molecules are marked with labels and some unidentified
  lines are marked as U.}
\label{fig_2mm_1G}
\end{figure*}

\begin{figure*}%f6
\includegraphics[angle=270,scale=.7]{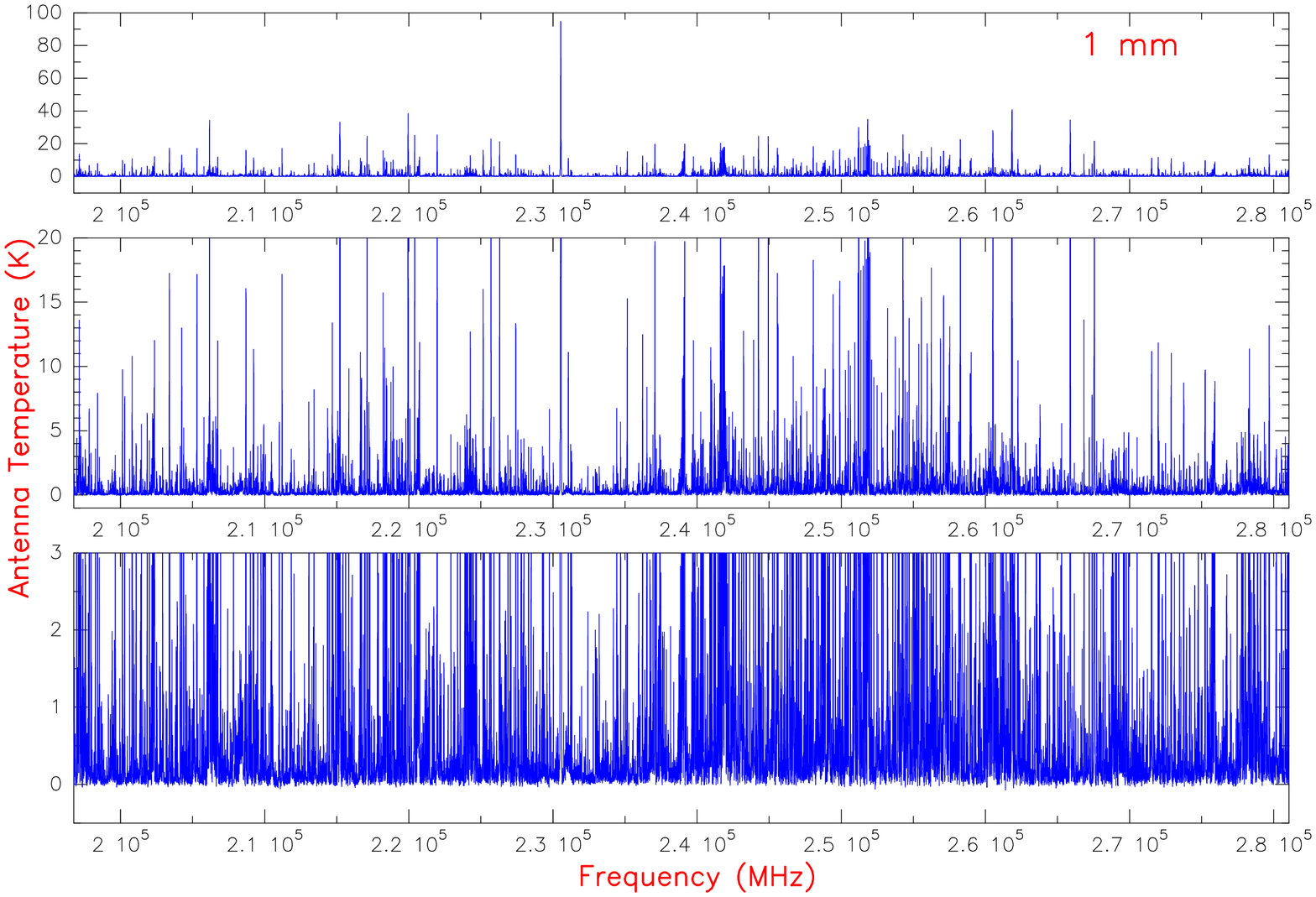}
\caption{Molecular line survey of Orion KL at 1.3 mm 
presented similarly to Fig. \ref{fig_3mm}.}
\label{fig_1mm}
\end{figure*}

\begin{figure*}%f7
\includegraphics[angle=270,scale=.7]{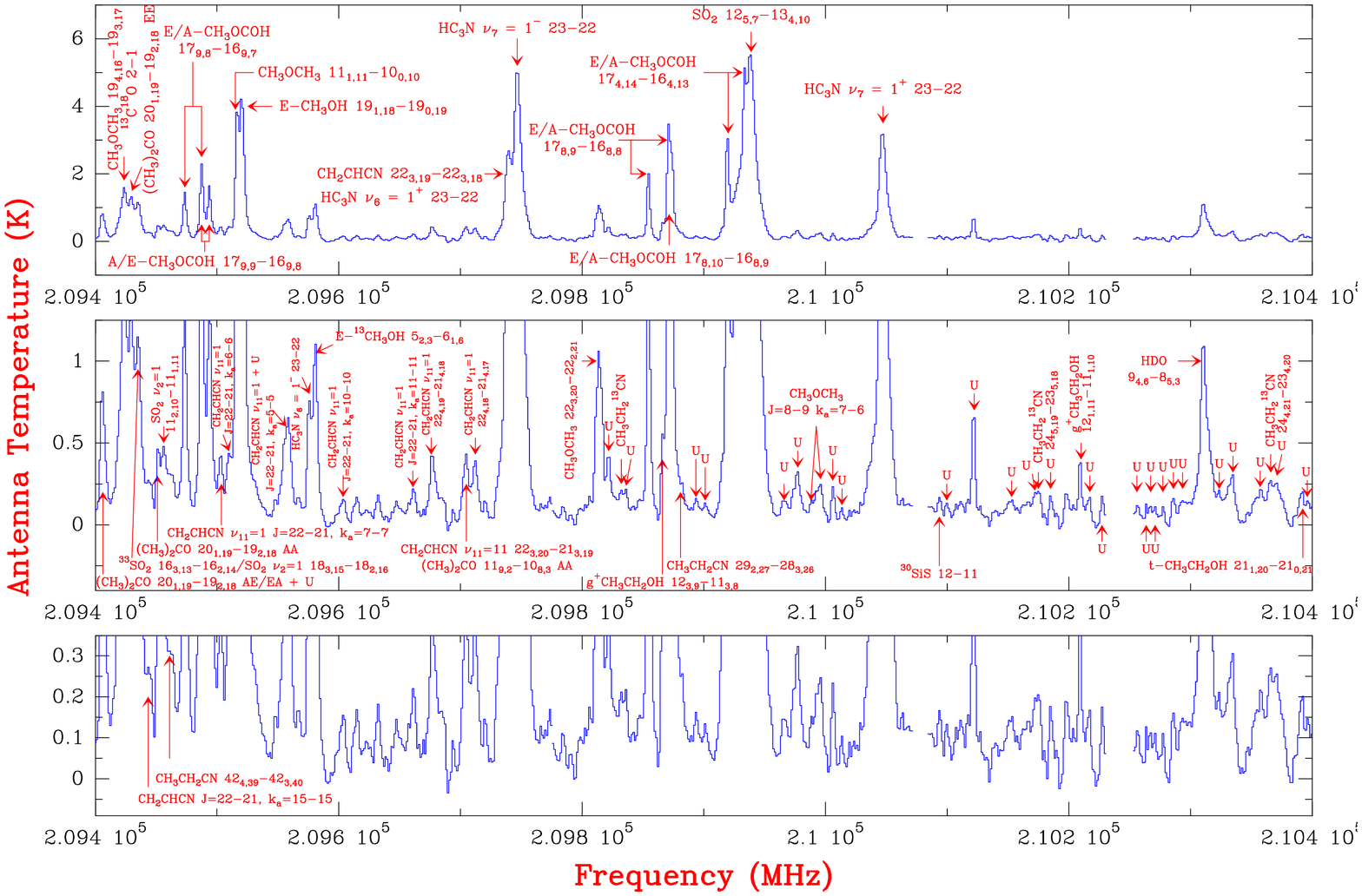}
\caption{Example of the Orion KL survey at 1.3 mm with 1 GHz
  bandwidth. The top panel
  shows the total intensity scale; the middle and the bottom panels
  show a zoom of the total intensity.
  Detected molecules are labeled and some unidentified
  lines are marked as U.}
\label{fig_1mm_1G}
\end{figure*}
%%%%%%------------------------------------------------------------------
Within the 168 GHz bandwidth covered, we detected more than 14400
spectral features of which 10040 were already identified and 
attributed to 43 molecules, including 148 different isotopologues
and vibrationally excited states.
Any feature covering more than 3-4 channels and of intensity
greater than 0.02 K is above 3$\sigma$ and is considered to be a line
in our survey. The noise was difficult to derive
from the data because of the high density of lines. We computed it from the
observing time and the system temperature.
In the 2 mm and 1.3 mm windows, the features weaker than 
0.1 K have not yet been systematically analyzed,
except when searching for isotopic species with good
laboratory frequencies.
We thus expect to considerably increase the number of both identified and
unidentified lines. We note that the number of U
lines was initially much larger. Identification of some isotopologues of most
abundant species (see below) allowed us to reduce the number of
U-lines at a rate of $\sim$ 500 lines per year.
We used standard procedures to identify lines above 0.2 K including
all possible contributions (taking into account the energy of the transition
and the line strength) from different species. 
Thanks to the wide frequency coverage of our survey, many rotational
lines were observed for each species, hence it is possible to estimate the
contribution of a given molecule to the intensity of a spectral feature
where several lines  from different species are blended.
We applied this procedure to all our previous
line surveys with the 30m telescope (e.g., \citealt{cer00}).
%{\bf{Due to the large
%frequency coverage of the line survey many lines have been observed for each
%molecular species. Hence, it is possible to estimate the expected 
%contribution of a given molecular
%species to the intensity of a particular feature consistent of a blend of
%several lines. We have been applying this procedure
%for all our previous works with the 30m telescope \citep{cer00}.}} 

As an example of the scope of this line survey in the field of
molecular spectroscopy, \citet{dem07}, \citet{mar09}, \citet{car09},
and \citet{mar10} 
identified 
more than 600, 100, 600, and 100
lines from the $^{13}$C and $^{15}$N isotopologues of CH$_3$CH$_2$CN,
the $^{13}$C isotopologues of HCOOCH$_3$, and CH$_3$OCOD, respectively. 
Many of the
remaining U-lines are certainly due to isotopologues of other abundant
species.

Figures \ref{fig_3mm}, \ref{fig_2mm}, and \ref{fig_1mm} show the whole 
data set of this Orion KL line survey at 3 mm, 2 mm and 1.3 mm 
respectively. Figures \ref{fig_3mm_1G}, \ref{fig_2mm_1G}, and 
\ref{fig_1mm_1G} show 1 GHz wide spectra as an example in each window. 
We have marked the identified features with labels (molecule and
transition quantum 
numbers) and the strongest unidentified ones as 'U'.  
In each figure, the top panels display the total intensity range,
while the middle 
and the bottom ones show different zoomed images of the intensity range. 

Because of the large amount of line features in the spectra, and
to follow the most efficient strategy for the line 
identification process, we decided to proceed 
in steps by studying the different molecular families 
including all possible isotopologues and vibrationally excited 
states. We continue to analyze our
line survey data, which we expect
to make public, with all line identifications, by 2011.

In agreement with previous works, four different spectral cloud components are
generally defined in the analysis of low angular resolution line
surveys where different physical components overlap in the beam. These
components are characterized by different physical and chemical conditions 
\citep{bla87, bla96}: 
(i) a narrow or 'spike' ($\lesssim$5 km s$^{-1}$ line-width)  
component at v$_{LSR}$ $\simeq$ 9 km s$^{-1}$ delineating a north-to-south 
\textit{extended ridge} or ambient cloud; (ii) a compact and quiescent 
region, the \textit{compact ridge}, (v$_{LSR}$ $\simeq$ 8 km s$^{-1}$, 
$\Delta$v $\simeq$ 3 km s$^{-1}$) identified for the first time by 
\citet{joh84}; (iii) the \textit{plateau} a mixture of outflows,
shocks, and interactions with the ambient cloud
(v$_{LSR}$ $\simeq$ 6-10 km s$^{-1}$, $\Delta$v $\gtrsim$25 km
s$^{-1}$); (iv) a \textit{hot core} component (v$_{LSR}$ $\simeq$ 
3-5 km s$^{-1}$, $\Delta$v $\lesssim$10-15 kms$^{-1}$) first detected 
in ammonia emission \citet{mor80}. Table \ref{tab_prop}
gives the physical parameters that we obtained for each
component from the modeling of the OCS, HCS$^+$, H$_2$CS, CS, CCS, 
and CCCS line emission
(Sect. \ref{sect_col}). The assumption of a single gas temperature and
density for each cloud component is 
the greatest simplification of our methodology. 
It is clear that the source
structure identified by much higher angular resolution interferometric  
%REVEALED BY MUCH HIGHER
%ANGULAR RESOLUTION INTERFEROMETRIC OBSERVATIONS
observations is far more complex than assumed in Table \ref {tab_prop}.
We attempted to use more complex structures using density and temperature
gradients, but the comparison with the data indicate that we do not have enough
information to fit these physical gradients, even when we have many lines for 
%data indicate that we DONT HAVE  ENOUGH information to fit these
%physical gradients
some species. Therefore, we fix the physical properties to be those given in
Table \ref{tab_prop}
(values derived from our data analysis) to ensure that we have only one
free parameter (the column density) when modeling the spectral lines.
%Still, it represents a
%major improvement compared with previous works based in simple LTE
%rotational diagram analysis where the physical conditions of each
%region are not taken into account.
Nevertheless, our multi-source excitation and radiative transfer approach 
%OUR MULTI-SOURCE EXCITATION AND RADIATIVE TRANSFER APPROACH
represents a major improvement on previous works based on LTE
analysis.
\section{Results}
\label{sect_res}

\subsection{OCS}
\label{sect_res_ocs}

\addtocounter{table}{1} %t3

%tabla de las componentes gaussianas de ocs.
\begin{table*}%t4
\begin{center}
\caption{OCS velocity components from Gaussian fits.\label{tab_gau1}}
\resizebox{0.9\textwidth}{!}{%
\begin{tabular}{llllllllll}
\hline 
\hline
Species/ & \multicolumn{3}{c}{Ridge} & \multicolumn{3}{c}{Plateau} & \multicolumn{3}{c}{Hot core}\\ 
Transition & v$_{LSR}$ (km s$^{-1}$) & $\Delta$v (km s$^{-1}$) & T$^*_A$ (K) &
v$_{LSR}$ (km s$^{-1}$) & $\Delta$v (km s$^{-1}$) & T$^*_A$ (K)&
v$_{LSR}$ (km s$^{-1}$) & $\Delta$v (km s$^{-1}$) & T$^*_A$ (K)\\
\hline
\\
 OCS 7-6 & 7.83$\pm$0.04 & 5.34$\pm$0.09 & 2.30 & 5.9$\pm$0.3 &
 26.3$\pm$0.6 & 0.49 & 5.6$\pm$0.2 & 12.5$\pm$0.3 & 1.02\\
 OCS 8-7 & 8.00$\pm$0.03 & 4.81$\pm$0.05 & 3.56 & 6.59$\pm$0.07 &
 25.8$\pm$0.4 & 0.85 & 5.16$\pm$0.04 & 11.7$\pm$0.2 & 1.68\\
 OCS 9-8 & 7.85$\pm$0.02 & 4.75$\pm$0.05 & 3.85 & 6.1$\pm$0.2 &
 25.8$\pm$0.5 & 1.08 & 5.41$\pm$0.13 & 11.73$\pm$0.06 & 2.29\\
 OCS 11-10 & 8.04$\pm$0.13 & 4.8$\pm$0.3 & 4.36 & 6.76$\pm$0.11 &
 25.8$\pm$1.2 & 1.80 & 4.5$\pm$0.3 & 9.0$\pm$0.7 & 2.53\\
 OCS 12-11 & 8.13$\pm$0.03 & 4.05$\pm$0.04 & 6.63 & 6.4$\pm$0.2 &
 23.3$\pm$0.3 & 3.29 & 5.42$\pm$0.08 & 9.49$\pm$0.08 & 2.82\\
 OCS 13-12 & 7.76$\pm$0.07 & 5.1$\pm$0.2 & 5.32 & 5.9$\pm$0.4 &
 25.8$\pm$1.5 & 2.67 & 5.0$\pm$0.2 & 9.9$\pm$0.8 & 4.16\\
 OCS 14-13 & 7.81$^1$$\pm$0.06 & 4.6$\pm$0.3 & 3.70 &
 ... & ... & ... & ... & ... & ...\\
 OCS 17-16 & 8.0$\pm$0.2 & 5.8$\pm$0.6 & 4.18 & 6.0$\pm$0.3 &
 28.6$\pm$1.0 & 2.83 & 4.3$\pm$0.4 & 8.8$\pm$0.3 & 5.83\\
 OCS 18-17 & 7.77$\pm$0.10 & 5.60$\pm$0.15 & 4.47 & 5.7$\pm$0.3 &
 25.8$\pm$0.8 & 3.06 & 3.71$\pm$0.12 & 8.05$\pm$0.08 & 4.77\\
 OCS 19-18 & 8.0$\pm$0.3 & 4.9$\pm$0.3 & 3.57 & 5.0$\pm$0.3 &
 24.1$\pm$1.2 & 4.39 & 4.66$\pm$0.14 & 8.1$\pm$0.6 & 6.14\\
 OCS 20-19 & 7.83$\pm$0.12 & 4.8$\pm$0.4 & 3.24 & 6.2$\pm$0.3 &
 31.2$\pm$1.2 & 2.75 & 4.3$\pm$0.4 & 10.2$\pm$0.6 & 5.10\\
 OCS 21-20 & 7.9$\pm$0.2 & 6.5$\pm$0.3 & 4.68 &
 3.7$^2$$\pm$0.8 & 28.0$\pm$2.2 & 1.46 &
 3.72$\pm$0.09 & 14.0$\pm$1.5 & 2.99$^2$\\
 OCS 23-22 & 8.1$\pm$0.2 & 5.5$\pm$0.4 & 4.34 &
 3.2$^2$$\pm$0.3 & 21.6$\pm$1.3 &
 3.09 & 3.9$\pm$0.4 & 7.7$\pm$0.9 & 2.62$^2$\\
\hline
\end{tabular}
%% Any table notes must follow the \end{tabular} command.
}
\end{center}
Note.-v$_{LSR}$, $\Delta$v and T$^*$$_A$ of the OCS lines shown in Fig.
\ref{fig_lin1} (see text, Sect. \ref{sect_res_ocs}) derived from three Gaussian fits.\\
$^1$ \textit{Due to line overlap only the ridge component can be fitted.}\\
$^2$ \textit{Although the line is clearly present, the derived
  parameters are biased by the presence of nearby lines from other species.}\\
\end{table*}

\begin{figure*}%f8
\includegraphics[angle=0,scale=.8]{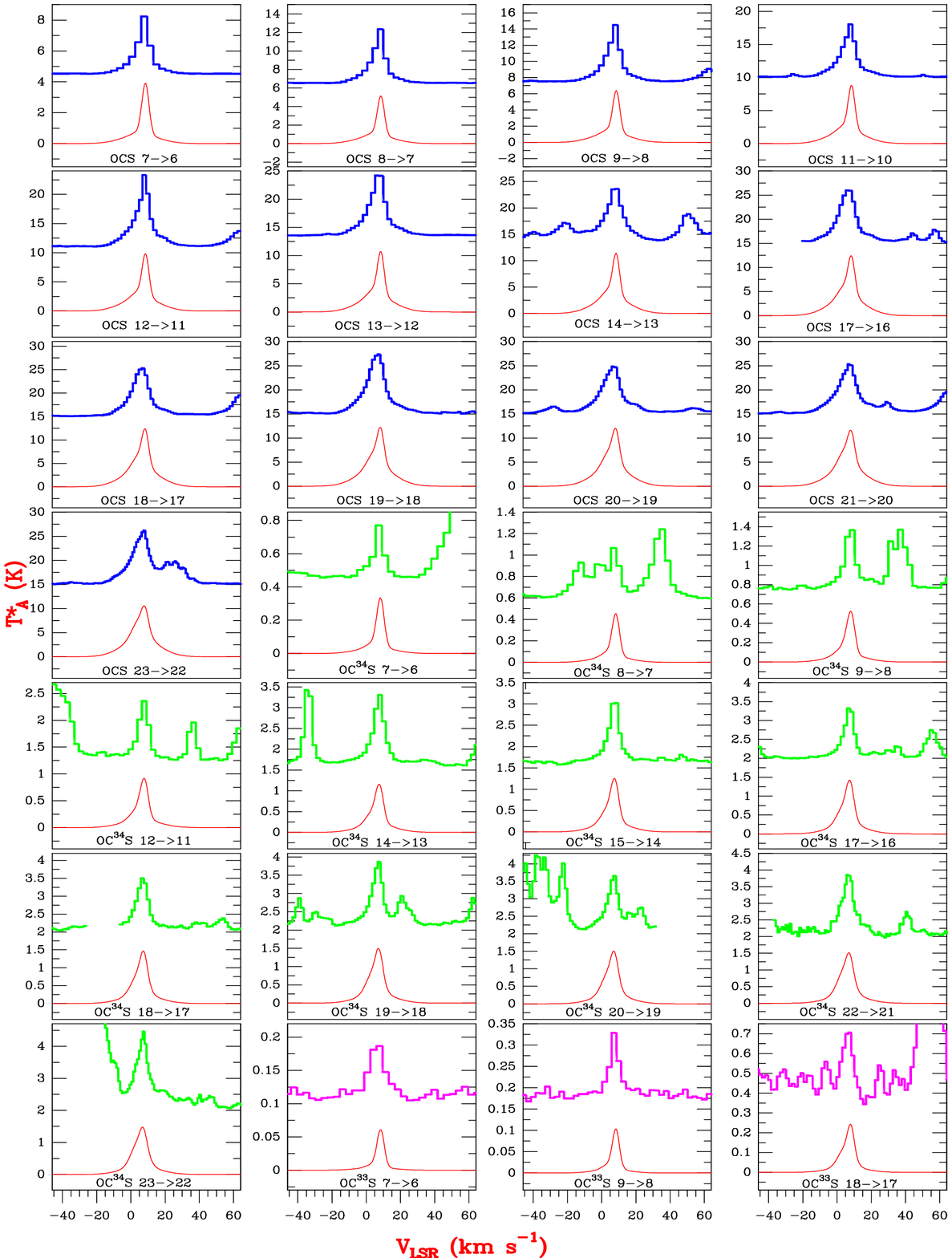}
\caption{Observed (offseted histogram) and model 
  (thin curves) OCS, OC$^{34}$S and OC$^{33}$S lines. 
A v$_{LSR}$ of 9 km s$^{-1}$ is
assumed.}
\label{fig_lin1}
\end{figure*}

\begin{figure*}%f10
\includegraphics[angle=0,scale=.8]{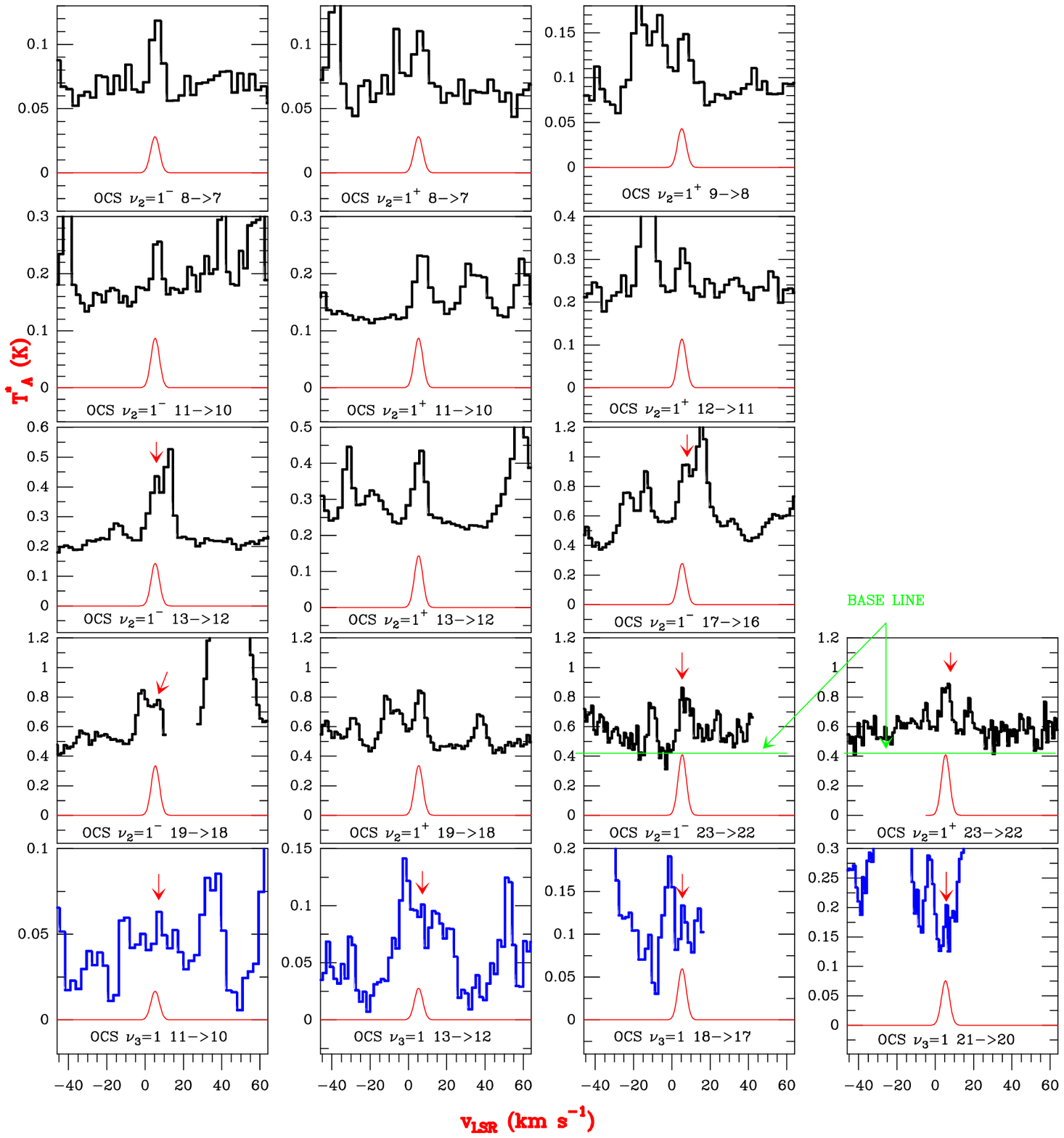}
\caption{Observed lines (offseted histogram) and model (thin
  curves) of OCS $\nu_2$ = 1 and OCS $\nu_3$ = 1. 
A v$_{LSR}$ of 9 km s$^{-1}$ is assumed.}
\label{fig_lin3}
\end{figure*}

\begin{figure*}%f12 
\includegraphics[angle=270,scale=.85]{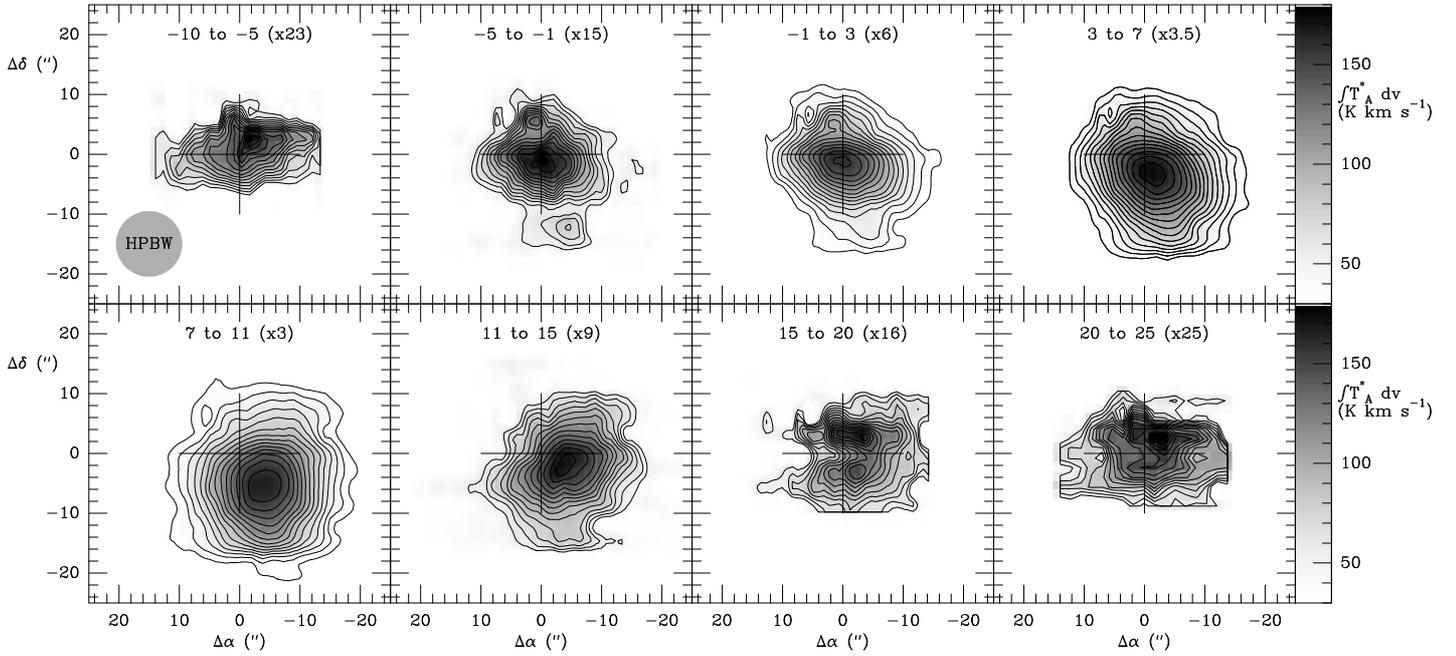}
\caption{OCS $J$ = 18-17 integrated line 
  intensity maps at different velocity
  ranges (indicated at the top of each
  panel). The integrated intensity of the maps has been multiplied by
  a scale factor (indicated in the panels) to maintain
  the same color dynamics for all maps. The interval of contours 
is 10 K km s$^{-1}$,
  the minimum contour is 30 K km s$^{-1}$ for the maps with
  velocities between -1
  and 11 km s$^{-1}$ and 50 K km s$^{-1}$ for the rest of the panels.}
\label{fig_ocs_map}
\end{figure*}

\begin{figure*}%f13
%\plottwo{fig_3mm_bn.eps}{fig_3mm.eps}
\includegraphics[angle=0,scale=.8]{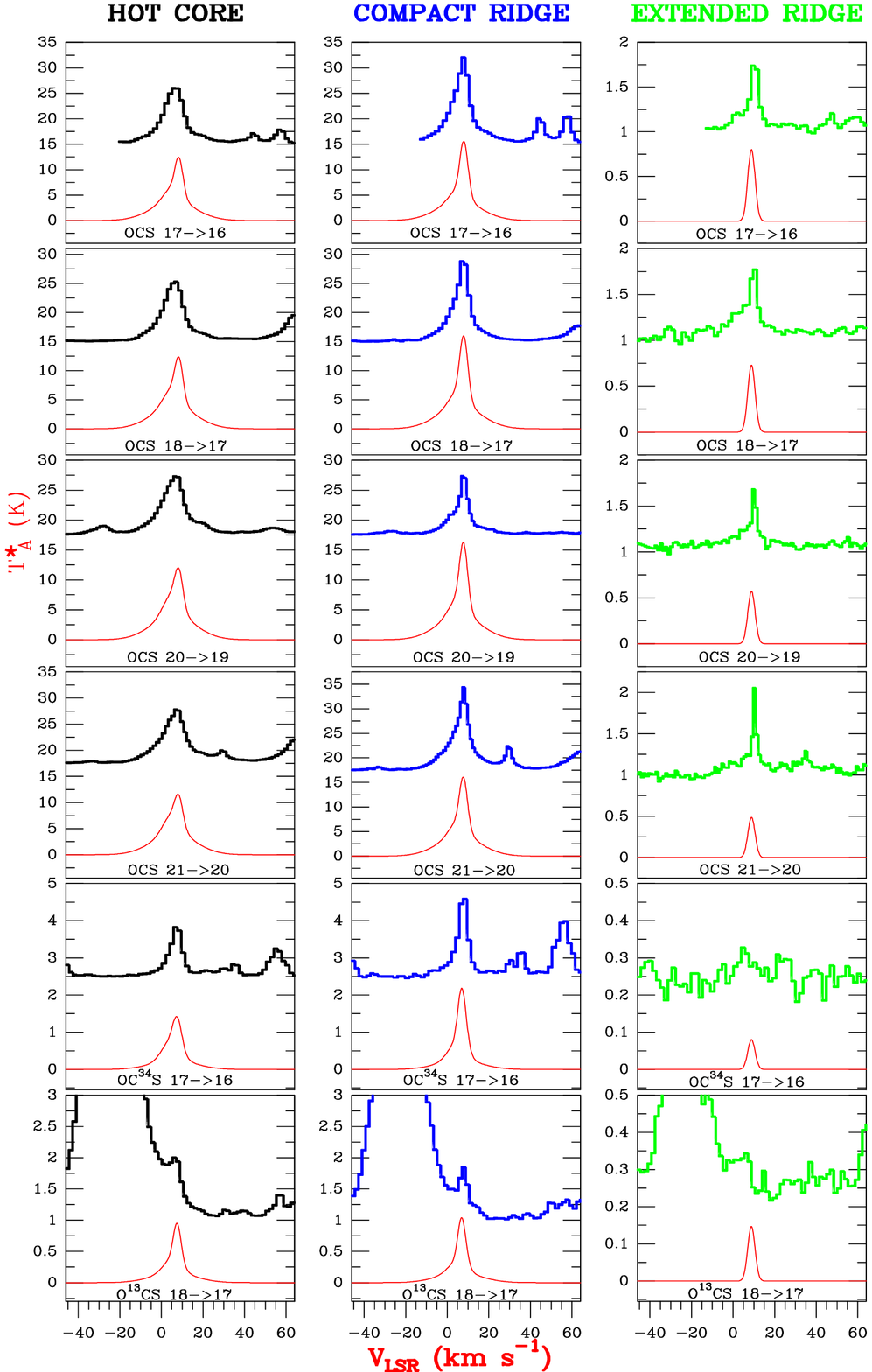}
\caption{Some lines of OCS, OC$^{34}$S
and O$^{13}$CS observed all different positions which correspond 
with different components of Orion KL. A v$_{LSR}$ of 9 km s$^{-1}$ is
assumed.}
\label{fig_com}
\end{figure*}

Carbonyl sulfide (OCS) has a linear structure and, because of its 
rotational constant (B=5932.83 MHz for $^{16}$O$^{12}$C$^{32}$S),
it harbours 
up to 15 transitions per vibrational state that can be observed in the 
covered frequency range. Line detections in our survey include the ground 
vibrational state of 6 isotopologues (OCS, OC$^{34}$S, OC$^{33}$S,
O$^{13}$CS, $^{18}$OCS, O$^{13}$C$^{34}$S), plus two vibrationally
excited states of the main 
isotopologue ($\nu_2 = 1$, $\nu_3 = 1$). The last two were detected 
here for the first time in space. Only a tentative detection is
presented for $^{17}$OCS and OC$^{36}$S because of the 
weakness of the features and/or their overlap with other spectral
lines.

The rotational constants used to derive the line frequencies 
were taken from 
\citet{gol05} (OCS), the NIST triatomic molecules database (OC$^{34}$S and 
O$^{13}$CS), \citet{bur81b} (OC$^{33}$S), \citet{bur81a} (all the others OCS 
isotopologues), and \citet{mor00} (OCS vibrationally excited states). The 
OCS dipole moment ($\mu$=0.7152D) was assumed to be that measured by 
\citet{tan85}. Observed line 
parameters and intensities are given in Table \ref{tab_lines}. Figures 
\ref{fig_lin1}, \ref{fig_lin2} and \ref{fig_lin3} show the lines 
that are not blended with features from other species and our
best-fit-model line profiles 
(see Sect. \ref{sect_col_ocs}). The line profiles and intensities show the 
contribution from the extended and compact molecular ridges,
the plateau, and the hot core. In previous line surveys, the 
extended ridge component was discarded as a significant source of OCS 
line emission. However, we include it here as a requirement to reproduce the 
observed intensities from $J$ = 7 - 6 up to $J$ = 23 - 22 (main and rare
isotopologues).  

To constrain the model more tightly, and determine the spatial
distribution of the OCS line emission, we obtained
a map of the OCS $J$=18-17 line and performed sensitive
observations of several lines at selected positions around
Orion IRc2. Figure \ref{fig_ocs_esp} shows the observed line profiles and
integrated line intensity spatial distribution.
Figure \ref{fig_ocs_map}
shows the line emission for different velocity ranges. 

The maximum integrated intensity lies approximately 3''
southwest of IRc2 (see Fig. \ref{fig_ocs_esp}) and is a mixture 
of compact ridge and hot core components, in agreement with the 
spatial distribution found by \citet{wri96}.
The velocity structure of the OCS emission depicted in Fig.
\ref{fig_ocs_map} shows all the cloud spectral components discussed
above. The spatial distribution of the red
wing ($v_{LSR}\simeq$15-20 km s$^{-1}$) of the $J$=18-17 line emission is
particularly interesting. It traces an elliptical expanding shell of
gas around IRc2, the low-velocity outflow.
The front of the
shell is traced by the emission at velocities from -10 to -1 
km s$^{-1}$ (blue wing),
while the red part of the shell appears at 20-25 km
s$^{-1}$.

The observed lines at
selected positions are shown in Fig. \ref{fig_com}. Altogether, these data
allow us to study the hot core, the compact ridge, and the extended
ridge. \citet{sut95} also observed the OCS $J$=28-27 line at
different positions. OCS line intensities are clearly 
brighter towards the 
compact ridge position than towards IRc2 (hot core). 
The antenna temperature measured towards the extended ridge 
position is $\simeq$ 1 K; however, the extended ridge contribution 
towards IRc2 should be larger to explain the data (see Sect.
\ref{sect_col_ocs}).

Table \ref{tab_gau1} gives the parameters of the 
OCS lines derived by fitting Gaussian profiles to all velocity
components with the CLASS 
software\footnote{http://www.iram.fr/IRAMFR/GILDAS}.  
The v$_{LSR}$ velocities derived 
for the hot core vary between 3.3 km s$^{-1}$ to 5.6 km s$^{-1}$
due to the overlap
with the other velocity components, in particular in
the blueshifted wing of the line profile  
(see Fig. \ref{fig_lin1}).
Table \ref{tab_gau2}, only available online, 
gives the parameters of the observed lines
of the most abundant isotopologues (OC$^{34}$S,
OC$^{33}$S, and O$^{13}$CS) and 
vibrationally excited OCS ($\nu_2$=1). Since these lines are 
much weaker than those of OCS $\nu$=0,
only a single velocity component has been fitted. Because of either 
their weakness or heavy blending, the analysis of the other  
OCS isotopologues and of OCS $\nu_3$=1 is not possible. 
We note that the line parameters
for OCS $\nu_2$=1 match those of the hot
core component. This behavior is as expected since the 
energy of the $\nu_2$=1 vibrational level is 749 K, which is populated for
the warmest gas component.  
%The velocity centroid and line widths  
%of all molecular lines observed towards the hot core have been fixed
%to those obtained for OCS $\nu_2$=1.

\subsection{HCS$^+$}
\label{sect_res_hcs+}

Four transitions of thioformyl cation (HCS$^+$) were detected in the
covered frequency range.
Line frequencies and observational parameters are given in Table
\ref{tab_hcs+lines}, only available online, which contains the following
information: Column 1
gives the observed (centroid) radial velocities, Col. 2 the
peak line temperature, Col. 3 the integrated line intensity, Col.
  4 the quantum numbers, Col. 5 the
assumed rest frequencies, Col. 6 the energy of the upper level, and
Col. 7 the line strength. Rotational constants were derived from the
rotational lines reported by \citet{mar03}.
The adopted dipole moment, $\mu$ = 1.958 D, is taken from \citet{bot85}.
Line profiles and our best-fit models (see Sect. \ref{sect_col_hcs+}) are shown
in Fig. \ref{fig_hcs+}. 

The HCS$^+$ line profiles display the four Orion's spectra components. 
In this case, the contribution of the extended
ridge component is very weak (see Sect. \ref{sect_col_hcs+}).

\subsection{H$_2$CS}
\label{sect_res_h2cs}

We detected several transitions of thioformaldehyde 
(45 transitions of ortho and para states). We also
detected H$_2$C$^{34}$S, H$_2$$^{13}$CS (both p- and o- states) and
HDCS isotopologues.

Line parameters are given in Table
\ref{tab_h2cslines}, 
only available online, which contains the following
information: Column 1 indicates the isotopologue or the vibrational
  state, Col. 2
gives the observed (centroid) radial velocities, Col. 3 the
peak line temperature, Col. 4 the integrated line intensity, Col.
  5 the quantum numbers, Col. 6 the
assumed rest frequencies, Col. 7 the energy of the upper level, and
Col. 8 the line strength. Figures 
\ref{fig_h2cs}, \ref{fig_h2c34s}, \ref{fig_h2_13cs}, and \ref{fig_hdcs} 
show the lines that are 
not blended with other species and our best-fit model
(see Sect. \ref{sect_col_h2cs}).
The rotational constants used to derive the line frequencies were 
taken from the CDMS 
Catalog\footnote{\citet{mul01}, \citet{mul05} 
http://www.astro.uni-koeln.de/site/vorhersagen/} for
H$_2$CS \citep{mae08} and H$_2$$^{13}$CS.
The H$_2$CS dipole moment, $\mu$=1.6491D, is the one measured by \citet{fab77}.
For H$_2$C$^{34}$S (ortho and para) and HDCS, the line parameters were
fitted from all rotational lines reported by \citet{min97}
and the observed lines towards B1 dark cloud by \citet{mar05}.
For HDCS, a small $\mu_b$ dipole moment is expected, which we
assumed to be identical to that of HDCO.

Line profiles and intensities indicate 
contributions from the extended ridge, compact ridge (very prominent),
the plateau, and the hot core. 

Table \ref{tab_gau_h2cs}, only available online, provides the parameters of selected 
lines of H$_2$CS and its isotopologues obtained assuming Gaussian
fits to the line profiles. We show only one narrow component fit
(a blend of
compact ridge, extended ridge, and hot core)
because the wide component (plateau) cannot be fitted due to blending with
other species. The main component contribution is the compact ridge. 
We note that for H$_2$CS (ortho and para), v$_{LSR}$, and $\Delta$v
tend to values similar to these of the hot core when the K$_a$ quantum 
number increases. 

\subsection{CS}
\label{sect_res_cs}

Three transitions ($J$ = 2-1,
3-2, 5-4) of carbon monosulfide substitutions C$^{32}$S, C$^{34}$S,
and C$^{33}$S along with four lines
of $^{13}$CS and $^{13}$C$^{34}$S ($J$ = 2-1, 3-2, 5-4, 6-5) 
were detected. 
For C$^{36}$S, $^{13}$C$^{33}$S, and vibrationally excited CS
($\textit{v}$=1), we present only tentative detections.
Line frequencies and observational 
parameters are given in Table \ref{tab_cslines}, only available online,
which contains the following information: Column 1 indicates the
isotopologue or the vibrational state, Col. 2
gives the observed (centroid) radial velocities, Col. 3 the
peak line temperature, Col. 4 the integrated line intensity, Col.
  5 the quantum numbers, Col. 6 the
assumed rest frequencies, Col. 7 the energy of the upper level, and
Col. 8 the line strength.
Line profiles for transitions that are not blended with other features
are shown in Fig. \ref{fig_cs}. The spectroscopic constants 
for CS and C$^{34}$S are taken from \citet{got03}, those of $^{13}$CS,
C$^{33}$S, C$^{36}$S, $^{13}$C$^{34}$S, and $^{13}$C$^{33}$S from
\citet{ahr98}, and those of CS $\textit{v}$=1 
come from \citet{kim03}. Dipole moments ($\mu$=1.958D for CS
$\textit{v}$=0 and $\mu$=1.936D for CS
$\textit{v}$=1) were taken from \citet{win68}.

Line profiles from the most abundant isotopologues display the 
four Orion KL spectral components.
At 3 mm and 2 mm,  the ridge and
the plateau emission dominate, at 1 mm the presence of the hot
core component in the line profile is very significant.
For the less abundant isotopologues (C$^{36}$S, $^{13}$C$^{34}$S and
$^{13}$C$^{33}$S), the compact ridge and hot core components are responsible
for most of the line emission. The emission of CS vibrationally
excited states comes mainly from the hot core component.

Line parameters for CS, C$^{34}$S, C$^{33}$S, $^{13}$CS,
and $^{13}$C$^{34}$S are given in Table \ref{tab_gau3}, only available online. 

\subsection{CCS}
\label{sect_res_ccs}

\begin{figure*}%f20
\includegraphics[angle=0,scale=.8]{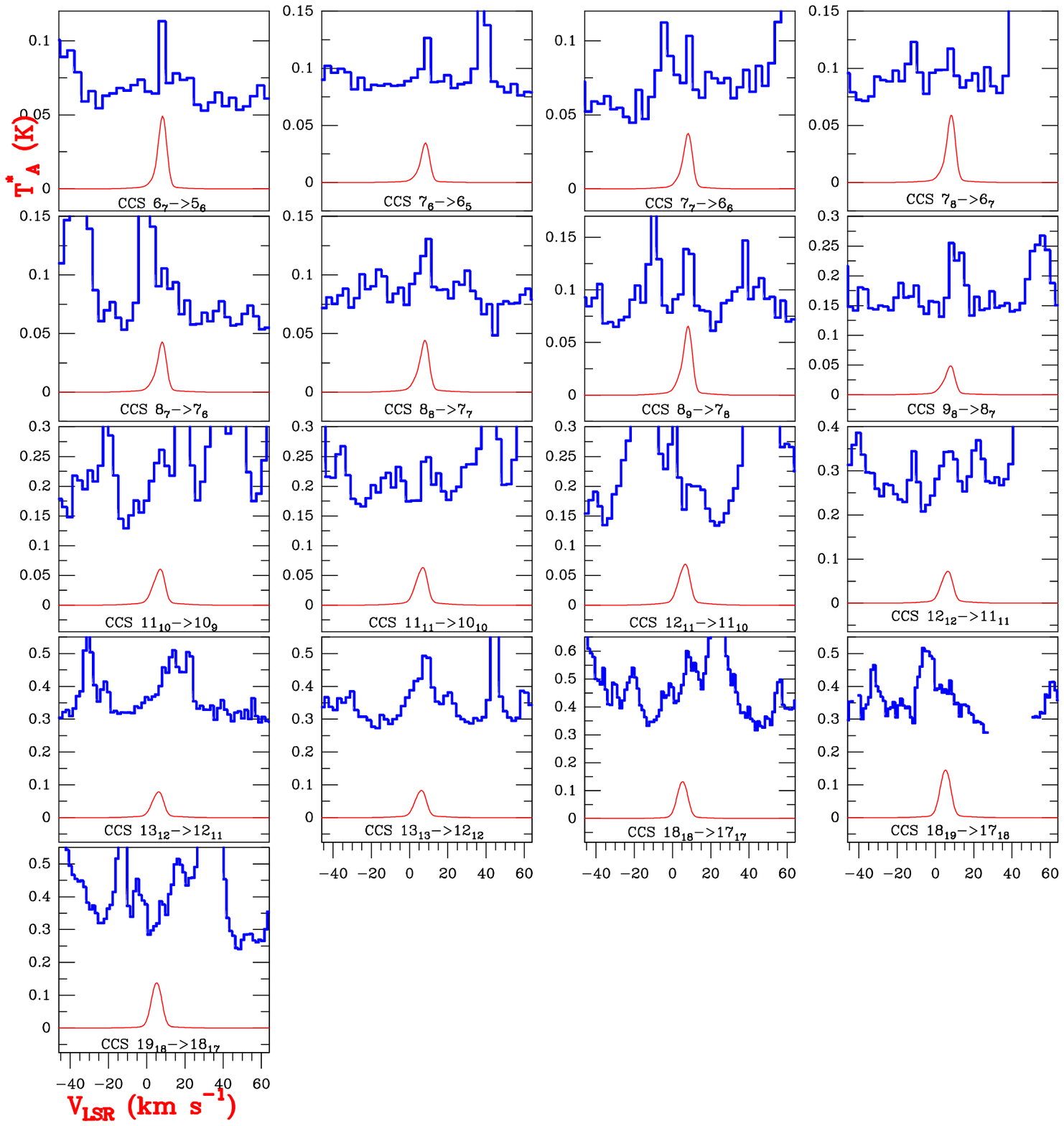}
\caption{Observed lines (offseted histogram) and model (thin
  curves) of CCS. A v$_{LSR}$ of 9 km s$^{-1}$ is
assumed.}
\label{fig_ccs}
\end{figure*}

The CCS radical ($^3\Sigma$ ground electronic state) has several transitions 
in the surveyed frequency range. Detected lines and main spectroscopic 
parameters are given in Table \ref{tab_ccs}, only available online,
containing the following information: Column 1
gives the observed (centroid) radial velocities, Col. 2 the
peak line temperature, Col. 3 the quantum numbers, Col. 4 the
assumed rest frequencies, Col. 5 the energy of the upper level, and
Col. 6 the line strength. Rotational constants were 
taken from \citet{yam90} and the dipole moment, $\mu$ = 2.88 D,
comes from \citet{lee97}. 
Figure \ref{fig_ccs} shows the detected transitions that are unaffected by
line overlap.

Owing to the line emission weakness, it is difficult to 
distinguish the different spectral cloud components in the observed line
profiles. To reproduce the line
intensities (see Sect. \ref{sect_col_ccs}), we assumed that the hot 
core component is
responsible for most of the observed emission. However, the line velocity 
centroid indicates that both the extended and compact ridge components
also contribute at 
the observed emission. \citet{ziu93} found the same
velocity components in their CCS observed lines 
(one detected and two tentative).
The much lower abundances of CCS isotopologues and of 
vibrationally excited CCS prevent us from detecting any of their transitions above the 
line confusion limit.

\subsection{C$_3$S}
\label{sect_res_c3s}

\begin{figure*}
\includegraphics[angle=0,scale=.8]{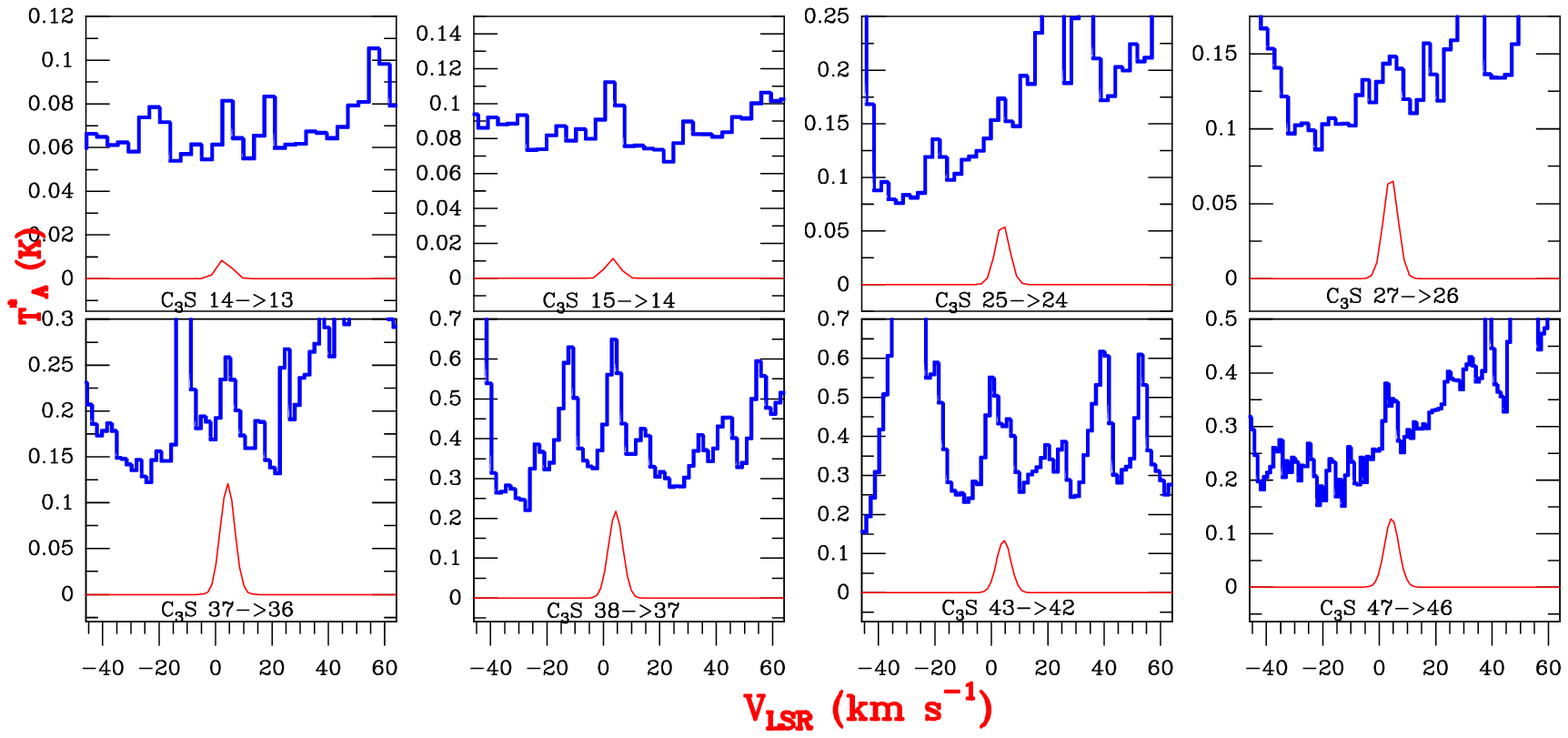}
\caption{Observed lines (offseted histogram) and model (thin
  curves) of C$_3$S. A v$_{LSR}$ of 9 km s$^{-1}$ is
assumed.}
\label{fig_cccs}
\end{figure*}

Previously, C$_3$S has been observed in cold dark clouds \citep{kai87} and in 
the envelopes of C-rich AGB stars
\citep{cer87a, bel93}. \citet{sut95} found a possible spectral line towards Orion hot core
and compact ridge positions, but its
identification as the C$_3$S $J$=58-57 line was discarded because of
the high energy level (475 K).

We report the first detection of C$_3$S in warm
clouds. We clearly 
identified 17 of the 29 rotational transitions covered in the survey.
The remaining transitions are blended 
with lines of other species. Figure \ref{fig_cccs} shows 
several C$_3$S detected lines. Line parameters are 
given in Table \ref{tab_cccs}, which is only available online, containing the
following information: Column 1
gives the observed (centroid) radial velocities, Col. 2 the
peak line temperature, Col.
  3 the quantum numbers, Col. 4 the
assumed rest frequencies, Col. 5 the energy of the upper level, and
Col. 6 the line strength. Rotational constants were taken 
from \citet{yam87} and the dipole moment ($\mu$ = 3.704 D) was assumed to be that 
measured by \citet{sue94}. The line centroid velocity 
indicates that the emission mainly arises from the hot core. As 
for CCS, we could not detect any C$_3$S isotopologues or vibrationally 
excited states.

\section{Determination of column densities}
\label{sect_col}

For all detected species column densities were calculated using
an excitation and radiative transfer code developed by J. Cernicharo
(Cernicharo 2010, in preparation). Depending on either the selected molecule
or physical conditions, we followed either a LVG (\citealt{sob58};
\citealt{sob60}) or LTE approach.
For each cloud component, we assumed uniform physical
conditions for the kinetic temperature, density, radial
velocity, and line width
(Table \ref{tab_prop}). 
We adopted these values from the data 
analysis (Gaussian fits and an attempt to simulate the line widths and
intensities with LTE and LVG codes) as representative parameters
for the different species. When a change in these 
values was required (e. g. C$_3$S analysis), we indicate this in the text.
Our modeling technique also takes into account the size
of each component and its offset position with respect IRc2. 
Corrections for 
beam dilution were applied to each line depending on their frequency. 
The only free parameter is therefore the column density 
of the corresponding observed species. Taking into account the compact
nature of most cloud components, the contribution from the error beam is
negligible except for the extended ridge, which has a small
contribution for all observed lines.

In addition to line opacity effects, other sources of uncertainty are
related to the following: 
\begin{itemize}
\item Adopting uniform physical conditions assumes that the
  physical structure of the cloud is simplified. However, parameters such as the
size, kinetic temperature, and density gradients of the different components of 
the cloud are difficult to assess from low resolution single-dish observations.
This problem can be partially overcome by analyzing many 
different molecular species and transitions covering a broad range
of excitation conditions, as allowed by our line survey.

%ADOPTING uniform physical conditions is a simplification of the
%physical structure of the cloud. However, parameters such as
%size, kinetic temperature and density GRADIENTS of the different
%components of the cloud are difficult to assess FROM LOW RESOLUTION
%SINGLE-DISH OBSERVATION. This problem
%can be partially overcome BY ANALYZING many different molecular
%species and transitions COVERING A BROAD RANGE OF EXCITATION CONDITIONS, 
%as allowed by OUR  line survey.
 
\item The angular resolution of any single-dish line survey is
  modest. Therefore, the emission from different physical components
  is usually blended and cannot be separated. 
However, important efforts have been made to separate them spectrally
thanks to the availability of a large number of lines from different 
isotopologues and vibrational states (different opacity regimes) and 
a wide frequency range (different source coupling regimes).

\item Pointing errors, as small as 2'', could introduce important
  changes in the contribution from each cloud component to the
  observed line profiles, especially at 1.3 mm. However,
  the modeled and observed line profiles never differ by more than
  20\%, which is compatible with the absolute calibration error of our
  line survey (estimated to be about 15 \%).

\end{itemize}

\subsection{OCS}
\label{sect_col_ocs}

\begin{table*} %t13
\begin{center}
\caption{Column densities - OCS \label{tab_cd}}
\begin{tabular}{lllll}
\hline 
\hline 
 Species & Extended ridge & Compact ridge & Plateau & Hot core \\
 & $N$ $\times$10$^{15}$ (cm$^{-2}$) & $N$ $\times$10$^{15}$ (cm$^{-2}$) &
$N$ $\times$10$^{15}$ (cm$^{-2}$) & $N$ $\times$10$^{15}$ (cm$^{-2}$)\\
\hline
OCS & 2.0$\pm$0.5 & 3.0$\pm$0.8 & 7.5$\pm$1.9 & 15$\pm$4 \\
OCS assuming $^{32}$S/$^{34}$S=20 & 2.0$\pm$0.5 & 14$\pm$4 &
10$\pm$3 & 60$\pm$15 \\
OCS assuming $^{12}$C/$^{13}$C=45 & 2.7$\pm$0.5 & 18$\pm$4 &
13.5$\pm$3 & 45$\pm$9 \\
OCS (average) & 2.4$\pm$0.5 & 16$\pm$4 &
11.8$\pm$3 & 53$\pm$10 \\ 
OC$^{34}$S & 0.15$\pm$0.03 & 0.70$\pm$0.18 &
0.50$\pm$0.13 & 3.0$\pm$0.8 \\
OC$^{33}$S & 0.050$\pm$0.025 & 0.090$\pm$0.045 &
0.10$\pm$0.05 & 0.30$\pm$0.15 \\
O$^{13}$CS & 0.060$\pm$0.015 & 0.40$\pm$0.10 & 0.30$\pm$0.08 & 1.0$\pm$0.3 \\
$^{18}$OCS & 0.010$\pm$0.005 & 0.070$\pm$0.035 & 0.030$\pm$0.015 & 0.10$\pm$0.05 \\
O$^{13}$C$^{34}$S & $\lesssim$0.010 & $\lesssim$0.050
&$\lesssim$0.050 & $\lesssim$0.070 \\
$^{17}$OCS & $\lesssim$0.005
&$\lesssim$0.020 & $\lesssim$0.010 & $\lesssim$0.020 \\
OC$^{36}$S & $\lesssim$0.005 & $\lesssim$0.030 &
$\lesssim$0.020 & $\lesssim$0.030 \\
OCS $\nu_2$ = 1 & ... & ... & ... & 1.5$\pm$0.4 \\
OCS $\nu_3$ = 1 & ... & ... & ... & 0.15$\pm$0.07 \\
\hline 
\end{tabular}
\end{center}
%% Any table notes must follow the \end{tabular} command.
Note.-Column densities for the different OCS isotopologues and OCS
vibrationally excited derived from our model of the different Orion's
components (see text, Sect. \ref{sect_col_ocs}).
\end{table*}

Detailed multi-source LVG excitation and radiative transfer 
calculations were performed to fit
%Detailed MULTI-SOURCE LVG excitation and radiative TRANSFER calculations
the OCS line emission from the extended ridge, compact ridge, and
plateau. Given the lower density in these components
(similar or lower than the critical densities of the observed OCS transitions),
the OCS level populations should be far from LTE, thus the LVG
calculation is
much more appropriately adapted to interpreting the data correctly.
%n(H2)_cr = 1e5 cm^-3 (para la 12-11)
%n(H2)_cr = 6e5 cm^-3 (para la 23-22)
Collisional cross-sections of OCS-H$_2$ are
taken from \citet{gre78}, which were calculated 
for temperatures in the range 10-100 K including levels up to
$J$=13. In addition, we included levels up to $J$=40 in our models. 
Collisional rates for $J$$>$13 levels
were derived using the energy sudden approximation \citep{gol77} 
and using the $\sigma$ (0 -$>$ $J$; $J$$\leq$13) rates. The 
LTE approximation was assumed for both the hot core and 
vibrationally excited OCS. 
For OCS $\nu_2$=1 and OCS $\nu_3$=1,
we changed the velocity width parameter
for the hot core component ($\Delta$v = 5 km s$^{-1}$) with respect 
to the value given in Table \ref{tab_prop} to provide 
a closer accurate fit to the line profiles.

The beam coupling strongly affects the observed OCS lines in the
different frequency ranges. At 1.3 mm, HPBW $\simeq$ 10'',
we lose most of the compact ridge 
emission when pointing to IRc2. 
Moreover, the different gas components are not always
centered on the beam. Our model takes into account all these spatial
structure effects. As an example, Fig.
\ref{fig_ocs_map} shows the OCS emission at different cloud positions, with
the result that at velocities of between
7 and 11 km s$^{-1}$, the OCS emission peak is out side the 
telescope beam at 1.3 mm.
 
Although the relatively low dipole moment of OCS (0.715 D,
\citealt{tan85}) 
helps to keep these lines optically thin, some of them, especially 
at the higher end of the explored $J$ range, may be optically thick
\citep{ziu93, sch97}. 
The opacities were taken into account
by the LVG and LTE codes. However, both LVG and LTE approximations
are more appropriate for optically thin
% LVG and LTE approximations WORK BETTER for
%optically thin emission
%(yo creo que el modelo LTE de Pepe tb incluye los efectos de
%la opacidad en el transfer, no?).
emission; hence, the column density for the main
isotopologue obtained with our LVG or LTE calculations should be
considered as a lower limit. The derived column densities 
from the lines shown in Figs. \ref{fig_lin1}, \ref{fig_lin2}, and 
\ref{fig_lin3} are given in Table \ref{tab_cd}. We also derived
the column density of OCS indirectly by means of the column density of its less
abundant isotopologues to assess the line opacity effect
(OC$^{34}$S and O$^{13}$CS assuming
isotopic abundances of $^{32}$S/$^{34}$S=20 and $^{12}$C/$^{13}$C=45;
the adopted isotopic abundances are an average of the values
obtained in this work, see
Sect. \ref{sect_iso}). 
Owing to the low intensity  
of the lines belonging to these other less abundant isotopologues, 
implying larger overlap 
problems, we can only get upper limits for their column density. We 
estimate the uncertainty to be in the range 20-30 \% for the results of
OCS, O$^{13}$CS,
OC$^{34}$S, and OCS $\nu_2$ = 1 and around 50 \% for 
OC$^{33}$S, $^{18}$OCS, and OCS $\nu_3$ = 1. 

The OCS column density derived from the isotopologue emission 
in the compact ridge and the hot core
is four times higher than the column densities obtained
from the lines of the main isotopologue. It appears 
that the OCS lines emerging from the hot core and the compact ridge are  
saturated, this is consistent with the optical depth 
estimation of \citet{sch97} for the 29-28 transition of OCS 
($\tau$=3.5 assuming $^{32}$S/$^{34}$S = 22.5).    
For the plateau and the extended ridge, we obtained similar
column densities using both methods indicating that the
OCS main isotopologue emission towards these components 
is optically thin.

The component with the highest OCS column density corresponds to the hot 
core with 
$N$(OCS)$_{hot\;\;core}$ $\simeq$ (5$\pm$1)$\times$10$^{16}$ cm$^{-2}$. 
In addition, to obtain 
a good fit to the line profiles we need to add a contribution from the
extended ridge component of 
$N$(OCS)$_{extended\;\;ridge}$ =(2.4$\pm$0.5)$\times$10$^{15}$ cm$^{-2}$.
However, the analysis of the emission from positions in the extended
ridge far away from IRc2 
(see Fig. \ref{fig_com}) implies a much lower column density, 
$N$(OCS)$_{extended\;\;ridge\;\;position}$ =
(2.0$\pm$0.5)$\times$10$^{14}$ cm$^{-2}$. Hence, it appears that the
extended ridge also undergoes either a volume density increase or an OCS
abundance increase in the direction of the hot core. Otherwise,
the strong emission emerging from the hot core
may affect the excitation of the OCS energy levels in the extended ridge
(radiative scattering, see the HCO$^{+}$ and HCN cases discussed by
\citealt{cer87} and \citealt{gon93}).
Our results infer OCS column densities more than ten times
higher than in many previous studies \citep{joh84, tur91, bla87}.
These works provided beam-average column densities and
do not address the determination of the column density for each cloud 
component. 
The beam-averaged results from \citet{sut95} can be converted into
source-averaged column densities after multiplying by the beam dilution factor
($\simeq$ 3 for the hot core component, assuming d$_{hot\;\;core}$=10''
and FWHM$_{JCMT}$=13.7'' at the observed frequencies).
We note that \citet{sut95} found a
corrected-source-averaged column density of $\simeq$3$\times$10$^{16}$
cm$^{-2}$ and \citet{per07} estimated a source-average column density
of $\simeq$1.7$\times$10$^{16}$ cm$^{-2}$ both for the hot core
position whose values are 2 and 3 times lower than our result, respectively.
These values compare well with our measurement above.

We estimated a difference varying from 5\% to 15\%,
depending on the molecule, between LTE or LVG (for molecules having
collisional rates available) results.

\subsection{HCS$^+$}
\label{sect_col_hcs+}

We determine the HCS$^+$ column density using collisional rates
HCS$^+$-He from \citet{mon84}.

To reproduce the line profiles more accurately, we changed the
v$_{LSR}$ of the compact ridge given in Table \ref{tab_prop}, 
adopting v$_{LSR}$ = 9 km s$^{-1}$.
The modeled lines are shown in Fig. \ref{fig_hcs+} (thin curves). 
We obtain the following column densities: (5$\pm$2)$\times$10$^{13}$, 
(5$\pm$1)$\times$10$^{13}$, 
(8$\pm$2)$\times$10$^{13}$, and (1.0$\pm$0.3)$\times$10$^{12}$ 
cm$^{-2}$ for the hot core, plateau, compact ridge, and
extended ridge, respectively. To reproduce the 3 mm line of
HCS$^+$, we had to significantly reduce the
extended ridge column density with respect to the values of the other
components. The HCS$^+$ column density towards the hot core has to be 
considered with caution because of its weak line emission
contribution.

Based on the observed values of v$_{LSR}$ ($\simeq$ 9 km s$^{-1}$) 
and $\Delta$v ($\simeq$ 4 km s$^{-1}$) and the reduced fractional 
ionization in the
high density gas, \citet{joh84}, \citet{bla86}, and \citet{sch97} exclusively 
attributed the emission of
this molecule to the extended ridge. These first two sets of authors reported  
beam-average column densities of 
4$\times$10$^{13}$ \citep{joh84} and 1.6$\times$10$^{13}$
cm$^{-2}$ \citep{bla86}. 
\citet{sut95} found emission of HCS$^+$ from the five positions of
their survey (extended ridge, hot core, compact ridge, northwest
plateau, and southeast plateau); they obtained beam-averaged column
densities of $N$(HCS$^+$) = 6$\times$10$^{13}$
cm$^{-2}$ for the extended ridge and $N$(HCS$^+$) $\simeq$ 10$^{13}$ cm$^{-2}$
for the remaining positions.
\citet{sch01} found a questionable assignment of HCS$^+$ $J$=16-15 (E$_{up}$
= 278.5 K, emission coming from the hot core).

To compare the abundances of different molecular ions,
we calculated the column density of
H$^{13}$CO$^+$, assuming the same physical conditions we adopted
for HCS$^+$. As a total column density (sum of all components), we derive 
$N$(H$^{13}$CO$^+$) = 5.3$\times$10$^{13}$
cm$^{-2}$;
considering an isotopic
abundance $^{12}$C/$^{13}$C = 45 (see Sect. \ref{sect_iso}), we obtain
$N$(HCO$^+$)/$N$(HCS$^+$) $\simeq$ 13. A similar value was given by
\citet{joh84} (in the extended ridge component), whereas \citet{bla86} 
obtained $N$(HCO$^+$)/$N$(HCS$^+$)
$\simeq$ 94 (the H$^{13}$CO$^+$ line was strongly blended with
HCOOCH$_3$ and its T$_A$$^*$ was estimated by subtracting the
contribution of methyl formate from the overall emission).
%However,
%recent chemical models of hot molecular cores (\citealt{nom04}) give N(HCO$^+$)/N(HCS$^+$)
%$\simeq$ 100, for time-scales of 10$^4$ years (in Orion KL is observed large
%abundances of both parent and daughter molecules will mean that the
%ages of the chemically rich hot core are about 10$^4$ years). 
Assuming that the main production and destruction mechanisms for
HCO$^+$ and HCS$^+$ are the reaction of H$_3$$^+$ with CO and CS and
the dissociative recombination of HCO$^+$ and HCS$^+$ with electrons,
we deduce that in chemical equilibrium $N$(CS)/$N$(CO) =
1.5$\times$10$^{-3}$$\times$[$N$(HCS$^+$)/$N$(HCO$^+$)] $\simeq$
1.5$\times$10$^{-4}$ 
(see Sect. \ref{sect_col_cs} for the
obtained CO/CS abundance ratio).

\subsection{H$_2$CS}
\label{sect_col_h2cs}

\begin{table*} %t14
\begin{center}
\caption{Column densities - H$_2$CS \label{tab_cd_h2cs}}
\begin{tabular}{lllll}
\hline 
\hline 
Species & Extended ridge & Compact ridge & Plateau & Hot core \\
 & $N$ $\times$10$^{14}$ (cm$^{-2}$) & $N$ $\times$10$^{14}$ (cm$^{-2}$) &
$N$ $\times$10$^{14}$ (cm$^{-2}$) & $N$ $\times$10$^{14}$ (cm$^{-2}$)\\
\hline
o-H$_2$CS & 4$\pm$1 & 10$\pm$3 & 7$\pm$2 & 10$\pm$3\\
p-H$_2$CS & 1.5$\pm$0.4 & 5$\pm$1 & 3.0$\pm$0.8 & 6$\pm$2\\
\hline
o-H$_2$C$^{34}$S & 0.20$\pm$0.05 & 0.40$\pm$0.10 & 0.20$\pm$0.05 & 0.7$\pm$0.2 \\
p-H$_2$C$^{34}$S & 0.07$\pm$0.02 & 0.20$\pm$0.05 & 0.08$\pm$0.02 & 0.35$\pm$0.09 \\
\hline
o-H$_2$$^{13}$CS & 0.10$\pm$0.03 & 0.20$\pm$0.05 & 0.15$\pm$0.04 & 0.50$\pm$0.13 \\
p-H$_2$$^{13}$CS & 0.035$\pm$0.009 & 0.10$\pm$0.03 & 0.065$\pm$0.016
& 0.30$\pm$0.08 \\
\hline
HDCS & 0.40$\pm$0.10 & 0.60$\pm$0.15 & 0.40$\pm$0.10 & 0.8$\pm$0.2 \\
\hline
o-D$_2$CS & $\lesssim$0.10 & $\lesssim$0.20
&$\lesssim$0.10 & $\lesssim$0.40 \\
p-D$_2$CS & $\lesssim$0.050
&$\lesssim$0.10 & $\lesssim$0.050 & $\lesssim$0.20 \\
\hline 
\end{tabular}
\end{center}
%% Any table notes must follow the \end{tabular} command.
Note.-Modeled column densities for the different H$_2$CS isotopologues
(see text, Sect. \ref{sect_col_h2cs}).\\
\end{table*}

\begin{table*} %t15
\begin{center}
\caption{Ortho/Para ratios - H$_2$CS \label{tab_ortho_para}}
%\resizebox{0.8\textwidth}{!}{%
\begin{tabular}{lllll}
\hline 
\hline 
Ratio & Extended ridge & Compact ridge & Plateau & Hot core \\
\hline
o-H$_2$CS/p-H$_2$CS & 2.6$\pm$0.7 & 2.0$\pm$0.7 &
2.3$\pm$0.9 & 1.7$\pm$0.7\\
o-H$_2$C$^{34}$S/p-H$_2$C$^{34}$S & 3$\pm$1 & 2.0$\pm$0.7 & 2.5$\pm$0.9 & 2.0$\pm$0.8 \\
o-H$_2$$^{13}$CS/p-H$_2$$^{13}$CS & 3$\pm$1 & 2.0$\pm$0.8 & 2.3$\pm$0.8 & 1.7$\pm$0.6 \\
\hline 
\end{tabular}
%}
\end{center}
Note.-Ortho/Para ratios for the different H$_2$CS isotopologues (see
text, Sect. \ref{sect_col_h2cs}).\\
\end{table*}

Owing to the lack of collisional rates for this molecule, 
we assumed LTE excitation in the H$_2$CS column density calculations. Figures
\ref{fig_h2cs}, \ref{fig_h2c34s}, \ref{fig_h2_13cs}, and \ref{fig_hdcs}
show the modeled line profiles (thin curves) for selected lines of H$_2$CS,
H$_2$C$^{34}$S, H$_2$$^{13}$CS, and HDCS. Results 
are given in Table \ref{tab_cd_h2cs}. 
The higher column densities correspond to the the compact ridge
and the hot core component (1.5$\times$10$^{15}$ and
1.6$\times$10$^{15}$, respectively). The hot core is primarily
responsible for the
line emission from transitions with K$_a$$>$3. Our column density
results agree with those
obtained in previous studies (\citealt{sch97}; \citealt{sut95}; 
\citealt{tur91}; \citealt{bla87}; \citealt{sut85}). 

Since we derived the ortho- and para- H$_2$CS column densities
independently, we also computed the ortho-to-para ratios
of this molecule for the different components (Table
\ref{tab_ortho_para}). The hottest, densest component
(hot core)
has an ortho-to-para ratio $\simeq$ 1.8$\pm$0.7, whereas the extended ridge (the
coldest, least dense component) has a ratio $\simeq$ 3$\pm$1. Taking
into account the uncertainties in these ratios, we conclude that the
ratio is compatible with the statistical weight of 3.

Assuming the same hypothesis than for H$_2$CS, we derived a
H$_2$$^{13}$CO column density of $\simeq$ 1.2 $\times$ 10$^{15}$
cm$^{-2}$ (sum of all components). Adopting the isotopic abundance 
$^{12}$C/$^{13}$C = 45 (see Sect. \ref{sect_iso}), we derive 
$N$(H$_2$CO)/$N$(H$_2$CS) $\simeq$ 12, very close to
the ratio $N$(HCO$^+$)/$N$(HCS$^+$) calculated in the previous
section. Unlike H$_2$CO, for which efficient gas-phase synthetic
pathways have been studied in the laboratory, analogous reactions
that might form thioformaldehyde do not occur. As an example, the
chemical model of
\citet{nom04} cannot reproduce, by several order of magnitudes, the
observed $N$(H$_2$CO)/$N$(H$_2$CS) abundance ratio in hot cores.

\subsection{CS}
\label{sect_col_cs}

\begin{table*} %t16
\begin{center}
\caption{Column densities of CS, CS isotopologues, and CS vibrationally
  excited \label{tab_cdcs}}
\begin{tabular}{lllll}
\hline 
\hline 
Species & Extended ridge & Compact ridge & Plateau & Hot core \\
 & $N$ $\times$10$^{15}$ (cm$^{-2}$) & $N$ $\times$10$^{15}$ (cm$^{-2}$) &
$N$ $\times$10$^{15}$ (cm$^{-2}$) & $N$ $\times$10$^{15}$ (cm$^{-2}$)\\
\hline
CS assuming $^{32}$S/$^{34}$S=20 & 0.60$\pm$0.15 & 8$\pm$2 &
2.0$\pm$0.5 & 14$\pm$4 \\
CS assuming $^{12}$C/$^{13}$C=45 & 0.9$\pm$0.2 & 7$\pm$1 &
2.7$\pm$0.5 & 27$\pm$5 \\
CS (Average) & 0.8$\pm$0.2 & 8$\pm$2 &
2.4$\pm$0.5 & 21$\pm$5 \\
C$^{34}$S & 0.030$\pm$0.008 & 0.40$\pm$0.10 &
0.10$\pm$0.03 & 0.70$\pm$0.18 \\
C$^{33}$S & 0.005$\pm$0.001 & 0.10$\pm$0.03 &
0.050$\pm$0.013 & 0.40$\pm$0.10 \\
$^{13}$CS & 0.020$\pm$0.005 & 0.15$\pm$0.04 & 0.060$\pm$0.015 & 0.60$\pm$0.15 \\
$^{13}$C$^{34}$S & ... & 0.025$\pm$0.006
& ... & 0.040$\pm$0.010 \\
$^{13}$C$^{33}$S & ... & $\lesssim$0.007$\pm$0.002
& ... & $\lesssim$0.020$\pm$0.005 \\
C$^{36}$S & ... & $\lesssim$0.007$\pm$0.002 &
... & $\lesssim$0.020$\pm$0.005 \\
CS $\textit{v}$ = 1 & ... & ... & ... & $\lesssim$0.050$\pm$0.013 \\
\hline 
\end{tabular}
\end{center}
%% Any table notes must follow the \end{tabular} command.
Note.-Modeled column densities of the different CS isotopologues and CS
vibrationally excited (see text, Sect. \ref{sect_col_cs}).\\
\end{table*}

Our CS column densities were derived 
using collisional CS-H$_2$ rates
from \citet{liq07}. They
are given in Table \ref{tab_cdcs} and the modeled
line profiles are shown in Fig. \ref{fig_cs}. 

The CS lines are optically thick and therefore the column density
for each cloud component may be significantly underestimated. Lines from
CS isotopologues 
are, however, optically thin so that we can estimate the
column density of CS by assuming a value for the isotopic ratios. Assuming   
$^{32}$S/$^{34}$S = 20 (see Sect. \ref{sect_iso}), the column density 
of CS in the hot core component is 1.4$\times$10$^{16}$ cm$^{-2}$. 
A value 2 times larger is obtained if we assume that 
$^{12}$C/$^{13}$C = 45 (see Sect. \ref{sect_iso}). On average, we
obtain $N$(CS)$_{hot\;\;core}=$2.1$\times$10$^{16}$ cm$^{-2}$.
This CS column density is about 10-30 times larger than 
found in many previous studies \citep{bla87, lee01, sch01}. 
In these earlier studies, the results were beam-averaged CS
column densities derived from a LTE
analysis. \citet{sut95} obtained a corrected-source-averaged
column density of 1.5$\times$10$^{16}$ cm$^{-2}$ for the hot core,
in agreement with both our result and the source-averaged
CS column density obtained by \citet{com05}.

For the less abundant isotopologues ($^{13}$C$^{33}$S, C$^{36}$S) 
and for CS vibrationally excited states, 
we can only derive upper limits due to the weakness of the lines and 
the large overlap with other features (see Fig. \ref{fig_cs}, bottom
panels. Among the three lines of CS $\textit{v}$=1, only one seems to be detected).

The components with the largest CS column density are the hot core
and the plateau, the latter having the larger value. However, in the 
emission of the CS isotopologues the hot core dominates (in agreement
with \citealt{sch01}).

To compare the CS and CO abundances, we
calculated the column density of C$^{18}$O in each component. We
obtain $N$(C$^{18}$O) of 1.5$\times$10$^{16}$, 1.5$\times$10$^{16}$,
1$\times$10$^{17}$, and 2$\times$10$^{17}$ cm$^{-2}$ for the 
extended ridge, compact ridge, plateau, and hot core, respectively.
We have to include a
high velocity plateau component with v$_{LSR}$=10 and $\Delta$v=55 km
s$^{-1}$ and a column density of 5$\times$10$^{16}$ cm$^{-2}$ to
reproduce the line profiles. Assuming the isotopic
abundance $^{16}$O/$^{18}$O=250 (see Sect. \ref{sect_iso}), 
we determine the column density of CO in each
component to be 3.75$\times$10$^{18}$, 3.75$\times$10$^{18}$,
2.5$\times$10$^{19}$, 5.0$\times$10$^{19}$, and 1.25$\times$10$^{19}$
for the extended ridge, compact ridge, plateau, hot core, and high
velocity plateau, respectively. Therefore, the corresponding
$N$(CS)/$N$(CO) ratio is 2.0$\times$10$^{-4}$, 2.0$\times$10$^{-3}$,
1.0$\times$10$^{-4}$, and 4.1$\times$10$^{-4}$ for the extended ridge,
the compact ridge, the plateau, and the hot core, respectively. 
In all cases, this ratio is in good
agreement with the $N$(CS)/$N$(CO)$\simeq$1.5$\times$10$^{-4}$, derived from
$N$(HCS$^+$)/$N$(HCO$^+$). 

When we fitted the line emission of CS, we found that it was difficult to 
distinguish the contribution of the high velocity plateau to the line profiles
from those of the other components. 
Assuming $N$(CS)/$N$(CO) = 1.5$\times$10$^{-4}$, the column density of
CS in the high velocity plateau would be $\simeq$2.0$\times$10$^{15}$
(peak T$_A$$\simeq$5 K).

\subsection{CCS}
\label{sect_col_ccs}

Collisional cross-sections of CCS-H$_2$ were
extrapolated from those of OCS \citep{gre78} using the IOS
approximation for a $^3\Sigma$ molecule (see \citealt{cor84a};
\citealt{cor84b}; \citealt{fue90}).
In this case, we changed the velocity parameters
for the hot core component with respect to the parameters given in
Table \ref{tab_prop} to reproduce the line 
profiles more accurately. 
The new values are v$_{LSR}$ = 5 km s$^{-1}$ and $\Delta$v = 6
km s$^{-1}$.   

The modeled lines are 
shown in Fig. \ref{fig_ccs} (thin lines).
The values of the column densities are:
(5.0$\pm$1.3)$\times$10$^{13}$, (7.0$\pm$1.8)$\times$10$^{12}$, 
(2.0$\pm$0.5)$\times$10$^{12}$, and (2.0$\pm$0.5)$\times$10$^{12}$ 
cm$^{-2}$ for the hot core, the compact ridge, the extended ridge, and 
the plateau, respectively. We note that \citet{tur91} reported the 
first tentative detection of CCS in this
source with a beam-averaged column density of 4.8$\times$10$^{12}$
cm$^{-2}$.

\subsection{C$_3$S}
\label{sect_col_c3s}

For this molecule, we considered that only the hot core component
is responsible for the emission, hence we assume 
LTE excitation. We chose the same physical conditions for this 
component as in the CCS analysis.
Figure \ref{fig_cccs} shows the modeled line profiles for some
selected lines, for $N$(C$_3$S) = (2.0$\pm$0.5)$\times$10$^{13}$
cm$^{-2}$. Taking into account the CCS
column density, we derive the ratio
CCS/C$_3$S = 2.5 which is similar to the value of $\simeq$ 3.5 found in the
dark cloud \object{TMC-1} by \citet{hir92} and the value of $\simeq$ 3 found 
in the envelope of the C-rich star \object{IRC+10216} by \citet{cer87a}
(note that we corrected this last value for the dipole moment of C$_3$S adopted 
in this study, 3.7
versus 2.6 in \citealt{cer87a}).

\subsection{Non-detected CS-bearing molecules}
\label{sect_col_non}

OC$_3$S.- The molecule OC$_3$S has not yet been detected in space. 
The rotational constants used to derive the line frequencies
were taken from \citet{win00}.
The dipole moment we used ($\mu$=0.63D) is quoted
in \citet{mat87}. Assuming the same
physical conditions as those derived for OCS, we obtain an
upper limit to its column density of 2$\times$10$^{13}$ cm$^{-2}$.
This result provides an OCS/OC$_3$S abundance ratio larger than 100.

H$_2$CCS.- For this molecule, we derived its line frequencies with
the rotational constants given in \citet{win80}; some distortion
constants were fixed to the value obtained from infrared data
by \citet{mcn96}.
The dipole moment ($\mu$=1.02D) was taken from \citet{geo79}.
We derive an upper limit to the column density of
thioketene of 2.4$\times$10$^{14}$ cm$^{-2}$, which infers a
H$_2$CS/H$_2$CCS abundance ratio near 20. This molecule has not
yet been detected in space. 

HNCS.-Isothiocyanic acid is a pseudolinear molecule with a large A
rotational constant, similar to that of isocyanic acid, HNCO. 
Only transitions up to Ka = 1 have been 
observed in the interstellar medium (in SgrB2 by \citealt{fre79}). 
However, this molecule has not yet
been detected in Orion. \citet{tur91} reported a tentative
detection of HNCS in Orion and listed five transitions as detected,
but three of them were not reliable. Turner derived an LTE column
density of 9.3 $\times$ 10$^{12}$ cm$^{-2}$ assuming T$_{rot}$ = 50 K 
based on a single transition. 
For frequency predictions, we used the rotational
constants presented by \citet{nie95}.
The a-dipole moment component ($\mu_a$=1.64D) was mentioned in
\citet{sza78}.
We derive an upper limit to the column density of 1.1 $\times$ 10$^{14}$
cm$^{-2}$. \citet{mar09a} calculated $N$(HNCO) towards Orion KL
from this survey, to be $N$(HNCO) $\simeq$ 9.4 $\times$ 10$^{15}$
cm$^{-2}$; these values imply a HNCO/HNCS ratio $>$85.  

HOCS$^+$.- Spectroscopic constants are taken from \citet{ohs96}.
The dipole moment ($\mu$=1.517D) was calculated by \citet{whe06}.
We obtain an upper limit to the column density of this
cation of $N$(HOCS$^+$) $\lesssim$ 3 $\times$ 10$^{13}$ cm$^{-2}$. This
result and the high column density of OCS may indicate
that this ion is efficiently destroyed by dissociative recombination
to produce OCS + H \citep{cha97}. 

NCS.- Thiocyanogen has not yet been detected in space. 
The rotational constants used to derive the line frequencies
were taken from CDMS Catalog. The dipole moment ($\mu$=2.45D)
is from an {\it{ab initio}} calculation by H. S. P. M\"uller (unpublished).
We derive here $N$(NCS) $\lesssim$ 7 $\times$ 10$^{13}$ cm$^{-2}$.

\section{Isotopic abundances}
\label{sect_iso}

From the derived column densities for OCS, H$_2$CS, CS, 
and their isotopologues, we can now estimate the
isotopic abundance ratios. They are given in Table \ref{tab_iso},
which is only available online.
The isotopic ratios that are not discussed in the following given but in Table
\ref{tab_iso} are consistent with the solar values (taking into account
a factor of 2 introduced by the $^{12}$C/$^{13}$C solar abundance, see
below).
 
Because of the large opacity of the OCS emission in the hot core and the
compact ridge, we can only provide a lower limit to 
the OCS column density ratios in these components. In the same way,
the column density ratios
O$^{13}$CS/O$^{13}$C$^{34}$S,
OC$^{34}$S/O$^{13}$C$^{34}$S, OCS/$^{17}$OCS, OCS/OC$^{36}$S, 
$^{13}$CS/$^{13}$C$^{34}$S, C$^{34}$S/$^{13}$C$^{34}$S, 
$^{13}$CS/$^{13}$C$^{33}$S, C$^{33}$S/$^{13}$C$^{33}$S 
represent lower limits due to the low intensity of the lines
and the strong blending overlap with other molecular lines.
From the remaining column density ratios of Table \ref{tab_iso}, we
estimated the following isotopic abundances:

$^{12}$C/$^{13}$C: From the OCS, lines we obtained a column
density ratio of $N$(OCS)/$N$(O$^{13}$CS) = 33$\pm$12 and 25$\pm$9 for the 
extended ridge and the
plateau, respectively. From H$_2$CS (o- and p-), we obtained
$N$(H$_2$CS)/$N$(H$_2$$^{13}$CS) = 42$\pm$16, 50$\pm$20, 47$\pm$18, and
20$\pm$9 for the extended ridge, compact ridge, plateau, and hot core,
respectively. The values estimated with the H$_2$CS lines are
slightly higher (except the value for the hot core) than those
derived from OCS, which is indicative of a low
opacity in the OCS lines coming from the plateau and
the extended ridge, and in the hot core emission of H$_2$CS. We
find an average value from our study of
$^{12}$C/$^{13}$C = 45$\pm$20.
Previous studies found that $N$(OCS)/$N$(O$^{13}$CS) 
$\simeq$ 30-40 \citep{joh84} and 
$^{12}$C/$^{13}$C $\simeq$ 30-40
(\citealt{bla87}, who used several molecules  
-CS, CO, HCN, HNC, OCS, H$_2$CO, CH$_3$OH- to achieve
tighter constraints), $N$(CN)/$N$($^{13}$CN) = 43$\pm$7
\citep{sav02}, and $N$(CH$_3$OH)/$N$($^{13}$CH$_3$OH) = 57$\pm$ 14
\citep{per07}.
The solar
isotopic abundance of $^{12}$C/$^{13}$C = 90
\citep{and89} is approximately a factor 2 higher than the values
obtained in Orion. This ratio is understood to be a sensitive indicator
of the degree of galactic chemical evolution and the solar isotope
value reflects conditions in the interstellar medium at an earlier
epoch (\citealt{sav02}; \citealt{wyc00}).    

$^{32}$S/$^{34}$S: From the values obtained of
$N$(OCS)/$N$(OC$^{34}$S) and p-/o- $N$(H$_2$CS)/$N$(H$_2$C$^{34}$S), we
estimate an average value $^{32}$S/$^{34}$S = 20$\pm$6, in agreement
with the solar isotopic abundance and with previous studies
($N$(OCS)/$N$(OC$^{34}$S) 
$\simeq$ 16 by \citealt{joh84}, $^{32}$S/$^{34}$S $\simeq$ 13-16 by
\citealt{bla87}, $N$($^{32}$SO)/$N$($^{34}$SO) = 21$\pm$6, and
$N$($^{32}$SO$_2$)/$N$($^{34}$SO$_2$) = 23$\pm$7 by \citealt{per07}).

$^{32}$S/$^{33}$S: From $N$(OCS)/$N$(OC$^{33}$S) in the extended ridge, 
we obtained a $^{32}$S/$^{33}$S ratio three times lower than the
solar abundance. It is
possible that we overestimated the column density of OC$^{33}$S in
the extended ridge because we had only three blending-free transitions 
to compare to the model (see Sect. \ref{sect_col_ocs}). 
For the plateau component, we obtained OCS/OC$^{33}$S =
75$\pm$29, in close agreement with the solar isotopic
abundance. \citet{per07} obtained $^{32}$S/$^{33}$S $\simeq$ 103-113.

$^{33}$S/$^{34}$S: The $^{33}$S/$^{34}$S abundance ratio is
$\simeq$ 0.22$\pm$0.10 and $\simeq$ 0.37$\pm$0.10 from OCS and CS
respectively, i.e., very close to the solar values and in agreement
with \citet{per07}.

$^{16}$O/$^{18}$O: This ratio was inferred from $N$(OCS)/$N$($^{18}$OCS). 
The ratio $^{16}$O/$^{18}$O obtained (250$\pm$135 for the plateau) is
two times lower than the solar value
for all the cloud components. 
However, taking into account the uncertainties in the column density of
both the plateau and the extended ridge, we consider that the
true values may be compatible with the solar one. Similar
conclusions can be obtained from the observed $^{18}$OCS/OC$^{34}$S
and $^{18}$OCS/OC$^{33}$S abundance ratios (see Table \ref{tab_iso}).

D/H: We found a $N$(HDCS)/$N$(H$_2$CS) column density ratio of 0.07$\pm$0.03,
0.040$\pm$0.012, 0.040$\pm$0.012, and 0.05$\pm$0.02 for the extended
ridge, compact ridge,
plateau, and hot core, respectively. 
\citet{par01b} found a $N$(HDO)/$N$(H$_2$O)
abundance ratio in the range 0.004-0.01 in
the plateau component, and \citet{per07} derived 0.005, 0.001, and 0.03 for
the large velocity plateau, the hot core, and the compact ridge, respectively. 
For $N$(HDCO)/$N$(H$_2$CO),
\citet{per07} derived a
value of 0.01 (for the compact ridge), whereas \citet{tur91} 
found $\simeq$0.14 (note that this value is too high and is incompatible
with the H$_2$CO column density reported in this work and by several
authors, see \citealt{sut95}). 
Water and formaldehyde present its higher deuterium fractionation in the
compact ridge component.
\citet{sch92} derived the DCN/HCN column density ratio of 0.001 and
0.01-0.06 for the hot core and the ridge region, respectively.  
Studies of hot core deuterium chemistry \citep{rod96} conclude
that the D/H ratios of molecules injected from the dust mantles to the
hot gaseous medium do not undergo significant modifications and should
represent those of the original mantles for molecules that were
efficiently deposited during the cold phase (such as water,
methanol, and formaldehyde). However, H$_2$CS is not considered to be a 
molecule deposited in the original mantles and its high deuteration 
may be caused by gas phase reactions.
Very high deuterium fractionation has been also found in cold
molecular clouds (\citealt{mar05}; \citealt{rob02}).

\section{Vibrational temperatures}
\label{sect_vib}

We report the first space detection of rotational line emission
from OCS $\nu_2$ = 1 and $\nu_3$ = 1 vibrational levels.
Given the high energy of these vibrational levels, the emission 
is dominated by the hot core component. 
From the column density obtained for OCS in the 
ground and the
vibrationally excited states, we can estimate a vibrational
temperature taking into account that
\begin{equation}
\frac{exp \left( \begin{array}{c} - \frac{E_{\nu_x}}{T_{vib}}\end{array} 
\right)}{f_{\nu}} = \frac{N
  (OCS\;\;\nu_x)}{N (OCS)},
\end{equation} 
where E$_{\nu_x}$ is the energy of the vibrational state (E$_{\nu_{2}}$
= 748.7 K; E$_{\nu_{3}}$
= 1235.9 K), T$_{vib}$ is the vibrational temperature,
f$_{\nu}$ is the vibrational partition function, $N$(OCS $\nu_x$) is the column
density of the vibrational state, and $N$(OCS) is the column density of
OCS in the ground state. The vibrational partition function can be 
approximated by

%% The eqnarray environment produces multi-line display math. The end of
%% each line is marked with a \\. Lines will be numbered unless the \\
%% is preceded by a \nonumber command.
%% Alignment points are marked by ampersands (&). There should be two
%% ampersands (&) per line.

\begin{eqnarray}
f_{\nu} = 1 + exp \left( \begin{array}{c} -
  \frac{E_{\nu_{3}}}{T_{vib}} \end{array} \right) + \nonumber \\
 + 2 exp \left( \begin{array}{c} -
  \frac{E_{\nu_{2}}}{T_{vib}} \end{array} \right) 
+ exp \left( \begin{array}{c} -
  \frac{E_{\nu_{1}}}{T_{vib}} \end{array} \right),  
\end{eqnarray}\\
which, for low T$_{vib}$ leads to f$_{\nu}$ $\simeq$ 1.

From the observed lines, we obtain T$_{vib}$ =
210$\pm$10 K for OCS $\nu_2$ = 1, and T$_{vib}$ = 
210$\pm$60 K for OCS $\nu_3$ = 1. 
These values are similar to the
averaged kinetic temperature we adopted for the hot core component (225 K). 
%These values are higher than the
%averaged kinetic temperature we have fixed for the hot core component (225 K). 
%One could expect that the line emission of OCS vibrationally excited
%comes from an inner and hotter part of the hot core due to the high 
%energies involved and the fact that lower line-widths are needed to reproduce
%the line profiles (see Sect. \ref{sect_col_ocs}). However, this result
%for the vibrational temperature do not show that scenario. 
%This result point to that the OCS vibrationally excited emission lines 
%probably come from an inner and hotter part of the hot core (also
%reflected in the lower line widths we have needed to fit the line
%profiles, see Sect. \ref{sect_col_ocs}).
%Concerning to the excitation mechanims for these lines,
%radiative pumping effects in the populations of 
%the $\nu_2$ and $\nu_3$ states could be important due to the high
%critical densities of the vibrationally excited lines.}} 
A direct comparison of the derived T$_{vib}$ for OCS with the average
T$_k$ assumed for the gas in the hot core is difficult. Vibrational
excitation is expected to depend strongly on temperature and density
gradients in that region. It is also difficult to ascertain if either IR dust
photons or molecular collisions dominate the vibrational excitation of OCS
given the lack of collision rates for that
species. Nevertheless, assuming that
ro-vibrational collision rates for OCS are similar to those
of SiO or SiS \citep{tob08}, $\sigma$($\nu_2$=0$\rightarrow$1,300 K)
$\simeq$ 10$^{-14}$ cm$^3$\,s$^{-1}$,
we find that, even for H$_2$ densities as high as 10$^9$ cm$^{-3}$,
the net collisional rate is well below the spontaneous de-excitation
rates from $\nu_2$ and $\nu_3$ to the ground state. Hence, the
population of these levels have to be mainly caused by IR photons
from the dust. That the OCS rotational lines are narrower
in vibrationally excited states than in $\nu$=0 may indicate that this
IR pumping operates in a more compact region with a shallower velocity
gradient.
%If the molecular ro-vibrational
%transitions have significant optical depths, the population of the
%energy levels can be modified by the strong IR
%continuum radiation. Radiative pumping does not
%affect significantly the populations of the rotational levels within
%the ground vibrational state of heavy molecules. However, the total 
%populations of the excited vibrational states of molecules such 
%as HC$_3$N, CH$_3$CN and CH$_3$OH are probably modified by this effect
%\citep{sut86, sut95}. 
%El fit de las líneas no es perfecto porque no hay excitación
%radiativa en el lvg
%All together suggests that the excitation temperature
%varies across the hot core, thus higher angular resolution is necessary
%to resolve this excitation gradient and the temperature profile in
%this component.
%This suggests that the excitation
%temperature varies across the hot core, thus higher angular
%resolution is necessary to RESOLVE this EXCITATION GRADIENT
%AND THE TEMPERATURE PROFILE in the hot core.
Higher angular resolution observations are necessary
to resolve any possible excitation gradient and temperature
profile in this component.

We also calculated
the vibrational temperature for the first vibrationally excited state of CS
(E$_{\textit{v}=1}$ = 1830.4 K). 
With the column density results 
of the CS hot core component (2.1$\times$10$^{16}$cm$^{-2}$) and 
of CS $\textit{v}$ = 1 ($\lesssim$5$\times$10$^{13}$cm$^{-2}$), we
obtain an upper limit to 
the vibrational temperature of $\simeq$300 K, which
agrees with the values obtained for vibrationally
excited OCS. However, in this case we could claim an inner and hotter emitting
region for vibrationally excited CS.

Nevertheless, since the vibrationally excited 
gas is not necessarily spatially coincident with the ground state gas,
the derived vibrational temperatures have to be considered as
lower limits.

\section{Discussion and conclusions}
\label{sect_dis}

The power of spectral line surveys at
different mm and sub-mm wavelengths to search
for new molecular species and derive the physical and chemical structure
of molecular sources
has been demonstrated (\citealt{bla87}; \citealt{sut95};
\citealt{cer00}; \citealt{sch01}; \citealt{par07}).
The main and final goal of our line survey is to provide a consistent set of
molecular abundances derived from a systematic analysis of the molecular
rotational transitions. Our line survey allows us to obtain with
unprecedented sensitivity and completeness the
census of the identified and
unidentified molecules in Orion KL.
These kinds of studies are necessary to understand the chemical evolution
of this archetypal star-forming region. Moreover, that 
many rotational transitions of the same
molecule have been observed in different frequency ranges
(the 3 mm window illustrates more clearly the extended ridge component, whereas
the 1.3 mm one identifies the
warmest gas at the hot core and along the compact ridge), provide
strong observational constraints on the source structure,
gas temperature, gas density, and molecular column densities.

\subsection{Molecular abundances}
\label{sect_dis_abu}

Molecular abundances were derived using the H$_2$ column density
calculated by means of the C$^{18}$O column
density provided in Sect. \ref{sect_col_cs}, 
assuming that CO is a robust tracer of
H$_2$ and therefore their abundance ratio is roughly constant,
ranging from CO/H$_2$
$\simeq$ 5$\times$10$^{-5}$ (for the ridge components) to
2$\times$10$^{-4}$ (for the hot core and the plateau). In spite of 
the large uncertainty in this calculation, we
include it as a more intuitive result for
the molecules described in the paper.
We obtained N(H$_2$) = 7.5$\times$10$^{22}$, 7.5$\times$10$^{22}$,
2.1$\times$10$^{23}$, and 4.2$\times$10$^{23}$ cm$^{-2}$ for the extended
ridge, compact ridge, plateau, and hot core, respectively. In addition, 
we assume that the
H$_2$ column density spatially coincides with the emission from the
species considered.
Our estimated source average abundances for each Orion KL component 
are summarized in
Table \ref{tab_abun} (only available online), 
together with comparison values from other
authors (\citealt{sut95} and \citealt{per07}). 
The differences between the abundances shown in Table \ref{tab_abun}
are mostly due to the different
H$_2$ column density considered, to the assumed cloud component of the
molecular emission and discrepancies in
the sizes of these components.  

\subsection{Column density ratios}
\label{sect_dis_rat}

\begin{table*} %t19
\begin{center}
\caption{Column density ratios\label{tab_rat}}
\resizebox{0.9\textwidth}{!}{%
\begin{tabular}{l|lllll|lllll}
\hline
\hline 
 Column & & This & work & & & $^{(1)}$ & $^{(1)}$ &
 $^{(2)}$ & Dark clouds & Hot core\\
 Density ratio & ER & CR & P & HC & Total & A & B &  &
 (TMC1) & \object{G327.3-0.6}\\
\hline
OCS/CS & 3 & 2 & 5 & 2 & 3 & 5 & 13 & 6(HC) & 0.5 & $<$4\\
CS/HCS$^+$ & 7700 & 100 & 50 & 420 & 180 & 1000 &
270 & ... & 10 & ...\\
CS/H$_2$CS & 14 & 50 & 2 & 12 & 7 & 25 & 8 & 4(CR) & 6 & $>5$\\
CS/CCS & 4000 & 1100 & 1200 & 420 & 530 & ... & ... & ... & 0.5 & 325\\ 
CS/C$_3$S & ... & ... & ... & 1050 & 1050 & ... & ... & ... & 4 & ...\\
CCS/C$_3$S & ... & ... & ... & 2.5 & 2.5 & ... & ... & ... & 8 & ...\\
CO/CS & 5000 & 500 & 10000 &
2500 & 2800 & 66700 & 133300 & 2400(P) & 20000 & 26200$^{(3)}$\\
HCO$^+$/HCS$^+$ & 270 & 4 & 9 & 27 & 13 & 12 & 10 & ... & 20 & ...\\
H$_2$CO/H$_2$CS & 12 & 8 & 18 & 11 & 12 & 6250 &
1250 & 15(CR) & 71 & $>3$\\
\hline 
\end{tabular}
}
\end{center}
$^{(1)}$: \citet{nom04}\\
$^{(2)}$: \citet{per07}\\
$^{(3)}$: assuming $^{16}$O/$^{17}$O = 2625\\
Note.-Derived column density ratios and comparison with other works
and sources. Column 1 gives the considered ratio, Cols. from 2 to 6
show the results obtained in this work in the different spectral cloud
components of Orion and the total value, Cols. 7 and 8 
ratios obtained by \citet{nom04} in their models of hot
cores (in model B trapping of mantle molecules in water ice is assumed), 
Col. 9 gives values in Orion by
\citet{per07}, Col. 10 in dark clouds (TMC-1) (\citealt{wal09} and
references within), and Col. 11 provides these ratios for the molecular
hot core G327.3-0.6, \citet{gib00}.\\
\end{table*}

To compare the chemistry of the different spectral cloud
components related to sulfur-bearing carbon chains molecules,
we derived the column density ratios showed in Table
\ref{tab_rat}. This table also shows the ratios found in chemical
models of hot cores, other results found in the literature for Orion, 
and other sources (the dark
cloud TMC-1 and the hot core G327.3-0.6).
We found good agreement between our ratios and those derived by
\citet{per07}, both set of values corresponding to Orion KL. For
the other molecular hot core, we noted a large difference in
the ratio CO/CS. This discrepancy also occurs with the chemical models
computed by \citet{nom04}. We note that the chemical models cannot
provide realistic values for the H$_2$CO/H$_2$CS column density ratio, 
as we have discussed in Sect. \ref{sect_col_h2cs}. 
TMC-1 exhibits ratios very different
by those of hot cores, as expected from their different chemical and physical
conditions.
 
We find $N$(C$^{34}$S/OC$^{34}$S)$\simeq$0.3, 0.6, 0.2, and 0.2 in the
extended ridge, the compact ridge, the plateau, and the hot core,
respectively. The chemical models for hot cores
computed by \citet{nom04} infer that $N$(CS)/$N$(OCS) = 0.2 (at 10$^{4}$ years). 
The $N$(CS)/$N$(CCS) abundance ratio
is 300, 1143, 1000, and 280 for the
extended ridge, the compact ridge, the plateau, and the hot core,
respectively. For the hot core, we also derive 
$N$(CS)/$N$(C$_3$S)$_{Hot\;\;Core}$=700. Both CCS 
and C$_3$S have
not been studied in the chemical models available for hot cores. As expected,
these values are very different from those derived in the dark
cloud TMC-1 for which $N$(CS)/$N$(CCS)=2.2 and $N$(CS)/$N$(C$_3$S)=7.8
\citep{hir92}. 
However, we obtain $N$(C$_2$S)/$N$(C$_3$S) = 2.5, very similar to the 
3.4 value derived by \citet{hir92} in TMC-1 (cyanopolyyne peak) and 
the value of
$\simeq$ 3 found in the envelope of the C-rich star IRC+10216 by 
\citet{cer87a}. 

This is a surprising result because CCS
is considered to be a typical molecule in cold dark clouds. Moreover, C$_3$S
is found only in the hot core, which is indicative of an enhancement in the production
of CCS and C$_3$S in the warm and dense gas. Although spectral confusion
is large when observing weak lines such as those of C$_3$S, thanks to
our survey, we detected
17 lines. They cover from the J=14-13 (E$_{up}$=29.1 K
with v$_{LSR}$=4.2 kms$^{-1}$)
up to J=47-46 (E$_{up}$=313 K with v$_{LSR}$=4.2 kms$^{-1}$),
thus, we are fully confident in its detection. In addition, the
observed velocities
correspond definitively to the hot core. 

Our results indicate that
C$_3$S is efficiently formed in warm regions. That the
C$_2$S/C$_3$S abundance ratio is similar to that of dark clouds or
evolved stars may indicate that these species formed in the gas
phase. Gas phase chemical models predict C$_2$S/C$_3$S $\simeq$ 2 and 0.3
in TMC-1 and IRC10216, respectively (\citealt{wal09}; \citealt{cor09}).
%In both models the effects of molecular anions have been included)

%The nature of the hot core has been widely discussed by several
%authors. The amount of -CN bearing molecules and the lack of
%many carbon-rich chains are proves that the hot core is dominated by a
%a O-rich chemistry (REFERENCIAS).  
%There are clear indications of that the hot core is composed of
%condensations with very different temperatrures (this work,
%\citealt{dev02}, \citealt{wri96}).
%However, there are controversies in different points: 
%Is the hot core is internally
%heated or the radio source I (considered the main heating source
%in the region \citealt{men95}, \citealt{gez98}, \citealt{bla96}) 
%heats this molecular component?. \citealt{bla96} have pointed out that
%there is no
%evidence of internal heating within the molecular hot core, 
%based on the distribution of the HC$_3$N J = 24-23 line in
%the 1v7 vibrationally excited state. However, \citealt{kau98} and
%\citealt{dev02} proposed that it is interanlly heated by young embedded
%proto-stars. The work of \citealt{dev02} was positive indicating that the
%radio souce I may not dominate
%the heating of the hot core, based on the distribution in the hot core of the J
%= 10-9 rotational line of HC$_3$N in the vibrationally excited level
%1v5.
\subsection{Orion KL cloud structure}
\label{sect_dis_str}

We have analyzed and discussed the emission
lines of the studied
molecules in terms of the four well-known Orion KL
cloud components (hot core, extended ridge, compact ridge, and
plateau). 
However, low angular resolution
does not enable us to detect any possible
%the vibrational temperature calculated here suggests
variation in the excitation temperature across the hot core and
the other Orion components. A more complex physical structure has 
been observed with sensitive interferometers 
\citep{wri96, beu05, pla09, zap09a}.
Further analysis of our survey indicates that at the position of IRc2, the
lines of both SiS and the SiO maser emission 
show a velocity component at 15.5 km s$^{-1}$, an additional cloud
component to those described above.
Owing to the high energies involved in some emission lines,
\citet{sch01} and \citet{com05} claimed that a hotter component exists
at the
hot core v$_{LSR}$ in their surveys at high frequency. In the
same way, we detected the emission of vibrationally excited OCS and
CS at the hot core LSR velocity.
%that cloud
%corresponds to the feature we have detected in SiS and SiO and in the
%vibrational states of SiO, CS and OCS.
In spite of the low angular resolution of our data, the amount
of molecules, the large number of
transitions, and the different vibrationally excited states found in
the survey permit us to derive realistic source-averaged physical
and chemical parameters.

\section{Summary}
\label{sect_sum}

We have presented an IRAM 30-m line survey of Orion KL with the
highest sensitivity  
achieved to date. Because of the wide frequency range covered and high data 
quality, we intent to present the line survey in a series of papers
focused in different molecular families. 
In this paper, we have presented the study of 
the emission from OCS, its isotopologues, and its vibrationally 
excited states ($\nu_2$ = 1 
and $\nu_3$ = 1), as well as HCS$^+$, H$_2$CS, CS, CCS, CCCS,  and 
different isotopic substitutions of them. The four well known 
components of Orion (hot core, plateau, extended ridge and compact
ridge) contribute to the observed emission from 
the main isotopologues of all these molecules except for C$_3$S,
which we only detected in the hot core component. 

Column densities have been calculated with radiative transfer codes 
based on either the LVG or the LTE approximations, taking into account
the physical structure of the source. Results are provided as source-averaged
column densities. In this way, our column density for
OCS are between 
4 to 10 times higher than the beam-averaged ones provided in 
previous surveys. 
Our column density derived for the hot core component compares well
with the values obtained by \citet{sut95} and \citet{com05}.
Among those
studied in this paper in all the cloud components,
OCS appears as the most abundant species.

We have also reported on the first detection in space of OCS 
in the vibrationally excited states $\nu_2$ = 1 and $\nu_3$ = 1. This 
emission arises mostly from the hot core component. 
The resulting 
vibrational temperature ($\simeq$ 210 K) is similar to the kinetic 
temperature of this component ($\simeq$ 225 K).

For HCS$^+$, we have to significantly reduce the contribution of the
extended ridge component to reproduce all lines arising in
our survey. The derived $N$(HCS$^+$)/$N$(HCO$^+$) ratio is in agreement
with the observed $N$(CS)/$N$(CO) ratio in terms of chemical equilibrium.  

The statistical value of 3 for the ortho-to-para ratio has been
confirmed by the study of the emission lines of H$_2$CS and its
isotopologues. We have derived an abundance ratio of H$_2$CS/HDCS
$\simeq$ 20.

For CS, we have analyzed the emission of 7 distinct isotopologues 
and the vibrational state $\textit{v}$ = 1. 
The lines of the main isotopologue 
are optically thick and therefore we have derived the column density by means 
of the isotopologues (assuming the derived isotopic abundances from
this work). Our results 
are in agreement with those of \citet{sut95} and \citet{com05}.
The vibrational temperature
calculated for CS $\textit{v}$ = 1 (tentative detection) is less than 300 K, 
in agreement with those values obtained for OCS $\nu_2$ = 1 and $\nu_3$=1.
However, this result may indicate that there is an inner, hotter region of
emission from vibrationally excited CS that is difficult to resolve with our low
angular resolution observations.

Since the intensity of the C$_2$S and C$_3$S lines is already weak, 
we could not detect either their isotopologues or their vibrationally excited
states. 
We derived column densities of 5$\times$10$^{13}$ cm$^{-2}$ and
2$\times$10$^{13}$ cm$^{-2}$, respectively in the hot
core component from the detected lines. We have detected C$_3$S for
the first time in warm regions. 
Finally, we have derived upper limits to the column density of
non-detected molecules (-CS bearing species), obtaining the maximum
value for thioketene $N$(H$_2$CCS) $\lesssim$ 2.4$\times$10$^{14}$ cm$^{-2}$.

From the column density results, we have derived several abundance
ratios that permit us to provide the following average isotopic abundances: 
$^{12}$C/$^{13}$C = 45$\pm$20, $^{32}$S/$^{34}$S
= 20$\pm$6, $^{32}$S/$^{33}$S = 75$\pm$29, 
and $^{16}$O/$^{18}$O = 250$\pm$135
(all of them in agreement with
the solar isotopic abundances, see Sect. \ref{sect_iso} and 
Table \ref{tab_iso}, only available online).

Orion KL has been observed and described by a large number of authors.
Single-dish observations cannot provide details about the complex physical
structure and true spatial distribution 
that interferometric observations do reveal. 
%Single dish observations cannot provide details
%on the complex physical structure AND TRUE SPATIAL DISTRIBUTION
%that interferometric observations do REVEAL
It is not possible, for example, to track the variation 
in the excitation temperature in the hot core region that appear
to exist according to the 
vibrational temperature derived from vibrationally excited CS. 
The need from a large contribution from the extended ridge component before
OCS can reproduce the line profiles towards the hot core, relative to the
contribution of the ambient molecular cloud (far away IRc2), 
implies that at the border between the hot core and the 
extended ridge, there is a physical structure with density and temperature 
gradients that the single-dish telescopes cannot resolve. 
Higher angular resolution observations are necessary to describe this source
in detail and to reveal its physical structure.
Interferometric instruments such as Plateau de Bure, SMA, or
the future ALMA are necessary to avoid spectral confusion related to varying
physical conditions inside large beams and to resolve the source structure.
Nevertheless, the 30m line survey of Orion will provide a consistent
determination of column densities of all species with emission above 0.1 K.
   
\begin{acknowledgements}
We thank the Spanish MEC for funding support through grants
AYA2003-2785, AYA2006-14876, AYA2009-07304, ESP2004-665 and
AP2003-4619 (M. A.), Consolider project CSD2009-00038
the DGU of the Madrid Community government for support under IV-PRICIT
project S-0505/ESP-0237 (ASTROCAM). Javier R. Goicoechea was
supported by a \textit{Ram\'on y Cajal} research contract
from the Spanish MICINN and co-financed by the European Social Fund. 
\end{acknowledgements}

\listofobjects

\longtab{3}{
\begin{longtable}{llllllll}%t3
\caption{\label{tab_lines} OCS observed line parameters}\\
\hline\hline
Molecule & Observed & $T^*_A$ & $\int T^*_A dv$ &
Transition & Rest & E$_{up}$ & S$_{ij}$\\ 
 & v$_{LSR}$ (km s$^{-1}$) & (K) & (K km s$^{-1}$) & $J$ & frequency (MHz) &
(K) & \\
\hline
\endfirsthead
\caption{continued.}\\
\hline\hline
Molecule & Observed & $T^*_A$ & $\int T^*_A dv$ &
Transition & Rest & E$_{up}$ & S$_{ij}$\\ 
 & v$_{LSR}$ (km s$^{-1}$) & (K) & (K km s$^{-1}$) & $J$ & frequency (MHz) &
(K) & \\
\hline
\endhead
\hline
\endfoot
OCS & 7.6 & 3.71 & 40.2$\pm$0.7 & 7-6 & 85139.104 & 16.3 & 7.00\\
  & 8.1 & 5.81 & 62.6$\pm$0.6 & 8-7 & 97301.208 & 21.0 & 8.00\\
  & 7.8 & 6.95 & 78$\pm$1 & 9-8 & 109463.063 & 26.3 & 9.00\\
  & 7.7 & 7.97 & 96$\pm$3 & 11-10 & 133785.898 & 38.5 & 11.0\\
  & 7.6 & 12.2 & 139$\pm$2 & 12-11 & 145946.815 & 45.5 & 12.0\\
  & 6.1 & 10.6 & 146$\pm$8 & 13-12 & 158107.358 & 53.1 & 13.0\\
  & 8.6 & 10.0 & ... & 14-13 & 170267.494 & 61.3 & 14.0\\
  & 7.5 & 10.9 & 167$\pm$10 & 17-16 & 206745.155 & 89.3 & 17.0\\
  & 7.0 & 10.2 & 151$\pm$3 & 18-17 & 218903.355 & 99.8 & 18.0\\
  & 7.2 & 12.3 & 184$\pm$10 & 19-18 & 231060.993 & 110.9 & 19.0\\
  & 6.5 & 9.73 & 163$\pm$9 & 20-19 & 243218.037 & 122.6 & 20.0\\
  & 6.7 & 10.2 & 121$\pm$3 & 21-20 & 255374.457 & 134.8 & 21.0\\
  & 7.7 & 20.3$^{(1)}$ & ... & 22-21 & 267530.220 & 
147.7 & 22.0\\
  & 7.7 & 11.1 & 118$\pm$6 & 23-22 & 279685.296 & 161.1 & 23.0\\
\hline\\
OC$^{34}$S & 7.1 & 0.32 & 2.3$\pm$0.2 & 7-6 & 83057.970  & 15.9 & 7.00\\
  & 6.8 & 0.46 & 5.2$\pm$0.7 & 8-7 & 94922.798 & 20.5 & 8.00\\
  & 8.7 & 0.61 & 5.3$\pm$0.6 & 9-8 & 106787.389 & 25.6 & 9.00\\
  & 7.2 & 1.46$^{(2)}$ & ... & 11-10 & 130515.735  & 
37.6 & 11.0\\
  & 7.6 & 1.15 & 10$\pm$1 & 12-11 & 142379.431 & 44.4 & 12.0\\
  & 1.9 & 3.06$^{(3)}$ & ... & 13-12 & 154242.770  & 
51.8 & 13.0\\
  & 7.8 & 1.70 & 15$\pm$1 & 14-13 & 166105.722 & 59.8 & 14.0\\
  & 6.9 & 1.41 & 8.3$\pm$0.6 & 15-14 & 177968.256 & 68.3 & 15.0\\
  & 6.6 & 1.31 & 11.4$\pm$0.8 & 17-16 & 201691.955 & 87.1 & 17.0\\
  & 6.4 & 1.49 & 15$\pm$1 & 18-17 & 213553.060  & 97.4 & 18.0\\
  & 7.3 & 1.75 & 15.5$\pm$0.9 & 19-18 & 225413.628 & 108.2 & 19.0\\
  & 7.4 & 1.79 & 18$\pm$2 & 20-19 & 237273.631 & 119.6 & 20.0\\
  & 7.7 & 2.77$^{(3)}$ & ... & 21-20 & 249133.037 & 
131.5 & 21.0\\
  & 5.9 & 1.84 & 18.9$\pm$0.9 & 22-21 & 260991.818 & 144.1 & 22.0\\
  & 7.4 & 2.03 & 23$\pm$1 & 23-22 & 272849.944 & 157.2 & 23.0\\
\hline\\
OC$^{33}$S & 7.5 & 0.086 & 1.27$\pm$0.10 & 7-6 & 84067.083 & 16.1 & 7.00\\
  & (4) & ... & ... & 8-7 & 96076.057 & 20.8 & 8.00\\
  & 7.0 & 0.15 & 1.12$\pm$0.11 & 9-8 & 108084.786 & 25.9 & 9.00\\
  & (4) & ... & ... & 11-10 & 132101.391 & 38.0 & 11.0\\
  & (5) & ... & ... & 12-11 & 144109.205 & 45.0 & 12.0\\
  & 7.2 & 0.61$^{(3)}$ & ... & 13-12 & 156116.654  & 
52.5 & 13.0\\
  & 8.7 & 0.56$^{(6)}$ & ... & 14-13 & 168123.706 & 
60.5 & 14.0\\
  & (1) & ... & ... & 17-16 & 204142.179 & 88.2 & 17.0\\
  & 7.3 & 0.27 & 2.2$\pm$0.2 & 18-17 & 216147.340  & 98.6 & 18.0\\
  & 6.8 & 0.66$^{(7)}$ & ... & 19-18 & 228151.952 & 
109.5 & 19.0\\
  & 7.2 & 0.73$^{(8)}$ & ... & 20-19 & 240155.985 & 
121.0 & 20.0\\
  & 10.1 & 1.03$^{(9)}$ & ... & 21-20 & 252159.408 & 
133.1 & 21.0\\
  & 7.1 & 0.79$^{(1)}$ & ... & 22-21 & 264162.190 & 
145.8 & 22.0\\
  & 6.7 & 0.75$^{(10)}$ & ... & 23-22 & 276164.302 & 
159.1 & 23.0\\
\hline\\
O$^{13}$CS & 7.8 & 0.14 & 1.09$\pm$0.14 & 7-6 & 84865.153 & 16.3 & 7.00\\
  & 7.8 & 0.26 & ... & 8-7 & 96988.123 & 20.9 & 8.00\\
  & 7.2 & 0.28 & 1.8$\pm$0.2 & 9-8 & 109110.844 & 26.2 & 9.00\\
  & 6.8 & 0.43$^{(11)}$ & ... & 11-10 & 133355.415 & 38.4 & 11.0\\
  & 8.2 & 0.65 & 5.5$\pm$0.4 & 12-11 & 145477.201 & 45.4 & 12.0\\
  & 7.3 & 0.68 & 5.4$\pm$0.5 & 13-12 & 157598.615 & 52.9 & 13.0\\
  & 7.5 & 0.89 & ... & 14-13 & 169719.623 & 61.1 & 14.0\\
  & (5) & ... & ... & 17-16 & 206079.906 & 89.0 & 17.0\\
  & 6.1 & 1.00$^{(13)}$ & ... & 18-17 & 218198.984 & 
99.5 & 18.0\\
  & 7.4 & 0.93 & 12$\pm$1 & 19-18 & 230317.500 & 110.5 & 19.0\\
  & (12) & ... & ... & 20-19 & 242435.425 & 122.2 & 20.0\\
  & 6.9 & 1.97$^{(14)}$ & ... & 21-20 & 254552.727 & 
134.4 & 21.0\\
  & 7.7 & 0.80 & 5.5$\pm$0.4 & 22-21 & 266669.375 & 147.2 & 22.0\\
  & 6.7 & 0.95 & 6.3$\pm$0.6 & 23-22 & 278785.337 & 160.6 & 23.0\\
\hline\\
$^{18}$OCS & noise level & ... & ... & 7-6 & 79866.447 & 15.3 & 7.00\\
  & noise level & ... & ... & 8-7 & 91275.396 & 19.7 & 8.00\\
  & 7.9 & 0.05 & ... & 9-8 & 102684.127 & 24.6 & 9.00\\
  & 9.3 & 0.18$^{(15)}$ & ... & 10-9 & 114092.613 & 
30.1 & 10.0\\
  & 7.3 & 0.31$^{(15)}$ & ... & 12-11 & 136908.743 & 
42.7 & 12.0\\
  & 9.1 & 0.11 & ... & 13-12 & 148316.332 & 49.8 & 13.0\\
  & 7.2 & 0.10 & ... & 14-13 & 159723.567 & 57.5 & 14.0\\
  & 9.6 & 0.17$^{(16)}$ & ... & 15-14 & 171130.422 & 
65.7 & 15.0\\
  & 3.2 & 0.65$^{(17)}$$^,$ $^{(18)}$ & ... & 18-17 & 
205348.431 & 93.6 & 18.0\\
  & 7.1 & 0.50$^{(3)}$ & ... & 19-18 & 216753.491 & 
104.0 & 19.0\\
  & (19) & ... & ... & 20-19 & 228158.034 & 115.0 & 20.0\\
  & 8.2 & 0.23$^{(2)}$ & ... & 21-20 & 239562.034 & 
126.5 & 21.0\\
  & 6.6 & 0.45$^{(4)}$ & ... & 22-21 & 250965.463 & 
138.5 & 22.0\\
  & 6.8 & 0.26$^{(15)}$ & ... & 23-22 & 262368.293 & 
151.1 & 23.0\\
  & (1) & ... & ... & 24-23 & 273770.498 & 164.3 & 24.0\\
\hline\\
O$^{13}$C$^{34}$S$^{(20)}$ & 5.3 & 0.25 & ... & 20-19 & 
236429.714 & 119.2 & 20.0\\
  & 7.9 & 0.15 & ... & 23-22 & 271879.480 & 156.6 & 23.0\\
\hline\\
$^{17}$OCS$^{(20)}$ & 7.3 & 0.09 & ... & 23-22 & 270589.988 & 
155.9 & 23.0\\
\hline\\
OC$^{36}$S$^{(20)}$ & 4.7 & 0.06 & ... & 12-11 & 
139184.392 & 43.4 & 12.0 \\
  & 7.1 & 0.03 & ... & 13-12 & 150781.546 & 50.7 & 13.0\\
  & 7.9 & 0.06 & ... & 14-13 & 162378.329 & 58.5 & 14.0\\
  & 8.3 & 0.17 & ... & 23-22 & 266727.978 & 153.6 & 23.0\\
\hline\\
OCS $\nu_2$ = 1$^+$ & 10.2 & 0.067$^{(21)}$ & ... & 7-6 & 
85331.836 & 15.8 & 6.86\\
  & 5.8 & 0.040 & 0.29$\pm$0.13 & 8-7 & 97521.462 & 20.5 & 7.88\\
  & 7.2 & 0.068 & 0.69$\pm$0.07 & 9-8 & 109710.834 & 25.7 & 8.89\\
  & 5.8 & 0.11 & 1.14$\pm$0.08 & 11-10 & 134088.688 & 38.0 & 10.9\\ 
  & 5.3 & 0.15 & 2.1$\pm$0.3 & 12-11 & 146277.106 & 45.0 & 11.9\\
  & 6.6 & 0.22 & 2.05$\pm$0.12 & 13-12 & 158465.143 & 52.7 & 12.9\\
  & 7.5 & 0.46$^{(8)}$ & ... & 14-13 & 170652.766 & 
60.8 & 13.9\\
  & 7.9 & 1.19$^{(22)}$ & ... & 17-16 & 207212.838 & 
88.9 & 16.9\\
  & (23) & ... & ... & 18-17 & 219398.490 & 99.5 & 17.9\\
  & 5.6 & 0.44 & 4.3$\pm$0.4 & 19-18 & 231583.570 & 110.6 & 18.9\\
  & 6.0 & 0.65$^{(8)}$ & ... & 20-19 & 243768.045 & 
122.3 & 19.9\\
  & (1) & ... & ... & 21-20 & 255951.885 & 134.6 & 21.0\\
  & 1.9 & 0.60$^{(8)}$ & ... & 22-21 & 268135.057 & 
147.4 & 22.0\\
  & 7.5 & 0.47 & 4.2$\pm$0.3 & 23-22 & 280317.529 & 160.9 & 23.0\\
\hline\\
OCS $\nu_2$ = 1$^-$ & 10.0 & 0.048$^{(3)}$ & ... & 7-6 & 
85242.782 & 15.8 & 6.86\\
  & 6.5 & 0.048 & 0.30$\pm$0.06 & 8-7 & 97419.688 & 20.5 & 7.88\\
  & (23) & ... & ... & 9-8 & 109596.341 & 25.7 & 8.89\\
  & 7.3 & 0.14 & 1.4$\pm$0.2 & 11-10 & 133948.759 & 38.0 & 10.9\\ 
  & (23) & ... & ... & 12-11 & 146124.461 & 45.0 & 11.9\\
  & 5.9 & 0.26 & 4.0$\pm$0.2 & 13-12 & 158299.783 & 52.6 & 12.9\\
  & (5) & ... & ... & 14-13 & 170652.766 & 60.8 & 13.9\\
  & 7.9 & 0.54 & 4$\pm$1 & 17-16 & 206996.634 & 88.9 & 16.9\\
  & (23) & ... & ... & 18-17 &  219169.579 & 99.4 & 17.9\\
  & 6.7 & 0.38 & ... & 19-18 & 231341.953 & 110.5 & 18.9\\
  & 8.8 & 0.50$^{(10)}$ & ... & 20-19 & 243513.725 &
122.1 & 19.9\\
  & (23) & ... & ... & 21-20 & 255684.863 & 134.4 & 21.0\\
  & (8) & ... & ... & 22-21 & 267855.336 & 147.3 & 22.0\\
  & 4.9 & 0.44 & 3.8$\pm$0.2 & 23-22 & 280025.112 & 160.7 & 23.0\\
\hline\\
OCS $\nu_3$ = 1$^{(20)}$ & 7.2 & 0.06 & ... & 11-10 & 
133386.789 & 38.4 & 11.0\\
  & 7.3 & 0.10 & ... & 13-12 & 157635.614 & 53.0 & 13.0\\
 & 5.5 & 0.13 & ... & 18-17 & 218249.861 & 99.5 & 18.0\\
 & 5.6 & 0.11 & ... & 21-20 & 254611.772 & 134.4 & 21.0\\ 
\hline
\resizebox{0.01\textwidth}{!}{%
-
}
\end{longtable}
}
\onecolumn
Note.-Observed transitions of OCS, OCS vibrationally excited and 
  OCS isotopologues in the frequency range of the Orion KL
  survey. Column 1 indicates the isotopologue or the vibrational
  state, Col. 2
gives the observed (centroid) radial velocity, Col. 3 the
peak line temperature, Col. 4 the integrated line intensity, Col.
  5 the quantum numbers, Col. 6 the
assumed rest frequencies, Col. 7 the energy of the upper level, and
Col. 8 the line strength.\\
\\
(1): blended with SO$_2$\\
(2): blended with CH$_3$OCH$_3$\\
(3): blended with CH$_3$CH$_2$CN b type\\
(4): blended with CH$_3$OCOH\\
(5): blended with CH$_3$CH$_2$CN a type\\
(6): blended with g$^+$CH$_3$CH$_2$OH\\
(7): blended with HC$_3$N $\nu_7$ = 2\\
(8): blended with U line\\
(9): blended with $^{34}$SHD\\
(10): blended with NH$_2$CHO\\
(11): blended with CH$_2$CHCN $\nu_{11}$ = 1\\
(12): blended with CH$_3$OH\\
(13): blended with p-H$_2$CO\\
(14): blended with SO\\
(15): blended with CH$_3$CH$_2$CN out of plane torsion
  (Tercero et al., in preparation)\\
(16): blended with CH$_2$CHCN $\nu_{15}$ = 1\\
(17): blended with CH$_2$CH$_3$$^{13}$CN\\
(18): blended with $^{13}$CH$_2$CH$_3$CN\\
(19): blended with CH$_2$CHCN\\
(20): only for non-blended lines\\
(21): blended with c-C$_2$H$_4$O\\
(22): blended with (CH$_3$)$_2$CO\\
(23): blended with HC$_3$N $\nu_7$ = 1\\
\twocolumn

\Online

\begin{appendix}
\section{Online Figures}

% Figure 9 available electronically only
\begin{figure*}%f9 %a.1
\includegraphics[angle=0,scale=.8]{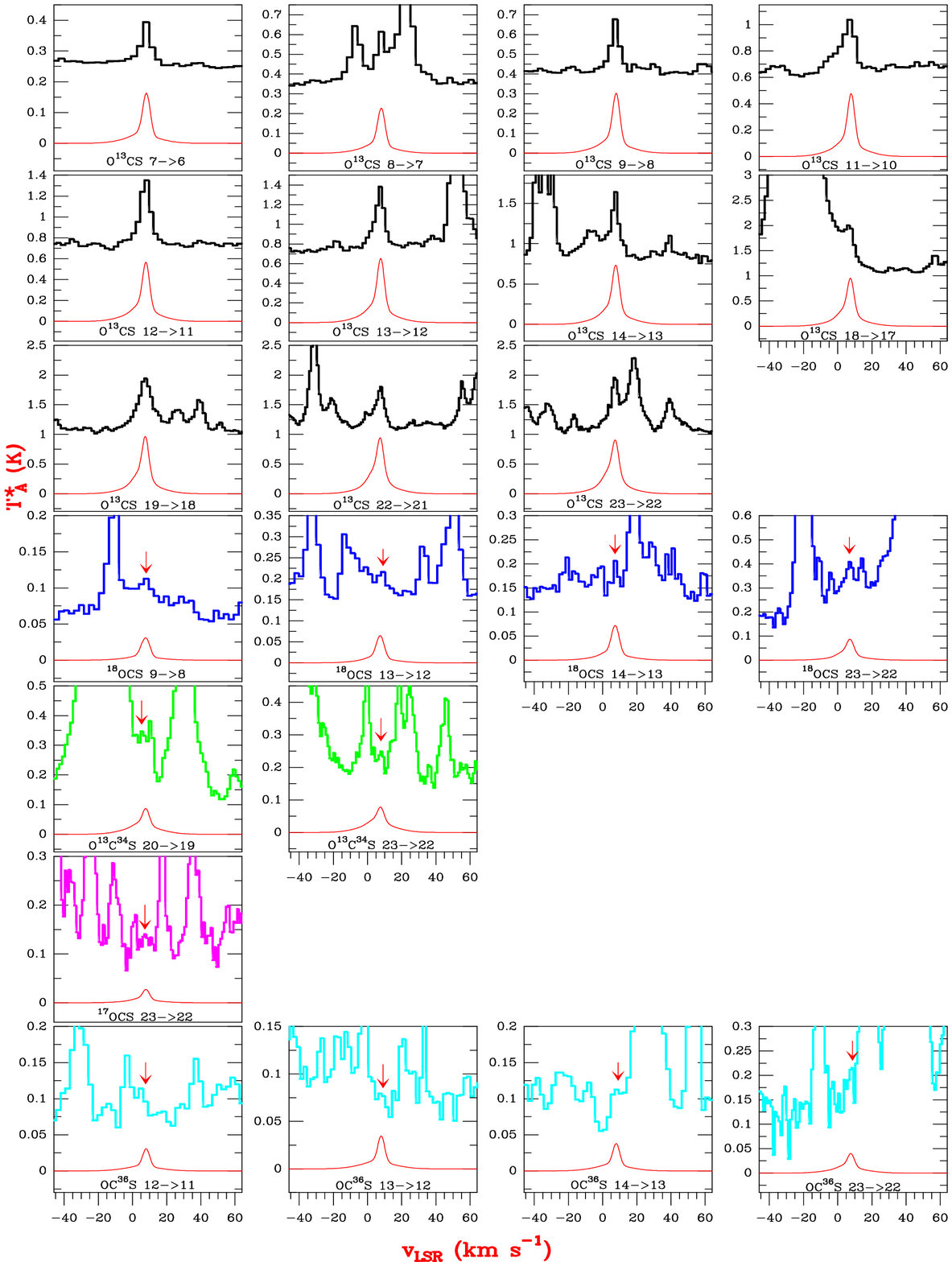}
\caption{Observed lines (offseted histogram) and model 
  (thin curves) of O$^{13}$CS, $^{18}$OCS, O$^{13}$C$^{34}S$,
  $^{17}$OCS, and OC$^{36}$S. A v$_{LSR}$ of 9 km s$^{-1}$ is
assumed.}
\label{fig_lin2}
\end{figure*}

% Figure 11 available electronically only
\begin{figure*}%f11 %a2
\includegraphics[angle=0,scale=.65]{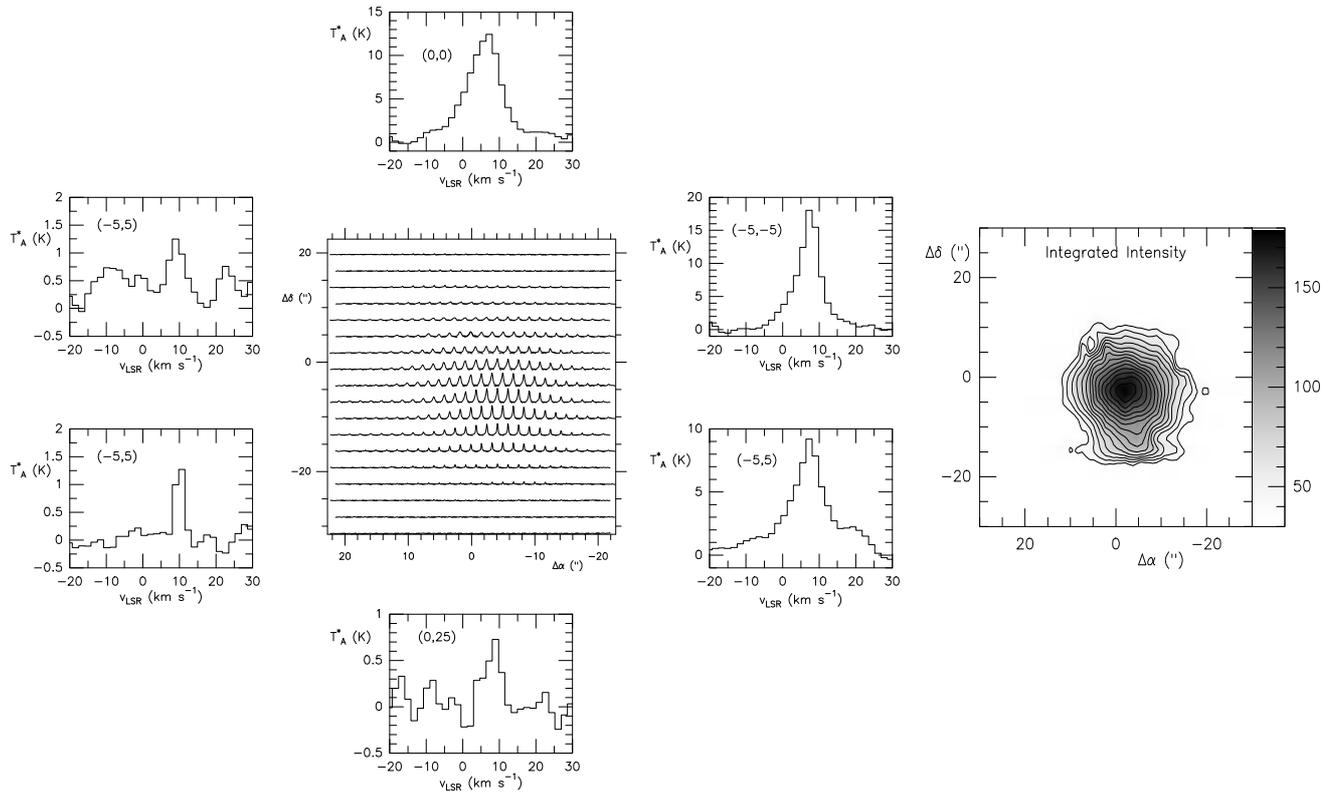}
\caption{Left panel shows the OCS J = 18-17
  line at different positions. The right panel map shows 
the total integrated intensity of this transition;
   the interval of contours is 10 K km s$^{-1}$,
  and the minimum contour is 40 K km s$^{-1}$.}
\label{fig_ocs_esp}
\end{figure*}

\begin{figure*}%f14 %a3
\includegraphics[angle=0,scale=.7]{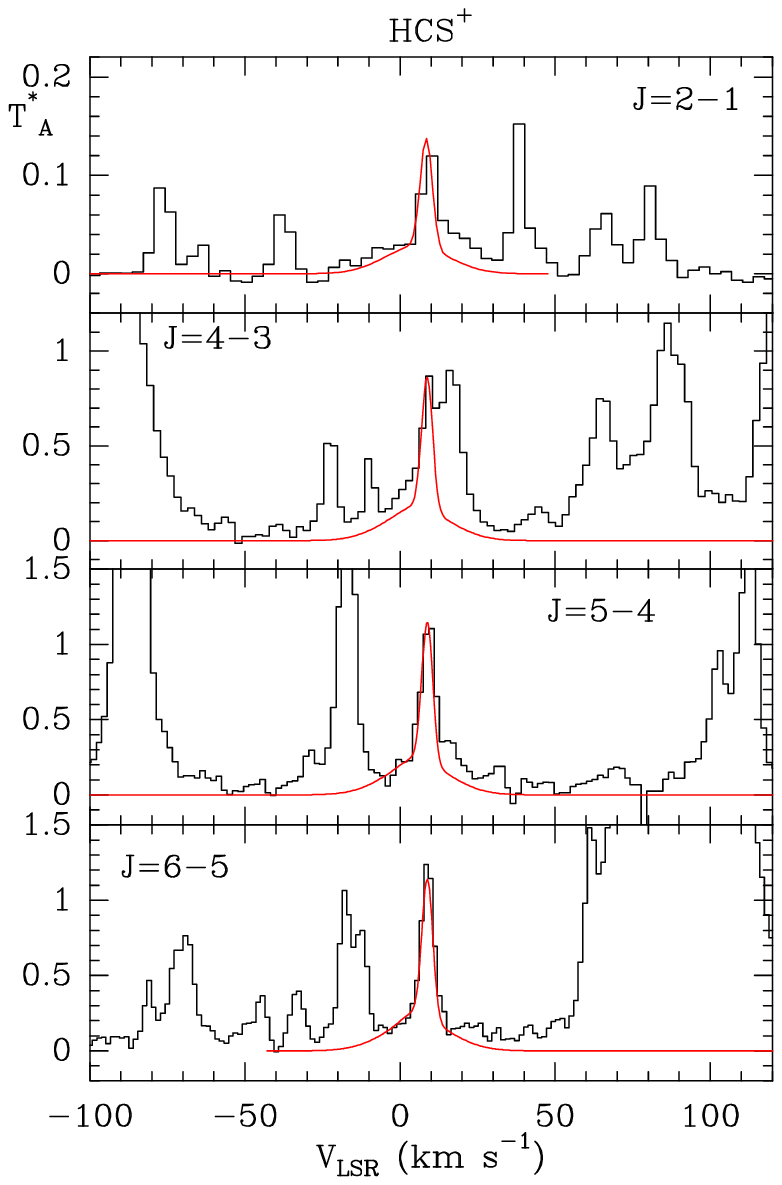}
\caption{Observed lines (histogram) and model (thin
  curves) of HCS$^+$. A v$_{LSR}$ of 9 km s$^{-1}$ is
assumed.}
\label{fig_hcs+}
\end{figure*}

\begin{figure*}%f15 %a4
\includegraphics[angle=0,scale=.8]{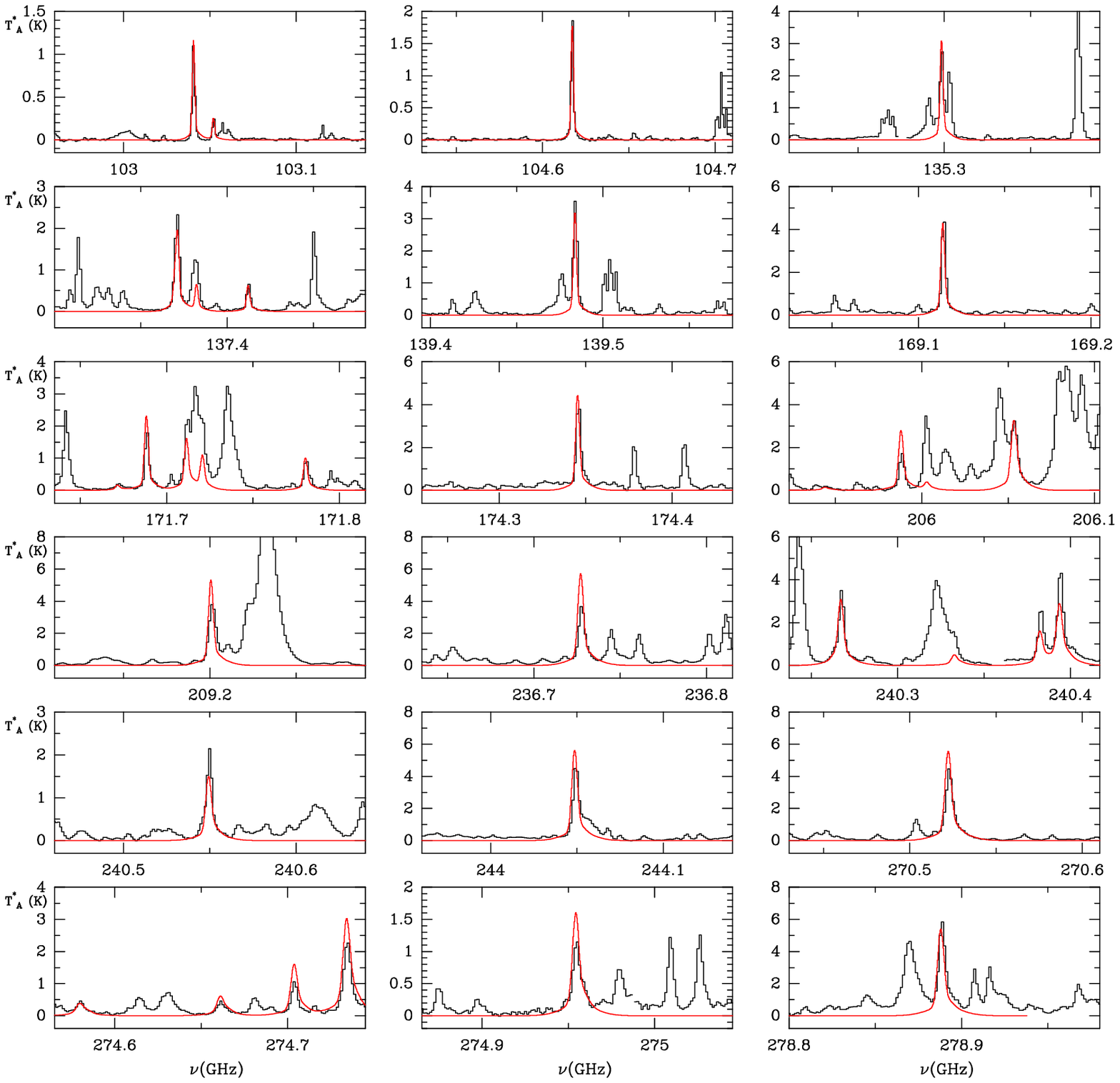}
\caption{Observed lines (histogram) and model (thin
  curves) of H$_2$CS. A v$_{LSR}$ of 9 km s$^{-1}$ is
assumed.}
\label{fig_h2cs}
\end{figure*}

\begin{figure*}%f16 %a5
\includegraphics[angle=0,scale=.8]{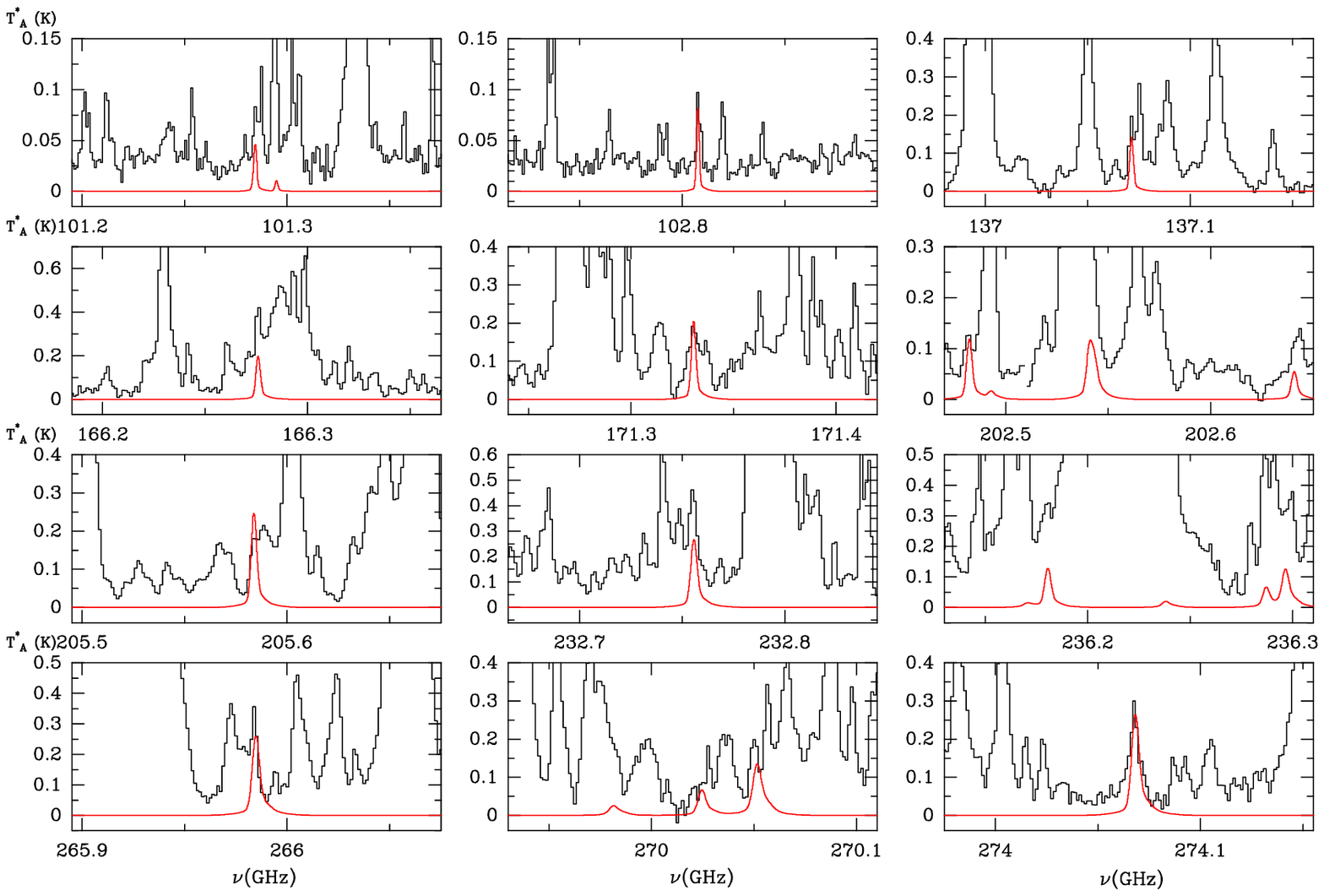}
\caption{Observed lines (histogram) and model (thin
  curves) of H$_2$C$^{34}$S. A v$_{LSR}$ of 9 km s$^{-1}$ is
assumed.}
\label{fig_h2c34s}
\end{figure*}

\begin{figure*}%f17 %a6
\includegraphics[angle=0,scale=.8]{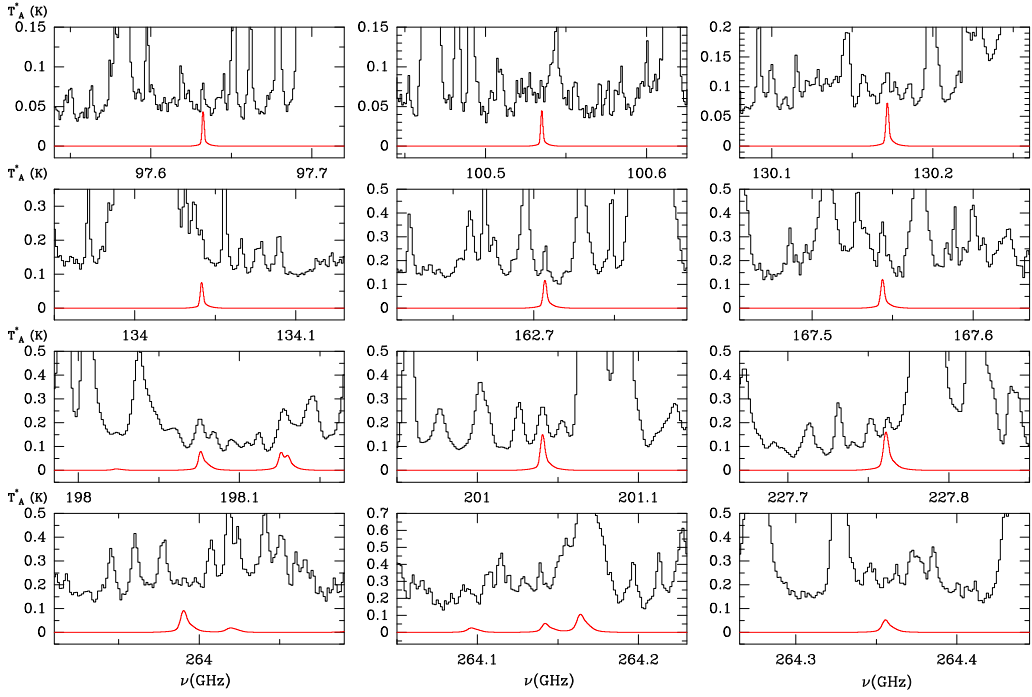}
\caption{H$_2$$^{13}$CS observed (offseted histogram) and modeled
  (continuum curves) lines. A v$_{LSR}$ of 9 km s$^{-1}$ is
assumed.}
\label{fig_h2_13cs}
\end{figure*}

\begin{figure*}%f18 %a7
\includegraphics[angle=0,scale=.8]{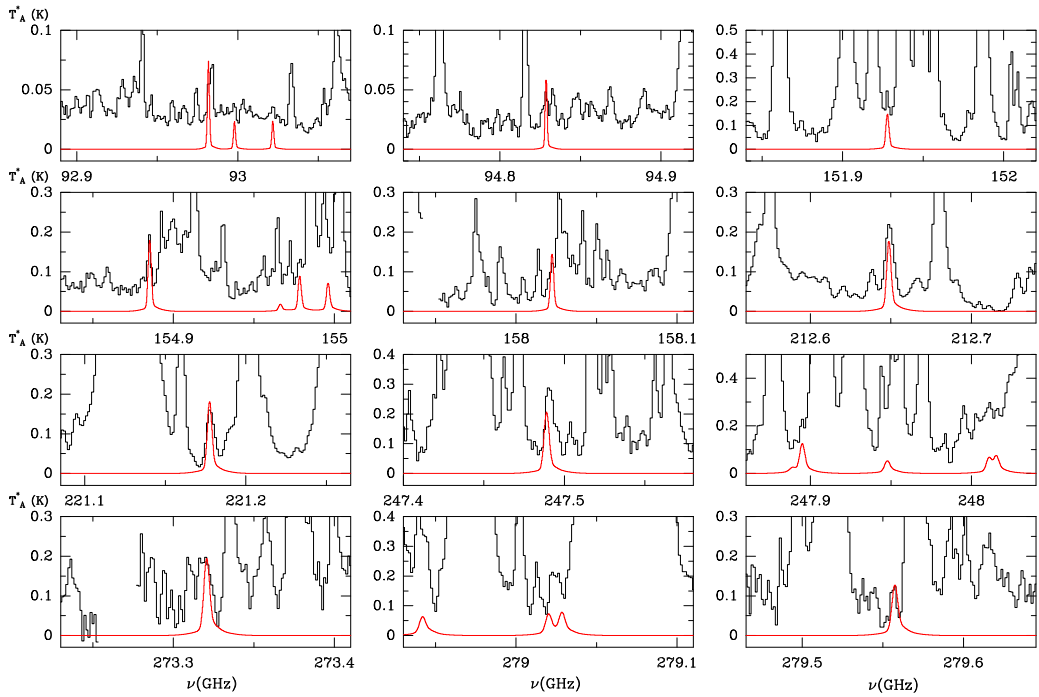}
\caption{HDCS observed (histogram) and modeled (thin
  curves) lines. A v$_{LSR}$ of 9 km s$^{-1}$ is
assumed.}
\label{fig_hdcs}
\end{figure*}

\begin{figure*} %f19 %a8
\includegraphics[angle=0,scale=.8]{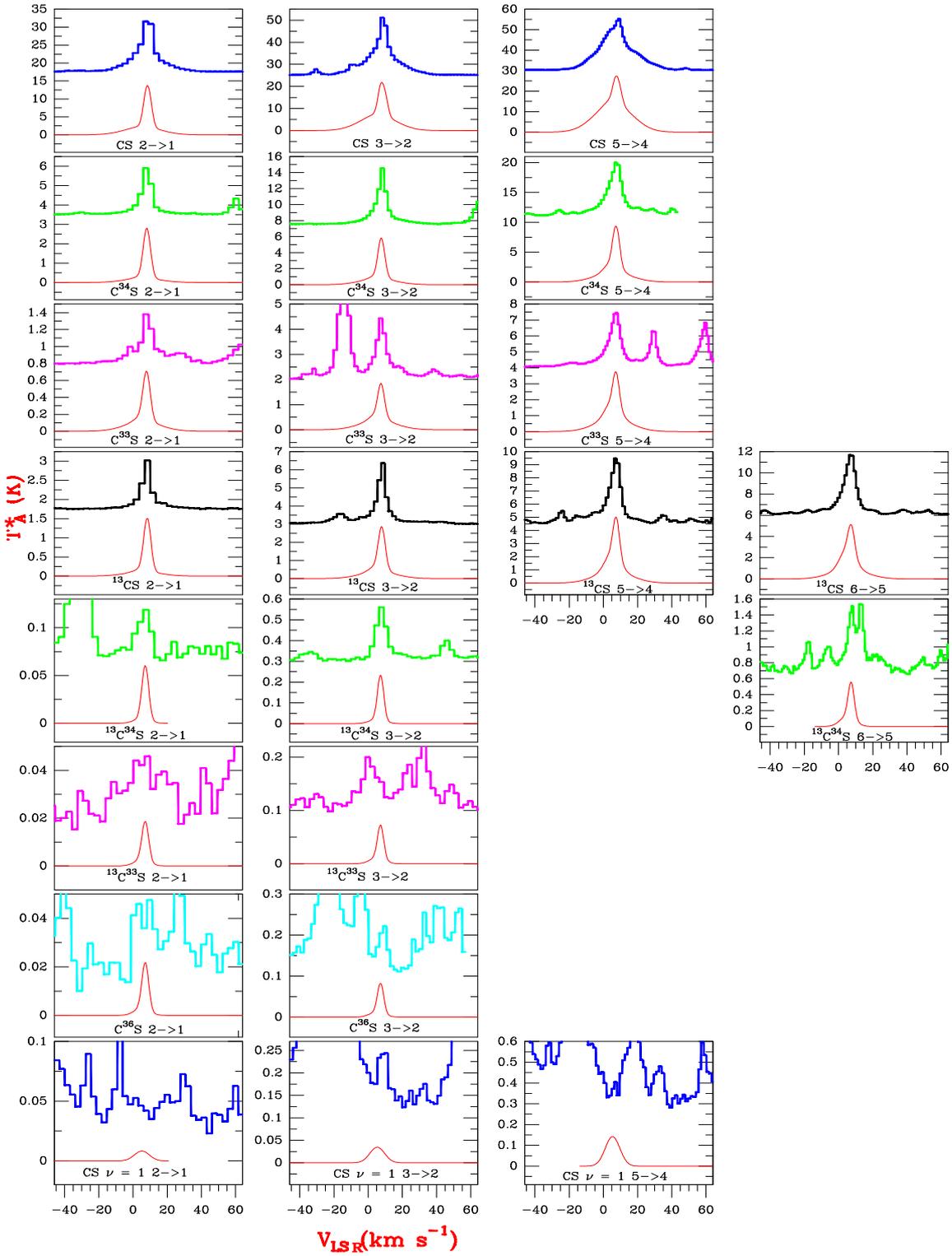}
\caption{Observed lines (offseted histogram) and modeled (thin
  curves) of CS, CS isotopologues, and CS $\textit{v}$ = 1. 
A v$_{LSR}$ of 9 km s$^{-1}$ is
assumed.}
\label{fig_cs}
\end{figure*}

\end{appendix}
\clearpage

\begin{appendix}
\section{Online Tables}
% table X available electronically only

%tabla de las componentes gaussianas de los isotopólogos de ocs.
\begin{table*}%t5 %A1
\begin{center}
\caption{OCS isotopologues and OCS vibrationally excited velocity components.\label{tab_gau2}}
\begin{tabular}{llll}
\hline 
\hline 
Species/ & \multicolumn{3}{c}{Blend of all components}\\ 
Transiion & v$_{LSR}$ (km s$^{-1}$) & $\Delta$v (km s$^{-1}$) & T$^*_A$ (K)\\
\hline 
\\
 OC$^{34}$S 7-6 & 7.1$\pm$0.3 & 7.3$\pm$0.8 & 0.30\\
 OC$^{34}$S 8-7 & 5.8$\pm$0.8 & 12.3$\pm$2.2 & 0.40\\
 OC$^{34}$S 9-8 & 7.5$\pm$0.5 & 8.3$\pm$1.3 & 0.60\\
 OC$^{34}$S 12-11 & 7.1$\pm$0.4 & 8.6$\pm$1.2 & 1.05\\
 OC$^{34}$S 14-13 & 7.1$\pm$0.3 & 9.3$\pm$0.9 & 1.52\\
 OC$^{34}$S 15-14 & 6.6$\pm$0.2 & 5.9$\pm$0.6 & 1.31\\
 OC$^{34}$S 17-16 & 6.8$\pm$0.2 & 7.7$\pm$0.5 & 1.25\\
 OC$^{34}$S 18-17 & 6.3$\pm$0.3 & 10.0$\pm$1.0 & 1.36\\
 OC$^{34}$S 19-18 & 6.3$\pm$0.2 & 9.1$\pm$0.7 & 1.61\\
 OC$^{34}$S 20-19 & 7.1$\pm$0.4 & 11.2$\pm$1.4 & 1.48\\
 OC$^{34}$S 22-21 & 6.2$\pm$0.2 & 11.1$\pm$0.7 & 1.60\\
 OC$^{34}$S 23-22 & 6.4$\pm$0.3 & 13.5$\pm$1.0 & 1.59\\
\hline \\
 OC$^{33}$S 7-6 & 6.1$\pm$0.5 & 15.1$\pm$1.7 & 0.08\\
 OC$^{33}$S 9-8 & 7.2$\pm$0.3 & 7.8$\pm$1.0 & 0.14\\
 OC$^{33}$S 18-17 & 6.6$\pm$0.4 & 9.4$\pm$1.5 & 0.22\\
\hline \\
 O$^{13}$CS 7-6 & 7.6$\pm$0.4 & 7.7$\pm$1.4 & 0.13\\
 O$^{13}$CS 8-7$^1$ & ... & ... & ...\\
 O$^{13}$CS 9-8 & 7.1$\pm$0.3 & 6.8$\pm$1.0 & 0.25\\
 O$^{13}$CS 11-10 & 6.9$\pm$0.2 & 5.7$\pm$0.6 & 0.25\\
 O$^{13}$CS 12-11 & 7.3$\pm$0.3 & 8.3$\pm$0.7 & 0.62\\
 O$^{13}$CS 13-12 & 6.8$\pm$0.4 & 9.0$\pm$1.2 & 0.56\\
 O$^{13}$CS 14-13 & 7.23$\pm$0.10 & 3.7$^1$$\pm$0.3 & 0.61\\
 O$^{13}$CS 18-17 & 4.1$^1$$\pm$0.6 & 13.9$\pm$1.4 & 1.00\\
 O$^{13}$CS 19-18 & 7.4$\pm$0.6 & 14.6$\pm$0.8 & 0.79\\
 O$^{13}$CS 22-21 & 7.3$\pm$0.2 & 7.3$\pm$0.5 & 0.72\\
 O$^{13}$CS 23-22 & 7.3$\pm$0.3 & 6.5$\pm$0.8 & 0.90\\
\hline \\
 OCS $\nu_2$=1$^-$ 8-7 & 5.7$\pm$0.5 & 5.3$\pm$1.2 & 0.05\\
 OCS $\nu_2$=1$^+$ 8-7 & 5.8$\pm$1.3 & 6.7$\pm$2.8 & 0.04\\
 OCS $\nu_2$=1$^+$ 9-8 & 5.9$\pm$0.5 & 10.0$\pm$1.3 & 0.07\\
 OCS $\nu_2$=1$^-$ 11-10 & 6.1$\pm$0.6 & 11.3$\pm$1.9 & 0.12\\
 OCS $\nu_2$=1$^+$ 11-10 & 7.2$\pm$0.3 & 9.4$\pm$0.9 & 0.11\\
 OCS $\nu_2$=1$^+$ 12-11 & 6.2$\pm$1.1 & 16.4$\pm$4.0 & 0.12\\
 OCS $\nu_2$=1$^-$ 13-12 & 7.7$\pm$0.3 & 15.5$\pm$0.7 & 0.24\\
 OCS $\nu_2$=1$^+$ 13-12 & 5.6$\pm$0.3 & 9.3$\pm$0.7 & 0.21\\
 OCS $\nu_2$=1$^-$ 17-16 & 5.3$\pm$0.6 & 8.9$\pm$1.8 & 0.47\\
 OCS $\nu_2$=1$^-$ 19-18$^1$ & ... & ... & ...\\
 OCS $\nu_2$=1$^+$ 19-18 & 5.3$\pm$0.4 & 10.3$\pm$1.3 & 0.39\\
 OCS $\nu_2$=1$^-$ 23-22 & 5.9$\pm$0.7 & 11.6$\pm$1.8 & 0.33\\
 OCS $\nu_2$=1$^+$ 23-22 & 5.7$\pm$0.3 & 9.2$\pm$0.8 & 0.43\\
\hline 
\end{tabular}
\end{center}
%% Any table notes must follow the \end{tabular} command.
Note.-v$_{LSR}$, $\Delta$v and T$^*$$_A$ of the lines of OC$^{34}$S, OC$^{33}$S,
O$^{13}$CS, and OCS $\nu_2$=1 shown in Figs. \ref{fig_lin1},
\ref{fig_lin2}, and \ref{fig_lin3} (see text, Sect. \ref{sect_res_ocs}) derived from one
Gaussian fit.\\
$^1$ \textit{Although the line is clearly detected, the obtained
  parameters are influenced by the presence of closely lines 
from other species.}\\
\end{table*}

\begin{table*} %t6 %A2
\begin{center}
\caption{HCS$^+$ Observed line parameters \label{tab_hcs+lines}}
\begin{tabular}{lllllll}
\hline
\hline 
Observed & $T^*_A$ & $\int T^*_A dv$ & Transition & Rest & E$_{up}$ &
S$_{ij}$\\ v$_{LSR}$ (km s$^{-1}$) & (K) & (K km s$^{-1}$) & $J$ & frequency (MHz) &
(K) &  \\
\hline
\\
7.7 & 0.13 & ... & 2-1 & 85347.875 & 6.1 & 2.00 \\
8.8 & 0.90 & ... & 4-3 & 170691.620 & 20.5 & 4.00 \\
8.9 & 0.74 & 9.2$\pm$0.3 & 5-4 & 213360.654 & 30.7 & 5.00 \\
8.7 & 0.65 & 4.3$\pm$0.4 & 6-5 & 256027.107 & 43.0 & 6.00 \\
\hline
%% Text for table notes should follow after the \enddata but before
%% the \end{deluxetable}. Make sure there is at least one \footnote
%% in the table for each \tablenotetext.
\end{tabular}
\end{center}
Note- Observed HCS$^+$ lines in the frequency range of the Orion KL
  survey. Column 1
gives the observed (centroid) radial velocities, Col. 2 the
peak line temperature, Col. 3 the integrated line intensity, Col.
  4 the quantum numbers, Col. 5 the
assumed rest frequencies, Col. 6 the energy of the upper level, and
Col. 7 the line strength.\\
\end{table*}

\clearpage

\longtab{3}{
\begin{longtable}{llllllll} %t7 %A3
\caption{\label{tab_h2cslines} H$_2$CS observed line parameters} \\
\hline
\hline
Molecule & Observed & $T^*_A$ & $\int T^*_A dv$ & Transition & Rest 
& E$_{up}$ &
S$_{ij}$\\  & v$_{LSR}$ (km s$^{-1}$) & (K) & (K km s$^{-1}$) &
$J$$_{K,K'}$ 
& frequency (MHz) &
(K) &  \\
\endfirsthead
\caption{continued.}\\
\hline\hline
Molecule & Observed & $T^*_A$ & $\int T^*_A dv$ & Transition & Rest 
& E$_{up}$ &
S$_{ij}$\\  & v$_{LSR}$ (km s$^{-1}$) & (K) & (K km s$^{-1}$) &
$J$$_{K,K'}$ 
& frequency (MHz) &
(K) &  \\
\hline
\endhead
\hline
\endfoot
\hline
o-H$_2$CS & 6.8 & 1.60$^{(1,2)}$ & 10.4$\pm$0.3 & 3$_{1,3}$-2$_{1,2}$ & 101477.750 & 8.1 & 2.67 \\
 & 7.5 & 1.85 & 9.5$\pm$0.3 & 3$_{1,2}$-2$_{1,1}$ & 104616.969 & 8.4 & 2.67 \\
 & 7.2 & 2.74$^{(3)}$ & 17.0$\pm$0.9 & 4$_{1,4}$-3$_{1,3}$ & 135298.094 & 14.6 & 3.75 \\
 & 4.5$^{(4)}$ & 2.32 & 18.4$\pm$0.9 & 4$_{3,2}$-3$_{3,1}$ & 137369.250 & 120.2 & 1.75 \\
 & 4.6$^{(5)}$ & ... & ... & 4$_{3,1}$-3$_{3,0}$ & 137369.281 & 120.2 & 1.75 \\
 & 7.9 & 3.54$^{(6)}$ & 24$\pm$3 & 4$_{1,3}$-3$_{1,2}$ & 139483.422 & 15.1 & 3.75 \\
 & 6.6 & 4.33 & 25$\pm$1 & 5$_{1,5}$-4$_{1,4}$ & 169113.719 & 22.7 & 4.80 \\
 & 6.0 & 2.31 & 18$\pm$1 & 5$_{3,3}$-4$_{3,2}$ & 171710.797 & 128.5 & 3.20 \\
 & 6.2$^{(5)}$ & ... & ... & 5$_{3,2}$-4$_{3,1}$ & 171710.906 & 128.5 & 3.20 \\
 & 6.6 & 3.78 & 31$\pm$2 & 5$_{1,4}$-4$_{1,3}$ & 174344.875 & 23.5 & 4.80 \\
 & 7.7 & 4.73$^{(7)}$ & 46$\pm$8 & 6$_{1,6}$-5$_{1,5}$ & 202923.516 & 32.5 & 5.83 \\
 & 7.6 & 0.15 & ... & 6$_{5,2}$-5$_{5,1}$ & 205942.672 & 349.1 & 1.83 \\
 & $^{(5)}$ & ... & ... & 6$_{5,1}$-5$_{5,0}$ & 205942.672 & 349.1 & 1.83 \\
 & 6.5$^{(4)}$ & 3.23 & 29$\pm$2 & 6$_{3,4}$-5$_{3,3}$ & 206051.859 & 138.4 & 4.50 \\
 & 6.9$^{(5)}$ & ... & ... & 6$_{3,3}$-5$_{3,2}$ & 206052.156 & 138.4 & 4.50 \\
 & 7.6 & 4.92 & 27$\pm$1 & 6$_{1,5}$-5$_{1,4}$ & 209200.094 & 33.5 & 5.83 \\
 & 7.4 & 3.89 & 30$\pm$1 & 7$_{1,7}$-6$_{1,6}$ & 236726.437 & 43.8 & 6.86 \\
 & $^{(4)}$ & ... & ... & 7$_{5,3}$-6$_{5,2}$ & 240261.406 & 360.6 & 3.43 \\
 & $^{(4)}$ & ... & ... & 7$_{5,2}$-6$_{5,1}$ & 240261.406 & 360.6 & 3.43 \\
 & 6.0 & 4.29 & 29$\pm$2 & 7$_{3,5}$-6$_{3,4}$ & 240392.422 & 149.9 & 5.71 \\
 & 6.9$^{(5)}$ & ... & ... & 7$_{3,4}$-6$_{3,3}$ & 240393.094 & 149.9 & 5.71 \\
 & 7.9$^{(8)}$ & 4.49 & 45$\pm$1 & 7$_{1,6}$-6$_{1,5}$ & 244047.922 & 45.2 & 6.86 \\
 & 7.7 & 4.38 & 37$\pm$1 & 8$_{1,8}$-7$_{1,7}$ & 270521.469 & 56.8 & 7.87 \\
 & 7.4 & 0.48 & 1.2$\pm$0.2 & 8$_{5,4}$-7$_{5,3}$ & 274578.000 & 373.8 & 4.88 \\
 & $^{(5)}$ & ... & ... & 8$_{5,3}$-7$_{5,2}$ & 274578.000 & 373.8 & 4.88 \\
 & 6.7 & 2.36 & 19$\pm$1 & 8$_{3,6}$-7$_{3,5}$ & 274732.437 & 163.1 & 6.88 \\
 & 8.2$^{(5)}$ & ... & ... & 8$_{3,5}$-7$_{3,4}$ & 274733.781 & 163.1 & 6.88 \\
 & 7.6 & 6.20 & 49$\pm$2 & 8$_{1,7}$-7$_{1,6}$ & 278887.156 & 58.6 & 7.87 \\
\hline
p-H$_2$CS & 7.1 & 1.10 & 5.79$\pm$0.08 & 3$_{2,2}$-2$_{2,1}$ & 103039.836 & 62.6 & 1.67 \\
 & 8.7$^{(5)}$ & ... & ... & 3$_{0,3}$-2$_{0,2}$ & 103040.398 & 9.9 & 3.00 \\
 & 6.9 & 0.25 & 1.4$\pm$0.3 & 3$_{2,1}$-2$_{2,0}$ & 103051.773 & 62.6 & 1.67 \\
 & 8.3 & 2.32$^{(9)}$ & 18.3$\pm$0.4 & 4$_{0,4}$-3$_{0,3}$ & 137371.000 & 16.5 & 4.00 \\
 & 10.1$^{(10)}$ & 1.23 & 14.4$\pm$0.5 & 4$_{2,3}$-3$_{2,2}$ & 137381.906 & 69.2 & 3.00 \\
 & 6.9 & 0.65 & 3.44$\pm$0.09 & 4$_{2,2}$-3$_{2,1}$ & 137411.750 & 69.2 & 3.00 \\
 & 5.7 & 0.21 & 2.3$\pm$0.2 & 5$_{4,2}$-4$_{4,1}$ & 171670.625 & 235.4 & 1.80 \\
 & $^{(5)}$ & ... & ... & 5$_{4,1}$-4$_{4,0}$ & 171670.625 & 235.4 & 1.80 \\
 & 7.7 & 2.12 & 25$\pm$1 & 5$_{0,5}$-4$_{0,4}$ & 171687.781 & 24.7 & 5.00 \\
 & 8.3 & 2.23$^{(11)}$ & ... & 5$_{2,4}$-4$_{2,3}$ & 171720.125 & 77.4 & 4.20 \\
 & 7.8 & 1.04$^{(1)}$ & 7.9$\pm$0.5 & 5$_{2,3}$-4$_{2,2}$ & 171779.812 & 77.4 & 4.20 \\
 & 6.7 & 1.83 & 10.0$\pm$0.4 & 6$_{0,6}$-5$_{0,5}$ & 205987.328 & 34.6 & 6.00 \\
 & $^{(12)}$ & ... & ... & 6$_{4,3}$-5$_{4,2}$ & 206001.812 & 245.3 & 3.33 \\
 & $^{(12)}$ & ... & ... & 6$_{4,2}$-5$_{4,1}$ & 206001.812 & 245.3 & 3.33 \\
 & 9.0 & 3.23$^{(9)}$ & 29$\pm$2 & 6$_{2,5}$-5$_{2,4}$ & 206053.594 & 87.3 & 5.33 \\
 & $^{(13)}$ & ... & ... & 6$_{2,4}$-5$_{2,3}$ & 206158.016 & 87.3 & 5.33 \\
 & $^{(14)}$ & ... & ... & 7$_{6,2}$-6$_{6,1}$ & 240178.672 & 520.2 & 1.86 \\
 & $^{(14)}$ & ... & ... & 7$_{6,1}$-6$_{6,0}$ & 240178.672 & 520.2 & 1.86 \\
 & 7.5 & 3.84 & 27$\pm$1 & 7$_{0,7}$-6$_{0,6}$ & 240266.328 & 46.1 & 7.00 \\
 & 7.9 & 1.57 & 10$\pm$2 & 7$_{4,4}$-6$_{4,3}$ & 240331.547 & 256.9 & 4.71 \\
 & $^{(5)}$ & ... & ... & 7$_{4,3}$-6$_{4,2}$ & 240331.547 & 256.9 & 4.71 \\
 & 6.4 & 2.53 & 9.2$\pm$0.3 & 7$_{2,6}$-6$_{2,5}$ & 240381.422 & 98.8 & 6.43 \\
 & 7.2 & 2.14 & 14.5$\pm$0.5 & 7$_{2,5}$-6$_{2,4}$ & 240548.437 & 98.8 & 6.43 \\
 & 6.3 & 0.29$^{(15)}$ & 2.6$\pm$0.3 & 8$_{6,3}$-7$_{6,2}$ & 274482.656 & 533.4 & 3.50 \\
 & $^{(5)}$ & ... & ... & 8$_{6,2}$-7$_{6,1}$ & 274482.656 & 533.4 & 3.50 \\
 & 7.7 & 2.62$^{(8)}$ & ... & 8$_{0,8}$-7$_{0,7}$ & 274521.469 & 59.3 & 8.00 \\
 & 7.0 & 0.48 & 3.5$\pm$0.2 & 8$_{4,5}$-7$_{4,4}$ & 274659.656 & 270.1 & 6.00 \\
 & $^{(5)}$ & ... & ... & 8$_{4,4}$-7$_{4,3}$ & 274659.656 & 270.1 & 6.00 \\
 & 7.2 & 1.19 & 5.7$\pm$0.6 & 8$_{2,7}$-7$_{2,6}$ & 274702.812 & 112.0 & 7.50 \\
 & 7.6 & 1.23 & 9.0$\pm$0.7 & 8$_{2,6}$-7$_{2,5}$ & 274953.187 & 112.0 & 7.50 \\
\hline
o-H$_2$C$^{34}$S & 10.3$^{(14)}$ & 0.25 & 1.56$\pm$0.02 & 3$_{1,3}$-2$_{1,2}$ & 99774.110 & 8.0 & 2.67 \\
 & 8.6 & 0.07 & 0.40$\pm$0.05 & 3$_{1,2}$-2$_{1,1}$ & 102807.380 & 8.3 & 2.67 \\
 & 6.1 & 0.32$^{(16)}$ & 2.1$\pm$0.4 & 4$_{1,4}$-3$_{1,3}$ & 133027.017 & 14.4 & 3.75 \\
 & $^{(9)}$ & ... & ... & 4$_{3,2}$-3$_{3,1}$ & 135027.885 & 120.0 & 1.75 \\
 & $^{(9)}$ & ... & ... & 4$_{3,1}$-3$_{3,0}$ & 135027.913 & 120.0 & 1.75 \\
 & 8.5 & 0.20 & 1.2$\pm$0.2 & 4$_{1,3}$-3$_{1,2}$ & 137071.093 & 14.9 & 3.75 \\
 & 7.4 & 0.42 & ... & 5$_{1,5}$-4$_{1,4}$ & 166275.541 & 22.3 & 3.20 \\
 & 8.5 & 0.51$^{(14)}$ & ... & 5$_{3,3}$-4$_{3,2}$ & 168784.197 & 128.1 & 3.20 \\
 & 8.6$^{(5)}$ & ... & ... & 5$_{3,2}$-4$_{3,1}$ & 168784.297 & 128.1 & 3.20 \\
 & 9.1 & 0.19 & 2.8$\pm$0.2 & 5$_{1,4}$-4$_{1,3}$ & 171330.153 & 23.1 & 4.80 \\
 & $^{(16)}$ & ... & ... & 6$_{1,6}$-5$_{1,5}$ & 199518.608 & 31.9 & 5.83 \\
 & $^{(8,16)}$ & ... & ... & 6$_{3,4}$-5$_{3,3}$ & 202540.043 & 137.9 & 4.50 \\
 & $^{(8,16)}$ & ... & ... & 6$_{3,3}$-5$_{3,2}$ & 202540.310 & 137.9 & 4.50 \\
 & 8.5 & 0.17 & 2.8$\pm$0.3 & 6$_{1,5}$-5$_{1,4}$ & 205583.378 & 32.9 & 5.83 \\
 & 9.8$^{(14)}$ & 0.21 & 1.7$\pm$0.5 & 7$_{1,7}$-6$_{1,6}$ & 232755.152 & 43.1 & 6.86 \\
 & $^{(17)}$ & ... & ... & 7$_{3,5}$-6$_{3,4}$ & 236295.303 & 149.2 & 5.71 \\
 & $^{(17)}$ & ... & ... & 7$_{3,4}$-6$_{3,3}$ & 236295.904 & 149.2 & 5.71 \\
 & $^{(7)}$ & ... & ... & 7$_{1,6}$-6$_{1,5}$ & 239829.578 & 44.5 & 6.86 \\
 & 8.6 & 0.37 & 3.3$\pm$0.2 & 8$_{1,8}$-7$_{1,7}$ & 265984.124 & 55.9 & 7.87 \\
 & 8.7 & 0.20 & 1.6$\pm$0.2 & 8$_{3,6}$-7$_{3,5}$ & 270049.842 & 162.2 & 6.88 \\
 & 10.0$^{(5)}$ & ... & ... & 8$_{3,5}$-7$_{3,4}$ & 270051.043 & 162.2 & 6.88 \\
 & 8.9 & 0.30 & 2.4$\pm$0.2 & 8$_{1,7}$-7$_{1,6}$ & 274067.548 & 57.6 & 7.87 \\
\hline
p-H$_2$C$^{34}$S & 5.78 & 0.05 & ... & 3$_{2,2}$-2$_{2,1}$ & 101283.413 & 62.5 & 1.67 \\
 & 8.6$^{(5)}$ & ... & ... & 3$_{0,3}$-2$_{0,2}$ & 101284.357 & 9.7 & 3.00 \\
 & $^{(12)}$ & ... & ... & 3$_{2,1}$-2$_{2,0}$ & 101294.548 & 62.5 & 1.67 \\
 & $^{(18)}$ & ... & ... & 4$_{0,4}$-3$_{0,3}$ & 135030.774 & 16.2 & 4.00 \\
 & $^{(1)}$ & ... & ... & 4$_{2,3}$-3$_{2,2}$ & 135040.380 & 68.9 & 3.00 \\
 & $^{(8)}$ & ... & ... & 4$_{2,2}$-3$_{2,1}$ & 135068.215 & 68.9 & 3.00 \\
 & $^{(19)}$ & ... & ... & 5$_{4,2}$-4$_{4,1}$ & 168745.511 & 235.1 & 1.80 \\
 & $^{(19)}$ & ... & ... & 5$_{4,1}$-4$_{4,0}$ & 168745.511 & 235.1 & 1.80 \\
 & $^{(19)}$ & ... & ... & 5$_{0,5}$-4$_{0,4}$ & 168764.309 & 24.3 & 5.00 \\
 & $^{(1)}$ & ... & ... & 5$_{2,4}$-4$_{2,3}$ & 168793.771 & 77.0 & 4.20 \\
 & 8.9 & 0.27 & ... & 5$_{2,3}$-4$_{2,2}$ & 168849.431 & 77.1 & 4.20 \\
 & 8.8 & 0.14 & 1.9$\pm$0.3 & 6$_{0,6}$-5$_{0,5}$ & 202481.755 & 34.0 & 6.00 \\
 & $^{(10)}$ & ... & ... & 6$_{4,3}$-5$_{4,2}$ & 202491.902 & 244.8 & 3.33 \\
 & $^{(10)}$ & ... & ... & 6$_{4,2}$-5$_{4,1}$ & 202491.902 & 244.8 & 3.33 \\
 & $^{(8)}$ & ... & ... & 6$_{2,5}$-5$_{2,4}$ & 202542.695 & 86.8 & 5.33 \\
 & 8.1 & 0.12$^{(20)}$ & ... & 6$_{2,4}$-5$_{2,3}$ & 202640.076 & 86.3 & 5.33 \\
 & 7.0 & 0.34$^{(21)}$ & ... & 7$_{0,7}$-6$_{0,6}$ & 236179.915 & 45.4 & 7.00 \\
 & $^{(6)}$ & ... & ... & 7$_{4,4}$-6$_{4,3}$ & 236236.815 & 256.1 & 4.71 \\
 & $^{(6)}$ & ... & ... & 7$_{4,3}$-6$_{4,2}$ & 236236.816 & 256.1 & 4.71 \\
 & 8.8 & 0.62$^{(22)}$ & 6$\pm$1 & 7$_{2,6}$-6$_{2,5}$ & 236286.259 & 98.1 & 6.43 \\
 & 6.7 & 0.71$^{(23)}$ & ... & 7$_{2,5}$-6$_{2,4}$ & 236442.013 & 98.1 & 6.43 \\
 & 10.2$^{(15)}$ & 0.64 & 6.3$\pm$0.3 & 8$_{0,8}$-7$_{0,7}$ & 269855.613 & 58.3 & 8.00 \\
 & 8.4 & 0.21 & ... & 8$_{4,5}$-7$_{4,4}$ & 269980.001 & 269.1 & 6.00 \\
 & $^{(5)}$ & ... & ... & 8$_{4,4}$-7$_{4,3}$ & 269980.003 & 269.1 & 6.00 \\
 & 7.3 & 0.19 & ... & 8$_{2,7}$-7$_{2,6}$ & 270023.571 & 111.1 & 7.50 \\
 & 9.8 & 0.71$^{(24,25)}$ & 6.10$\pm$0.15 & 8$_{2,6}$-7$_{2,5}$ & 270257.089 & 111.1 & 7.50 \\
\hline
o-H$_2$$^{13}$CS & 8.2 & 0.03 & 0.187$\pm$0.013 & 3$_{1,3}$-2$_{1,2}$ & 97632.227 & 7.8 & 2.67 \\
 & 9.8 & 0.03 & ... & 3$_{1,2}$-2$_{1,1}$ & 100534.773 & 8.1 & 2.67 \\
 & 9.3 & 0.03 & 0.198$\pm$0.008 & 4$_{1,4}$-3$_{1,3}$ & 130171.461 & 14.1 & 3.75 \\
 & $^{(1,26)}$ & ... & ... & 4$_{3,2}$-3$_{3,1}$ & 132084.812 & 119.9 & 1.75 \\
 & $^{(1,26)}$ & ... & ... & 4$_{3,1}$-3$_{3,0}$ & 132084.828 & 119.9 & 1.75 \\
 & 8.7 & 0.13 & ... & 4$_{1,3}$-3$_{1,2}$ & 134041.265 & 14.5 & 3.75 \\
 & 7.3 & 0.19 & 1.2$\pm$0.2 & 5$_{1,5}$-4$_{1,4}$ & 162706.562 & 21.9 & 4.80 \\
 & $^{(12)}$ & ... & ... & 5$_{3,3}$-4$_{3,2}$ & 165105.172 & 127.8 & 3.20 \\
 & $^{(12)}$ & ... & ... & 5$_{3,2}$-4$_{3,1}$ & 165105.250 & 127.8 & 3.20 \\
 & 8.1 & 0.27 & 2.8$\pm$0.2 & 5$_{1,4}$-4$_{1,3}$ & 167543.375 & 22.6 & 4.80 \\
 & 5.3$^{(26)}$ & 0.22 & 1.16$\pm$0.04 & 6$_{3,4}$-5$_{3,3}$ & 198124.937 & 137.3 & 4.50 \\
 & 5.6$^{(5)}$ & ... & ... & 6$_{3,3}$-5$_{3,2}$ & 198125.172 & 137.3 & 4.50 \\
 & 7.2 & 0.24 & 1.29$\pm$0.08 & 6$_{1,5}$-5$_{1,4}$ & 201039.969 & 32.2 & 5.83 \\
 & 7.8 & 0.32 & 1.3$\pm$0.2 & 7$_{1,7}$-6$_{1,6}$ & 227760.297 & 42.2 & 6.86 \\
 & 9.4 & 0.11 & ... & 7$_{3,5}$-6$_{3,4}$ & 231143.984 & 148.4 & 5.71 \\
 & 10.1$^{(5)}$ & ... & ... & 7$_{3,4}$-6$_{3,3}$ & 231144.516 & 148.4 & 5.71 \\
 & 8.7 & 0.31$^{(1)}$ & 4.2$\pm$0.9 & 7$_{1,6}$-6$_{1,5}$ & 234529.969 & 43.5 & 6.86 \\
 & $^{(8)}$ & ... & ... & 8$_{1,8}$-7$_{1,7}$ & 260276.937 & 54.7 & 7.87 \\
 & $^{(6,27)}$ & ... & ... & 8$_{3,6}$-7$_{3,5}$ & 264162.156 & 161.1 & 6.88 \\
 & $^{(6,27)}$ & ... & ... & 8$_{3,5}$-7$_{3,4}$ & 264163.219 & 161.1 & 6.88 \\
 & $^{(11)}$ & ... & ... & 8$_{1,7}$-7$_{1,6}$ & 268012.219 & 56.3 & 7.87 \\
\hline
p-H$_2$$^{13}$CS & 7.1 & 0.02 & ... & 3$_{0,3}$-2$_{0,2}$ & 99077.867 & 9.5 & 3.00 \\
 & $^{(14)}$ & ... & ... & 4$_{0,4}$-3$_{0,3}$ & 132089.922 & 15.9 & 4.00 \\
 & $^{(1,27)}$ & ... & ... & 4$_{2,3}$-3$_{2,2}$ & 137381.906 & 68.7 & 3.00 \\
 & $^{(17)}$ & ... & ... & 4$_{2,2}$-3$_{2,1}$ & 132123.062 & 68.7 & 3.00 \\
 & $^{(12)}$ & ... & ... & 5$_{0,5}$-4$_{0,4}$ & 165090.062 & 23.8 & 5.00 \\
 & $^{(12)}$ & ... & ... & 5$_{2,4}$-4$_{2,3}$ & 165115.641 & 76.6 & 4.20 \\
 & 7.1 & 0.53$^{(11)}$ & ... & 5$_{2,3}$-4$_{2,2}$ & 165166.531 & 76.6 & 4.20 \\
 & 9.7$^{(14)}$ & 0.18 & 0.86$\pm$0.12 & 6$_{0,6}$-5$_{0,5}$ & 198075.344 & 33.3 & 6.00 \\
 & $^{(26)}$ & ... & ... & 6$_{2,5}$-5$_{2,4}$ & 198129.422 & 86.1 & 5.33 \\
 & $^{(1,8)}$ & ... & ... & 6$_{2,4}$-5$_{2,3}$ & 198218.469 & 86.1 & 5.33 \\
 & $^{(28)}$ & ... & ... & 7$_{0,7}$-6$_{0,6}$ & 231042.781 & 44.4 & 7.00 \\
 & $^{(29)}$ & ... & ... & 7$_{2,6}$-6$_{2,5}$ & 231138.109 & 97.2 & 6.43 \\
 & $^{(12)}$ & ... & ... & 7$_{2,5}$-6$_{2,4}$ & 231280.531 & 97.2 & 6.43 \\
 & 6.5 & 0.09 & ... & 8$_{0,8}$-7$_{0,7}$ & 263989.437 & 57.0 & 8.00 \\
 & 7.0 & 0.15 & ... & 8$_{2,7}$-7$_{2,6}$ & 264140.875 & 109.9 & 7.50 \\
 & 7.5 & 0.19 & ... & 8$_{2,6}$-7$_{2,5}$ & 264354.406 & 109.9 & 7.50 \\
\hline
HDCS & 7.2 & 0.03 & ... & 3$_{1,3}$-2$_{1,2}$ & 91171.039 & 17.7 & 2.67 \\
 & 9.3 & 0.03 & ... & 3$_{0,3}$-2$_{0,2}$ & 92981.592 & 8.9 & 3.00 \\
 & 9.0 & 0.02 & 0.12$\pm$0.05 & 3$_{1,2}$-2$_{1,1}$ & 94828.495 & 18.1 & 2.67 \\
 & 9.1 & 0.25$^{(16)}$ & ... & 5$_{1,5}$-4$_{1,4}$ & 151927.558 & 30.9 & 4.80 \\
 & 8.1 & 0.22 & 1.4$\pm$0.2 & 5$_{0,5}$-4$_{0,4}$ & 154885.030 & 22.3 & 5.00 \\
 & 8.3 & 0.15 & ... & 5$_{2,4}$-4$_{2,3}$ & 154978.123 & 58.2 & 4.20 \\
 & $^{(11)}$ & ... & ... & 5$_{3,3}$-4$_{3,2}$ & 154995.479 & 103.1 & 3.20 \\
 & $^{(11)}$ & ... & ... & 5$_{3,2}$-4$_{3,1}$ & 154995.858 & 103.1 & 3.20 \\
 & 7.6 & 0.08 & 0.66$\pm$0.05 & 5$_{2,3}$-4$_{2,2}$ & 155096.966 & 58.3 & 4.20 \\
 & 8.2 & 0.12 & ... & 5$_{1,4}$-4$_{1,3}$ & 158022.086 & 31.7 & 4.80 \\
 & $^{(2)}$ & ... & ... & 7$_{1,7}$-6$_{1,6}$ & 212648.337 & 49.8 & 6.86 \\
 & 5.6$^{(6)}$ & 0.48 & ... & 7$_{0,7}$-6$_{0,6}$ & 216662.428 & 41.6 & 7.00 \\
 & $^{(16)}$ & ... & ... & 7$_{2,6}$-6$_{2,5}$ & 216931.367 & 77.6 & 6.43 \\
 & 10.2$^{(1,14)}$ & 0.22 & ... & 7$_{3,5}$-6$_{3,4}$ & 217003.261 & 122.5 & 5.71 \\
 & 13.3$^{(5)}$ & ... & ... & 7$_{3,4}$-6$_{3,3}$ & 217005.536 & 122.5 & 5.71 \\
 & $^{(1,30,31)}$ & ... & ... & 7$_{2,5}$-6$_{2,4}$ & 217263.690 & 77.6 & 6.43 \\
 & 8.6 & 0.19 & 1.58$\pm$0.11 & 7$_{1,6}$-6$_{1,5}$ & 221177.077 & 41.5 & 6.86 \\
 & 8.7 & 0.39 & ... & 8$_{1,8}$-7$_{1,7}$ & 242991.143 & 61.5 & 7.87 \\
 & 7.0$^{(16)}$ & 0.29 & 1.4$\pm$0.3 & 8$_{0,8}$-7$_{0,7}$ & 247488.469 & 53.5 & 8.00 \\
 & 6.9 & 0.17 & ... & 8$_{5,4}$-7$_{5,3}$ & 247887.785 & 277.8 & 4.88 \\
 & $^{(5)}$ & ... & ... & 8$_{5,3}$-7$_{5,2}$ & 247887.785 & 277.8 & 4.88 \\
 & 7.7 & 0.23$^{(16)}$ & ... & 8$_{2,7}$-7$_{2,6}$ & 247894.427 & 89.5 & 7.50 \\
 & $^{(1)}$ & ... & ... & 8$_{4,5}$-7$_{4,4}$ & 247946.853 & 197.1 & 6.00 \\
 & $^{(1)}$ & ... & ... & 8$_{4,4}$-7$_{4,3}$ & 247946.868 & 197.1 & 6.00 \\
 & 7.6 & 0.20 & ... & 8$_{3,6}$-7$_{3,5}$ & 248010.224 & 134.4 & 6.88 \\
 & 8.1 & 0.15 & ... & 8$_{3,5}$-7$_{3,4}$ & 248014.773 & 134.4 & 6.88 \\
 & $^{(10)}$ & ... & ... & 8$_{2,6}$-7$_{2,5}$ & 248392.331 & 89.5 & 7.50 \\
 & 8.4 & 0.40$^{(14)}$ & ... & 8$_{1,7}$-7$_{1,6}$ & 252735.000 & 63.6 & 7.87 \\
 & 9.0 & 0.20 & ... & 9$_{1,9}$-8$_{1,8}$ & 273320.111 & 74.6 & 8.89 \\
 & $^{(11)}$ & ... & ... & 9$_{0,9}$-8$_{0,8}$ & 278264.807 & 66.9 & 9.00 \\
 & 6.6 & 0.14 & ... & 9$_{6,4}$-8$_{6,3}$ & 278797.261 & 389.5 & 5.00 \\
 & $^{(5)}$ & ... & ... & 9$_{6,3}$-8$_{6,2}$ & 278797.261 & 389.5 & 5.00 \\
 & $^{(8)}$ & ... & ... & 9$_{2,8}$-8$_{2,7}$ & 278846.646 & 102.9 & 8.56 \\
 & $^{(11)}$ & ... & ... & 9$_{5,5}$-8$_{5,4}$ & 278870.647 & 291.1 & 6.22 \\
 & $^{(11)}$ & ... & ... & 9$_{5,4}$-8$_{5,3}$ & 278870.647 & 291.1 & 6.22 \\
 & 7.2 & 0.34 & ... & 9$_{4,6}$-8$_{4,5}$ & 278940.873 & 210.5 & 7.22 \\
 & 7.3$^{(5)}$ & ... & ... & 9$_{4,5}$-8$_{4,4}$ & 278940.909 & 210.5 & 7.22 \\
 & 6.8$^{(32)}$ & 0.35 & ... & 9$_{3,7}$-8$_{3,6}$ & 279019.418 & 147.8 & 8.00 \\
 & 8.2 & 0.52 & ... & 9$_{3,6}$-8$_{3,5}$ & 279027.753 & 147.8 & 8.00 \\
 & 8.2 & 0.14 & ... & 9$_{2,7}$-8$_{2,6}$ & 279556.774 & 103.0 & 8.56 \\
\hline
\resizebox{0.009\textwidth}{!}{%
-
}
\end{longtable}
}
\onecolumn
Note.-Observed transitions of H$_2$CS and
  H$_2$CS isotopologues in the frequency range of the Orion KL
  survey. Column 1 indicates the isotopologue or the vibrational
  state, Col. 2
gives the observed (centroid) radial velocities, Col. 3 the
peak line temperature, Col. 4 the integrated line intensity, Col.
  5 the quantum numbers, Col. 6 the
assumed rest frequencies, Col. 7 the energy of the upper level, and
Col. 8 the line strength.\\
\\
(1): blended with CH$_3$OCOH\\
(2): blended with (CH$_3$)$_2$CO\\
(3): blended with CH$_3$OD\\
(4): blended with p-H$_2$CS\\
(5): blended with the last one\\
(6): blended with SO$_2$\\
(7): blended with CH$_3$CN $\nu_8$=1\\
(8): blended with CH$_3$CH$_2$CN in the plane torsion\\
(9): blended with o-H$_2$CS\\
(10): blended with CH$_3$OCH$_3$\\
(11): blended with CH$_3$CH$_2$CN\\
(12): blended with CH$_3$OH\\
(13): blended with SO\\
(14): blended with U line\\
(15): blended with CH$_3$CH$_2$CN out of plane torsion\\
(16): blended with CH$_2$CHCN\\
(17): blended with $^{34}$SO$_2$\\
(18): blended with H$_2$CO\\
(19): blended with SH$_2$\\
(20): blended with H$_2$$^{13}$C$^{18}$O\\
(21): blended with CH$_3$CH$_2$$^{13}$CN\\
(22): blended with $^{34}$S$^{17}$O\\
(23): blended with HCC$^{13}$CN $\nu_7$=1$^+$\\
(24): blended with HCOOH\\
(25): blended with CH$_3$CHO\\
(26): blended with $^{33}$SO$_2$\\
(27): blended with OC$^{33}$S\\
(28): blended with OCS\\
(29): blended with o-H$_2$$^{13}$CS\\
(30): blended with g-CH$_3$CH$_2$OH\\
(31): blended with $^{13}$CH$_3$CH$_2$CN\\
(32): blended with CH$_3$CH$_2$C$^{15}$N\\
\twocolumn

\clearpage

%tabla de las componentes gaussianas de h2cs.
\begin{table*} %t8 %A4
\begin{center}
\caption{H$_2$CS and its isotopologues velocity components.\label{tab_gau_h2cs}}
\begin{tabular}{llll}
\hline 
\hline 
Species/ & \multicolumn{3}{c}{Narrow component}\\ 
Transition & v$_{LSR}$ (km s$^{-1}$) & $\Delta$v (km s$^{-1}$) & T$^*_A$ (K)\\
\hline 
 o-H$_2$CS 3$_{1,2}$-2$_{1,1}$ & 7.97$\pm$0.08 & 4.7$\pm$0.2 & 1.90\\
 o-H$_2$CS 4$_{1,4}$-3$_{1,3}$ & 7.74$\pm$0.13 & 5.9$\pm$0.4 & 2.71\\
 o-H$_2$CS 5$_{1,5}$-4$_{1,4}$ & 7.25$\pm$0.14 & 5.2$\pm$0.4 & 4.55\\
 o-H$_2$CS 5$_{1,4}$-4$_{1,3}$ & 7.27$\pm$0.07 & 4.7$\pm$0.2 & 3.45\\
 o-H$_2$CS 6$_{1,5}$-5$_{1,4}$ & 7.02$\pm$0.11 & 5.5$\pm$0.2 & 4.61\\
 o-H$_2$CS 7$_{1,7}$-6$_{1,6}$ & 7.31$\pm$0.05 & 4.4$\pm$0.2 & 3.09\\
 o-H$_2$CS 8$_{1,8}$-7$_{1,7}$ & 7.43$\pm$0.03 & 4.25$\pm$0.12 & 3.53\\
 o-H$_2$CS 8$_{5,4}$-7$_{5,3}$ &  &  & \\
 and 8$_{5,3}$-7$_{5,2}$ & 6.70$\pm$0.10 & 3.9$\pm$0.4 & 0.29\\
 o-H$_2$CS 8$_{1,7}$-7$_{1,6}$ & 7.36$\pm$0.04 & 3.79$\pm$0.14 & 4.47\\
\hline 
 p-H$_2$CS 3$_{0,3}$-2$_{0,2}$ & 8.34$\pm$0.03 & 4.90$\pm$0.07 & 1.11\\
 p-H$_2$CS 3$_{2,1}$-2$_{2,0}$ & 7.5$\pm$0.5 & 5$\pm$1 & 0.25\\
 p-H$_2$CS 4$_{2,2}$-3$_{2,1}$ & 7.34$\pm$0.05 & 5.54$\pm$0.08 & 0.58\\
 p-H$_2$CS 5$_{0,5}$-4$_{0,4}$ & 7.65$\pm$0.03 & 3.66$\pm$0.09 & 1.89\\
 p-H$_2$CS 6$_{0,6}$-5$_{0,5}$ & 7.26$\pm$0.11 & 5.8$\pm$0.3 & 1.64\\
 p-H$_2$CS 7$_{0,7}$-6$_{0,6}$ & 7.38$\pm$0.04 & 3.52$\pm$0.13 & 3.10\\
 p-H$_2$CS 7$_{4,4}$-6$_{4,3}$ & & & \\
 and 7$_{4,3}$-6$_{4,2}$ & 7.6$\pm$0.6 & 8$\pm$1 & 1.17\\
 p-H$_2$CS 7$_{2,6}$-6$_{2,5}$ & 7.02$\pm$0.03 & 4.10$\pm$0.09 & 2.11\\
 p-H$_2$CS 7$_{2,5}$-6$_{2,4}$ & 7.27$\pm$0.05 & 3.6$\pm$0.2 & 1.55\\
 p-H$_2$CS 8$_{0,8}$-7$_{0,7}$ & 7.67$\pm$0.09 & 5.13$\pm$0.03 & 2.07\\
 p-H$_2$CS 8$_{4,5}$-7$_{4,4}$ & & & \\
 and 8$_{4,4}$-7$_{4,3}$ & 6.7$\pm$0.2 & 8.6$\pm$0.6 & 0.38\\
 p-H$_2$CS 8$_{2,7}$-7$_{2,6}$ & 7.62$\pm$0.07 & 3.9$\pm$0.2 & 0.95\\
 p-H$_2$CS 8$_{2,6}$-7$_{2,5}$ & 7.48$\pm$0.08 & 3.1$\pm$0.3 & 0.71\\
\hline 
 o-H$_2$C$^{34}$S 3$_{1,2}$-2$_{1,1}$ & 7.6$\pm$0.4 & 6$\pm$1 & 0.06\\
 o-H$_2$C$^{34}$S 4$_{1,3}$-3$_{1,2}$ & 8.9$\pm$0.5 & 6$\pm$1 & 0.18\\
 o-H$_2$C$^{34}$S 6$_{1,5}$-5$_{1,4}$ & 9.0$\pm$0.2 & 2.9$\pm$0.5 & 0.10\\
 o-H$_2$C$^{34}$S 8$_{1,8}$-7$_{1,7}$ & 9.38$\pm$0.08 & 2.1$\pm$0.2 & 0.23\\
 o-H$_2$C$^{34}$S 8$_{1,7}$-7$_{1,6}$ & 9.1$\pm$0.4 & 10$\pm$1 & 0.23\\
\hline 
 p-H$_2$C$^{34}$S 6$_{0,6}$-5$_{0,5}$ & 7.6$\pm$0.9 & 16$\pm$3 & 0.12\\
\hline 
 o-H$_2$$^{13}$CS 3$_{1,3}$-2$_{1,2}$ & 9.3$\pm$0.2 & 4.2$\pm$0.5 & 0.04\\
 o-H$_2$$^{13}$CS 5$_{1,5}$-4$_{1,4}$ & 8.1$\pm$0.4 & 7$\pm$2 & 0.16\\
 o-H$_2$$^{13}$CS 5$_{1,4}$-4$_{1,3}$ & 8.3$\pm$0.3 & 10$\pm$1 & 0.25\\
 o-H$_2$$^{13}$CS 6$_{1,5}$-5$_{1,4}$ & 8.4$\pm$0.2 & 6.6$\pm$0.4 & 0.18\\
\hline
 HDCS 3$_{1,2}$-2$_{1,1}$ & 8$\pm$1 & 5$\pm$2 & 0.02\\
 HDCS 5$_{0,5}$-4$_{0,4}$ & 8.7$\pm$0.4 & 7$\pm$1 & 0.19\\
 HDCS 5$_{2,3}$-4$_{2,2}$ & 8.4$\pm$0.3 & 7.6$\pm$0.6 & 0.08\\
 HDCS 7$_{1,6}$-6$_{1,5}$ & 7.5$\pm$0.3 & 8.7$\pm$0.8 & 0.17\\
\hline 
\end{tabular}
\end{center}
%% Any table notes must follow the \end{tabular} command.
Note.-v$_{LSR}$, $\Delta$v, and T$^*$$_A$ of selected lines of H$_2$CS and its
isotopologues derived from one Gaussian fit.
\end{table*}

\begin{table*}%t9 %A5
\begin{center}
\caption{CS observed line parameters \label{tab_cslines}}
\begin{tabular}{llllllll}
\hline
\hline 
Molecule & Observed & $T^*_A$ & $\int T^*_A dv$ & Transition & Rest 
& E$_{up}$ &
S$_{ij}$\\  & v$_{LSR}$ (km s$^{-1}$) & (K) & (K km s$^{-1}$) & $J$ 
& frequency (MHz) &
(K) &  \\
\hline
\\
CS & 7.3 & 14.0 & 196$\pm$6 & 2-1 & 97980.953 & 7.1 & 2.00 \\
 & 8.0 & 26.0 & 374$\pm$11 & 3-2 & 146969.026 & 14.1 & 3.00 \\
 & 9.1 & 25.1 & 519$\pm$21 & 5-4 & 244935.556  & 35.3 & 5.00 \\
\hline
\\
C$^{34}$S & 7.3 & 2.39 & 18.1$\pm$0.7 & 2-1 & 96412.952 & 6.9 & 2.00 \\
 & 8.2 & 7.00 & 59$\pm$3 & 3-2 & 144617.101 & 13.9 & 3.00 \\
 & 7.4 & 8.95 & 110$\pm$5 & 5-4 & 241016.089 & 34.7 & 5.00 \\
\hline \\
C$^{33}$S & 7.4 & 0.58 & 3.5$\pm$0.4 & 2-1 & 97171.991 & 7.0 & 2.00 \\
 & 7.2 & 2.43 & 21$\pm$2 & 3-2 & 145755.623 & 14.0 & 3.00 \\
 & 7.6 & 3.43 & 32$\pm$2 & 5-4 & 242913.430 & 35.0 & 5.00 \\
\hline \\
$^{13}$CS & 8.3 & 1.26 & 8.3$\pm$0.4 & 2-1 & 92494.270 & 6.7 & 2.00 \\
 & 8.5 & 3.34 & 28$\pm$2 & 3-2 & 138739.263 & 13.3 & 3.00 \\
 & 6.8 & 4.96 & 51$\pm$8 & 5-4 & 231220.685 & 33.3 & 5.00 \\
 & 6.6 & 5.64 & 57$\pm$3 & 6-5 & 277455.401 & 46.6 & 6.00 \\
\hline \\
$^{13}$C$^{34}$S & 7.4 & 0.048 & 0.46$\pm$0.04 & 2-1 & 90926.002 & 6.5 & 2.00 \\
 & 7.3 & 0.26 & 2.1$\pm$0.2 & 3-2 & 136386.933 & 13.1 & 3.00 \\
 & 6.5 & 1.57$^{(1)}$ & ... & 5-4 & 227300.518 & 32.7 & 5.00 \\
 & 7.9 & 0.91 & 1.70$\pm$0.03 & 6-5 & 272751.516 & 45.8 & 6.00 \\
\hline \\
$^{13}$C$^{33}$S & 8.2 & 0.021 & ... & 2-1 & 91685.264 & 6.6 & 2.00 \\
 & -0.4 & 0.10$^{(2)}$ & ... & 3-2 & 137525.791 & 13.2 & 3.00 \\
 & (3) & ... & ... & 5-4 & 229198.427 & 33.0 & 5.00 \\
 & (4) & ... & ... & 6-5 & 272751.516 & 45.8 & 6.00 \\
\hline \\
C$^{36}$S & 9.5 & 0.024 & ... & 2-1 & 95016.677 & 6.8 & 2.00 \\
 & 9.3 & 0.17${^(5)}$ & ... & 3-2 & 142522.755 & 13.7 & 3.00 \\
 & (6) & ... & ... & 5-4 & 237525.869 & 34.2 & 5.00 \\
\hline \\
CS $\textit{v}$ = 1 & 1.4 & 0.023 & ... & 2-1 & 97271.021 & 7.0 & 2.00 \\
 & 10.4 & 0.12 & ... & 3-2 & 145904.163 & 14.0 & 3.00 \\
 & 6.9 & 0.23 & ... & 5-4 & 243160.971 & 35.0 & 5.00 \\
\hline
\end{tabular}
\end{center}
Note- Observed transitions of CS, CS vibrationally excited, and
  CS isotopologues in the frequency range of the Orion KL
  survey. Column 1 indicates the isotopologue or the vibrational state, Col. 2
gives the observed (centroid) radial velocities, Col. 3 the
peak line temperature, Col. 4 the integrated line intensity, Col.
  5 the quantum numbers, Col. 6 the
assumed rest frequencies, Col. 7 the energy of the upper level, and
Col. 8 the line strength.\\
(1): blended with CH$_3$CH$_2$CN in the plane torsion\\
(2): blended with U line\\
(3): blended with H$^{13}$CCCN\\
(4): blended with CH$_3$OCOH\\
(5): blended with CH$_3$CH$_2$$^{13}$CN\\
(6): blended with $^{34}$SO$_2$\\ 
\end{table*}

%tabla de las componentes gaussianas de cs.
\begin{table*} %t10 %A6
\begin{center}
\caption{CS and CS isotopologues velocities \label{tab_gau3}}
\resizebox{0.9\textwidth}{!}{%
\begin{tabular}{llllllllll}
\hline 
\hline
Species & \multicolumn{3}{c}{Ridge} & \multicolumn{3}{c}{Plateau} & \multicolumn{3}{c}{Hot core}\\ 
 & v$_{LSR}$ (km s$^{-1}$) & $\Delta$v (km s$^{-1}$) & T$^*_A$ (K) &
v$_{LSR}$ (km s$^{-1}$) & $\Delta$v (km s$^{-1}$) & T$^*_A$ (K)&
v$_{LSR}$ (km s$^{-1}$) & $\Delta$v (km s$^{-1}$) & T$^*_A$ (K)\\
\hline
\\
 CS 2-1 & 8.63$\pm$0.07 & 5.7$\pm$0.2 & 10.3 & 6.7$\pm$0.3 &
 23.6$\pm$1.0 & 5.30 & (1) & ... & ...\\
 CS 3-2 & 8.63$\pm$0.15 & 4.8$\pm$0.4 & 16.5 & 7.0$\pm$0.4 &
 27.3$\pm$1.4 & 9.93 & (1) & ... & ...\\
 CS 5-4 & 9.02$\pm$0.12 & 3.8$\pm$0.4 & 8.11 & 7.9$\pm$0.3 & 30.1$\pm$0.8 & 12.5 & 4.4$\pm$0.5 & 10.4$\pm$1.1 & 7.6\\
\hline
\\
 C$^{34}$S 2-1$^{(2)}$ & 7.75$\pm$0.13 & 7.3$\pm$0.4 & 2.33 & ... &
 ... & ... & ... & ... & ...\\
 C$^{34}$S 3-2 & 8.10$\pm$0.08 & 4.2$\pm$0.3 & 5.07 & 6.8$\pm$0.4 &
 18.3$\pm$1.8 & 1.88 & (1) & ... & ...\\
 C$^{34}$S 5-4$^{(3)}$ & 7.29$\pm$0.14 & 6.3$\pm$0.4 & 5.95 &
 5.1$\pm$0.2 & 21.5$\pm$1.7 & 3.08 & ... & ... & ...\\
\hline
\\
 C$^{33}$S 2-1$^{(2)}$ & 8.1$\pm$0.3 & 6.6$\pm$0.8 & 0.50 & ... &
 ... & ... & ... & ... & ...\\
 C$^{33}$S 3-2$^{(2)}$ & 7.5$\pm$0.4 & 9.6$\pm$1.3 & 2.02 & ... &
 ... & ... & ... & ... & ...\\
 C$^{33}$S 5-4$^{(3)}$ & 6.96$\pm$0.07 & 5.6$\pm$0.2 & 2.37 &
 5.0$\pm$0.3 & 20.2$\pm$1.1 & 1.10 & ... & ... & ...\\
\hline
\\
 $^{13}$CS 2-1$^{(2)}$ & 7.78$\pm$0.14 & 6.4$\pm$0.4 & 1.23 & ... &
... & ... & ... & ... & ...\\
 $^{13}$CS 3-2 & 7.97$\pm$0.07 & 4.3$\pm$0.3 & 2.67 & 6.0$\pm$0.7 &
20.7$\pm$2.5 & 0.72 & (1) & ... & ...\\
 $^{13}$CS 5-4$^{(3)}$ & 7.1$\pm$0.2 & 5.8$\pm$0.6 & 3.82 &
3.4$\pm$2.1 & 20.6$\pm$6.5 & 1.29 & ... & ... & ...\\
 $^{13}$CS 6-5$^{(3)}$ & 7.06$\pm$0.09 & 5.7$\pm$0.3 & 3.82 &
4.6$\pm$0.2 & 15.6$\pm$0.8 & 2.06 & ... & ... & ...\\
\hline
\\
 $^{13}$C$^{34}$S 2-1$^{(2)}$ & 6.0$\pm$0.4 & 9.2$\pm$0.8 & 0.047 &
... & ... & ... & ... & ... & ...\\
 $^{13}$C$^{34}$S 3-2$^{(2)}$ & 7.5$\pm$0.3 & 8.3$\pm$0.9 & 0.24 &
... & ... & ... & ... & ... & ...\\
 $^{13}$C$^{34}$S 6-5$^{(4)}$ & 7.399$\pm$0.007 & 3.40$\pm$0.13 & 0.47
& ... & ... & ... & ... & ... & ...\\
\hline
\end{tabular}
}
%% Any table notes must follow the \end{tabular} command.
\end{center}
Note.-v$_{LSR}$, $\Delta$v, and T$^*$$_A$ of the CS and most abundant CS
isotopologues lines shown in Fig. \ref{fig_cs} (see text,
Sect. \ref{sect_res_cs}) derived from Gaussian fits.\\ 
$^{(1)}$ \textit{At 3 mm and 2 mm, the hot core component is diluted
  with the other components in the line profile.}\\
$^{(2)}$ \textit{The parameters we present have been obtained by the
  Gaussian fit of the single line.}\\
$^{(3)}$ \textit{The parameters of the ridge and plateau components are affected by the
  hot core component.}\\
$^{(4)}$ \textit{Due to line overlap only the ridge component can be fitted.}
\end{table*}

\begin{table*} %t11 %A7
\begin{center}
\caption{CCS observed lines parameters\label{tab_ccs}}
\begin{tabular}{llllll}
\hline
\hline 
Observed & $T^*_A$ & Transition & Rest & E$_{up}$ &
S$_{ij}$\\ v$_{LSR}$ (km s$^{-1}$) & (K) & N$_J$ & frequency (MHz) &
(K) &  \\
\hline
\\
7.9 & 0.053 & 6$_7$-5$_6$ & 81505.211 & 15.4 & 6.97 \\
9.4 & 0.061 & 7$_6$-6$_5$ & 86181.413 & 23.3 & 5.84 \\
8.6 & 0.043 & 7$_7$-6$_6$ & 90686.386 & 26.1 & 6.86 \\
7.7 & 0.042 & 7$_8$-6$_7$ & 93870.101 & 19.9 & 7.97 \\
9.0 & 0.040 & 8$_7$-7$_6$ & 99866.510 & 28.1 & 6.87 \\
9.7 & 0.055 & 8$_8$-7$_7$ & 103640.754 & 31.1 & 7.87 \\
6.9 & 0.063 & 8$_9$-7$_8$ & 106347.743 & 25.0 & 8.97 \\
8.2 & 0.15 & 9$_8$-8$_7$ & 113410.206 & 33.6 & 7.89 \\
7.6 & 0.17 & 10$_{11}$-9$_{10}$ & 131551.974 & 37.0 & 11.0 \\
7.6 & 0.11 & 11$_{10}$-10$_{9}$ & 140180.751 & 46.4 & 9.92 \\
9.6 & 0.15 & 11$_{11}$-10$_{10}$ & 142501.703 & 49.7 & 10.9 \\
8.5 & 0.25$^{(1)}$ & 11$_{12}$-10$_{11}$ & 144244.837 & 43.9 & 12.0 \\
8.4 & 0.10 & 12$_{11}$-11$_{10}$ & 153449.782 & 53.8 & 10.9 \\
7.8 & 0.14 & 12$_{12}$-11$_{11}$ & 155454.496 & 57.2 & 11.9 \\
7.4 & 0.58$^{(2)}$ & 12$_{13}$-11$_{12}$ & 156981.666 & 51.5 & 13.0 \\
8.7 & 0.21 & 13$_{12}$-12$_{11}$ & 166662.354 & 61.8 & 11.9 \\
7.6 & 0.24 & 13$_{13}$-12$_{12}$ & 168406.791 & 65.3 & 12.9 \\
(3) & ... & 13$_{14}$-12$_{13}$ & 169753.461 & 59.6 & 14.0 \\
7.8 & 0.40$^{(2)}$ & 16$_{15}$-15$_{14}$ & 206063.973 & 89.5 & 15.0 \\
5.9 & 0.56$^{(3)}$ & 16$_{16}$-15$_{15}$ & 207260.275 & 93.3 & 15.9 \\
10.4 & 0.50$^{(4)}$ & 16$_{17}$-15$_{16}$ & 208215.906 & 87.8 & 17.0 \\
11.2 & 0.16 & 17$_{16}$-16$_{15}$ & 219142.681 & 100.1 & 16.0 \\
8.5 & 0.090 & 17$_{17}$-16$_{17}$ & 220210.164 & 103.8 & 16.9 \\
(2) & ... & 17$_{18}$-16$_{17}$ & 221071.130 & 98.4 & 18.0 \\
(6) & ... & 18$_{17}$-17$_{16}$ & 232201.872 & 111.2 & 17.0 \\
7.5 & 0.30$^{(5)}$ & 18$_{18}$-17$_{17}$ & 233159.349 & 115.0 & 17.9 \\
8.9 & 0.17 & 18$_{19}$-17$_{18}$ & 233938.428 & 109.6 & 19.0 \\
7.1 & 0.23 & 19$_{18}$-18$_{17}$ & 245244.868 & 123.0 & 18.0 \\
(7) & ... & 19$_{19}$-18$_{18}$ & 246107.787 & 126.8 & 18.9 \\
6.7 & 0.40$^{(2)}$ & 19$_{20}$-18$_{19}$ & 246815.623 & 121.4 & 20.0 \\
(h) & ... & 20$_{19}$-19$_{18}$ & 258274.296 & 135.4 & 19.0 \\
7.9 & 0.10 & 20$_{20}$-19$_{19}$ & 259055.437 & 139.3 & 19.9 \\
10.7 & 0.15$^{(6)}$ & 20$_{21}$-19$_{20}$ & 259700.952 & 133.9 & 21.0 \\
14.0 & 0.14$^{(7)}$ & 21$_{20}$-20$_{19}$ & 271292.251 & 148.4 & 20.0 \\
(j) & ... & 21$_{21}$-20$_{20}$ & 272002.258 & 152.3 & 21.0 \\
8.4 & 0.13 & 21$_{22}$-20$_{21}$ & 272592.978 & 147.0 & 22.0 \\
\hline
\end{tabular}
\end{center}
%% Text for table notes should follow after the \enddata but before
%% the \end{deluxetable}. Make sure there is at least one \footnote
%% in the table for each \tablenotetext.
Note-  Observed transitions of CCS in the frequency range of the Orion KL
  survey. Column 1
gives the observed (centroid) radial velocities, Col. 2 the
peak line temperature, Col. 3 the quantum numbers, Col. 4 the
assumed rest frequencies, Col. 5 the energy of the upper level, and
Col. 6 the line strength.\\
(1): blended with CH$_3$CH$_2$$^{13}$CN\\
(2): blended with CH$_3$OCOH\\
(3): blended with (CH$_3$)$_2$CO\\
(4): blended with g$^+$CH$_3$CH$_2$OH\\
(5): blended with CH$_3$CH$_2$CN in the plane torsion\\
(6): blended with g$^-$CH$_3$CH$_2$OH\\
(7): blended with CH$_3$OCH$_3$\\
\end{table*}

\begin{table*} %t12 %A8
\begin{center}
\caption{CCCS observed lines parameters\label{tab_cccs}}
\begin{tabular}{llllll}
\hline
\hline 
Observed & $T^*_A$ & Transition & Rest & E$_{up}$ &
S$_{ij}$\\ v$_{LSR}$ (km s$^{-1}$) & (K) & $J$ & frequency (MHz) &
(K) &  \\
\hline
\\
4.2 & 0.021 & 14-13 & 80928.183 & 29.1 & 14.0 \\
2.1 & 0.047 & 15-14 & 86708.378 & 33.3 & 15.0 \\
(1) & ... & 16-15 & 92488.491 & 37.7 & 16.0 \\
9.0 & 0.12 & 17-16 & 98268.518 & 42.4 & 17.0 \\
(2) & ... & 18-17 & 104048.454 & 47.4 & 18.0 \\
(3) & ... & 19-18 & 109828.292 & 52.7 & 19.0 \\
7.7 & 0.046 & 20-19 & 115608.029 & 58.3 & 20.0 \\
7.0 & 0.060 & 23-22 & 132946.571 & 76.6 & 23.0 \\
4.4 & 0.12 & 24-23 & 138725.845 & 83.2 & 24.0 \\
3.5 & 0.11 & 25-24 & 144504.989 & 90.2 & 25.0 \\
9.2 & 0.20$^{(4)}$ & 26-25 & 150283.999 & 97.4 & 26.0 \\
4.7 & 0.073 & 27-26 & 156062.869 & 104.9 & 27.0 \\
(2) & ... & 28-27 & 161841.594 & 112.6 & 28.0 \\
4.9 & 0.24 & 29-28 & 167620.168 & 120.7 & 29.0 \\
(2) & ... & 30-29 & 173398.586 & 129.0 & 30.0 \\
(5) & ... & 35-34 & 202288.148 & 174.8 & 35.0 \\
(2) & ... & 36-35 & 208065.518 & 184.8 & 36.0 \\
4.3 & 0.15 & 37-36 & 213842.693 & 195.0 & 37.0 \\
3.5 & 0.40 & 38-37 & 219619.670 & 205.6 & 38.0 \\
(6), (7) & ... & 39-38 & 225396.443 & 216.4 & 39.0 \\
6.6 & 0.64 & 40-39 & 231173.006 & 227.5 & 40.0 \\
7.6 & 0.89 & 41-40 & 236949.354 & 238.8 & 41.0 \\
(8) & ... & 42-41 & 242725.482 & 250.5 & 42.0 \\
5.9 & 0.44 & 43-42 & 248501.384 & 262.4 & 43.0 \\
(9) & ... & 44-43 & 254277.056 & 274.6 & 44.0 \\
4.5 & 0.49 & 45-44 & 260052.490 & 287.1 & 45.0 \\
(1) & ... & 46-45 & 265827.684 & 299.9 & 46.0 \\
4.7 & 0.20 & 47-46 & 271602.630 & 312.9 & 47.0 \\
(10) & ... & 48-47 & 277377.324 & 326.2 & 48.0 \\
\hline
\end{tabular}
\end{center}
%% Text for table notes should follow after the \enddata but before
%% the \end{deluxetable}. Make sure there is at least one \footnote
%% in the table for each \tablenotetext.
Note- Observed transitions of CCS in the frequency range of the Orion KL
  survey. Column 1
gives the observed (centroid) radial velocities, Col. 2 the
peak line temperature, Col.
  3 the quantum numbers, Col. 4 the
assumed rest frequencies, Col. 5 the energy of the upper level, and
Col. 6 the line strength.\\
(1): blended with CH$_3$OCOH\\
(2): blended with CH$_3$CH$_2$CN\\
(3): blended with $^{33}$SO$_2$\\
(4): blended with c-C$_3$H$_2$\\
(5): blended with CH$_3$CN\\
(6): blended with g$^+$CH$_3$CH$_2$OH\\
(7): blended with $^{13}$CH$_3$OH\\
(8): blended with CH$_3$CH$_2$CN out of plane torsion\\
(9): blended with SO$_2$\\
(10): blended with t-CH$_3$CH$_2$OH\\
\end{table*}

\begin{table*} %t17 %A9
\begin{center}
\caption{Isotopologue ratios\label{tab_iso}}
\begin{tabular}{lllll|l}
\hline
\hline 
Ratio & Extended ridge & Compact ridge & Plateau & Hot core & \textit{Solar
 isotopic abundance$^{1}$} \\
\hline
OCS / OC$^{34}$S & 20$\pm$8 & $\gtrsim$4 & 15$\pm$5 & $\gtrsim$5 & $^{32}$S/$^{34}$S $\simeq$ 22.5  \\
OCS / OC$^{33}$S & 40$\pm$22 & $\gtrsim$33 & 75$\pm$29 & $\gtrsim$50 & $^{32}$S/$^{33}$S $\simeq$ 127 \\
OCS / O$^{13}$CS & 33$\pm$12 & $\gtrsim$7 & 25$\pm$9 & $\gtrsim$15 & $^{12}$C/$^{13}$C $\simeq$ 90\\
OCS / $^{18}$OCS & 200$\pm$112 & $\gtrsim$42 & 250$\pm$135 & $\gtrsim$150 & $^{16}$O/$^{18}$O $\simeq$ 499 \\
O$^{13}$CS / O$^{13}$C$^{34}$S & $\gtrsim$6 & $\gtrsim$6 & $\gtrsim$2 & $\gtrsim$14 & $^{32}$S/$^{34}$S $\simeq$ 22.5\\
OC$^{34}$S / O$^{13}$C$^{34}$S & $\gtrsim$10 & $\gtrsim$14 &
$\gtrsim$2 & $\gtrsim$42 & $^{12}$C/$^{13}$C $\simeq$ 90\\
OCS / $^{17}$OCS & $\gtrsim$400 & $\gtrsim$150 & $\gtrsim$750 & $\gtrsim$750 & $^{16}$O/$^{17}$O $\simeq$ 2625 \\
OCS / OC$^{36}$S & $\gtrsim$400 & $\gtrsim$100 & $\gtrsim$375 & $\gtrsim$500 & $^{32}$S/$^{36}$S $\simeq$ 4747 \\
O$^{13}$CS / OC$^{34}$S & 0.6$\pm$0.2 & 0.6$\pm$0.2 &
 0.6$\pm$0.2 & 0.33$\pm$0.13 & ($^{13}$C/$^{12}$C)$\times$($^{32}$S/$^{34}$S) $\simeq$ 0.25  \\
O$^{13}$CS / OC$^{33}$S & 1.2$\pm$0.7 & 4$\pm$2 & 3$\pm$2 & 3$\pm$2 &
 ($^{13}$C/$^{12}$C)$\times$($^{32}$S/$^{33}$S) $\simeq$ 1.4  \\
O$^{13}$CS / $^{18}$OCS & 6$\pm$3 & 6$\pm$3 & 10$\pm$6 & 10$\pm$6 & ($^{13}$C/$^{12}$C)$\times$($^{16}$O/$^{18}$O) $\simeq$ 5.5  \\
OC$^{33}$S / OC$^{34}$S & 0.5$\pm$0.3 & 0.1$\pm$0.2 & 0.20$\pm$0.11 & 0.1$\pm$0.2 &
 $^{33}$S/$^{34}$S $\simeq$ 0.2  \\
$^{18}$OCS / OC$^{34}$S & 0.10$\pm$0.06 & 0.10$\pm$0.06 &
 0.16$\pm$0.03 & 0.03$\pm$0.02 & ($^{18}$O/$^{16}$O)$\times$($^{32}$S/$^{34}$S) $\simeq$ 0.05  \\
$^{18}$OCS / OC$^{33}$S & 0.20$\pm$0.10 & 0.6$\pm$0.5 & 0.8$\pm$0.2 & 0.3$\pm$0.2 & ($^{18}$O/$^{16}$O)$\times$($^{32}$S/$^{33}$S) $\simeq$ 0.25  \\
\hline
o-H$_2$CS / o-H$_2$C$^{34}$S & 20$\pm$7 & 25$\pm$10 &
35$\pm$13 & 14$\pm$6 & $^{32}$S/$^{34}$S $\simeq$ 22.5\\
p-H$_2$CS / p-H$_2$C$^{34}$S & 21$\pm$8 & 25$\pm$8 &
38$\pm$14 & 17$\pm$7 & $^{32}$S/$^{34}$S $\simeq$ 22.5\\
o-H$_2$CS / o-H$_2$$^{13}$CS & 40$\pm$16 & 50$\pm$20 &
47$\pm$18 & 20$\pm$8 & $^{12}$C/$^{13}$C $\simeq$ 90\\
p-H$_2$CS / p-H$_2$$^{13}$CS & 43$\pm$16 & 50$\pm$18 &
46$\pm$17 & 20$\pm$9 & $^{12}$C/$^{13}$C $\simeq$ 90\\
HDCS / (o-H$_2$CS+p-H$_2$CS) & 0.07$\pm$0.03 & 0.040$\pm$0.012 &
0.040$\pm$0.012 & 0.050$\pm$0.02 & D/H $\simeq$ 3.4$\times$10$^{-5}$\\
o-H$_2$C$^{34}$S / o-H$_2$$^{13}$CS & 2.0$\pm$0.8 & 2.0$\pm$0.7 &
1.3$\pm$0.5 & 1.4$\pm$0.5 & ($^{34}$S/$^{32}$S)$\times$($^{12}$/C$^{13}$C) $\simeq$ 4\\
p-H$_2$C$^{34}$S / p-H$_2$$^{13}$CS & 2.0$\pm$0.8 & 2.0$\pm$0.8 &
1.2$\pm$0.4 & 1.2$\pm$0.4 & ($^{34}$S/$^{32}$S)$\times$($^{12}$/C$^{13}$C) $\simeq$ 4\\
(o-H$_2$C$^{34}$S+p-H$_2$C$^{34}$S)/ HDCS & 0.7$\pm$0.2 & 1.0$\pm$0.3 &
0.7$\pm$0.2 & 1.3$\pm$0.4 & ($^{34}$S/$^{32}$S)$\times$H/D $\simeq$
 2$\times$ 10$^{-6}$\\
(o-H$_2$$^{13}$CS+p-H$_2$$^{13}$CS)/ HDCS & 0.34$\pm$0.12 & 0.5$\pm$0.2 &
0.5$\pm$0.2 & 1.0$\pm$0.3 & ($^{13}$C/$^{12}$C)$\times$H/D $\simeq$ 3.8$\times$10$^{-7}$\\
\hline
$^{13}$CS / $^{13}$C$^{34}$S & ... & $\gtrsim$6  & ... &
$\gtrsim$15 & $^{32}$S/$^{34}$S $\simeq$ 22.5\\
C$^{34}$S / $^{13}$C$^{34}$S & ... & $\gtrsim$16 & ... & 
$\gtrsim$18 & $^{12}$C/$^{13}$C $\simeq$ 490\\
$^{13}$CS / $^{13}$C$^{33}$S & ... & $\gtrsim$21 & ... & 
$\gtrsim$30 & $^{32}$S/$^{33}$S $\simeq$ 127\\
C$^{33}$S / $^{13}$C$^{33}$S & ... & $\gtrsim$14 & ... & 
$\gtrsim$20 & $^{12}$C/$^{13}$C $\simeq$ 90\\
$^{13}$CS / C$^{34}$S & 0.7$\pm$0.2 & 0.37$\pm$0.14 &
0.6$\pm$0.2 & 0.9$\pm$0.3 & ($^{13}$C/$^{12}$C)$\times$($^{32}$S/$^{34}$S) $\simeq$ 0.25\\
$^{13}$CS / C$^{33}$S & 4.0$\pm$1.3 & 1.5$\pm$0.7 & 1.2$\pm$0.4
& 1.5$\pm$0.5 & ($^{13}$C/$^{12}$C)$\times$($^{33}$S/$^{34}$S) $\simeq$ 1.4\\
C$^{33}$S / C$^{34}$S & 0.17$\pm$0.03 & 0.25$\pm$0.10 & 0.5$\pm$0.2 & 0.6$\pm$0.2 &
$^{33}$S/$^{34}$S $\simeq$ 0.2\\
\hline 
\end{tabular}
\end{center}
%% Any table notes must follow the \end{tabular} command.
Note.-Isotopologue ratios for the OCS, H$_2$CS, and CS species obtained 
with the column density results of the Tables \ref{tab_cd}, 
\ref{tab_cd_h2cs}, and \ref{tab_cdcs}.\\
$^{1}$ \citet{and89}
\end{table*}

\begin{table*} %t18 %A10
\begin{center}
\caption{Molecular abundances\label{tab_abun}}
\begin{tabular}{lllll}
\hline
\hline 
 Region & Species & X (This work) & X$^5$ & X$^6$\\
 & & ($\times$10$^{-8}$) & ($\times$10$^{-8}$) &
 ($\times$10$^{-8}$)\\
\hline
Extended & OCS & 11.7 & $<$0.3 & ...\\
ridge$^1$ & HCS$^+$ & 0.005 & 0.02 & ...\\
 & H$_2$CS & 2.7 & 0.11 & ...\\
 & CS & 3.7 & 1.1 & 2.1\\
 & CCS & 0.01 & ... & ...\\
\hline
Compact & OCS & 5.5 & 3.0 & ...\\
ridge$^2$ & HCS$^+$ & 0.04 & 0.0013 & ...\\
 & H$_2$CS & 0.7 & 0.12 & 0.65\\
 & CS & 3.6 & 1.0 & 4.0\\
 & CCS & 0.0035 & ... & ...\\
\hline
Plateau$^3$ & OCS & 3.9 & 1.4 & ...\\
 & HCS$^+$ & 0.017 & 0.004 & ...\\
 & H$_2$CS & 0.3 & 0.08 & ...\\
 & CS & 0.7 & 0.4 & 1.2\\
 & CCS & 0.0007 & ... & ...\\
\hline
Hot & OCS & 5.2 & 1.1 & 1.7\\
core$^4$ & HCS$^+$ & 0.005 & 0.0013 & ...\\
 & H$_2$CS & 0.1 & 0.08 & ...\\
 & CS & 2.0 & 0.6 & 0.29\\
 & CCS & 0.005 & ... & ...\\
 & CCCS & 0.002 & ... & ...\\
\hline 
\end{tabular}
\end{center}
Note.-Derived molecular abundances and comparation with other works.\\
$^1$ Assuming $N$(H$_2$)=7.5$\times$10$^{22}$ cm$^{-2}$. 
$^2$ Assuming $N$(H$_2$)=7.5$\times$10$^{22}$ cm$^{-2}$.
$^3$ Assuming $N$(H$_2$)=2.1$\times$10$^{23}$ cm$^{-2}$.
$^4$ Assuming $N$(H$_2$)=4.2$\times$10$^{23}$ cm$^{-2}$.\\
$^5$ From \citet{sut95}.
$^6$ From \citet{per07}.
\end{table*}

\end{appendix}

\end{document}